\documentclass[onecolumn,sort&compress,numbers]{els-mrw} 

\usepackage{amsmath,amssymb,amsfonts,amsthm,makeidx,graphicx}
\usepackage{txfonts}
\usepackage{helvet}
\usepackage{slashed}

\begin{document}


\chapter{Hadronic tau decays}\label{chap1}

\author[1]{Antonio Rodríguez Sánchez}%

\address[1]{\orgname{Instituto de Física Corpuscular}, \orgdiv{Universitat de València — CSIC}, \orgaddress{Parque Científico, Catedrático José Beltrán 2, E-46980 Paterna, Spain}}

\maketitle


\begin{abstract}[Abstract]
	We give a pedagogical introduction to the rich phenomenology of hadronic tau decays. These decays provide a unique window into the interplay of electroweak and strong interactions at low energies, as they occur primarily via $W$ exchange after the electroweak quark current hadronizes. In this manuscript, we summarize the basic ingredients required to perform precision physics studies in this sector. We detail the derivation of the different distributions within the Standard Model, discuss how to parametrize the non-perturbative QCD dynamics, and present methods commonly used to achieve clean theoretical predictions. Additionally, we briefly review how these distributions generalize in the presence of relatively heavy particles from beyond the Standard Model. This overview thus aims to serve as a useful starting point for readers interested in understanding how the only lepton capable of decaying into hadrons does so.
\end{abstract}

\begin{keywords}
 	Tau Physics\sep Phenomenology of Particle Physics\sep Low-energy QCD \sep EW interactions \sep EFTs.
\end{keywords}


\begin{figure}[h]
	\centering
	\includegraphics[width=0.7\textwidth]{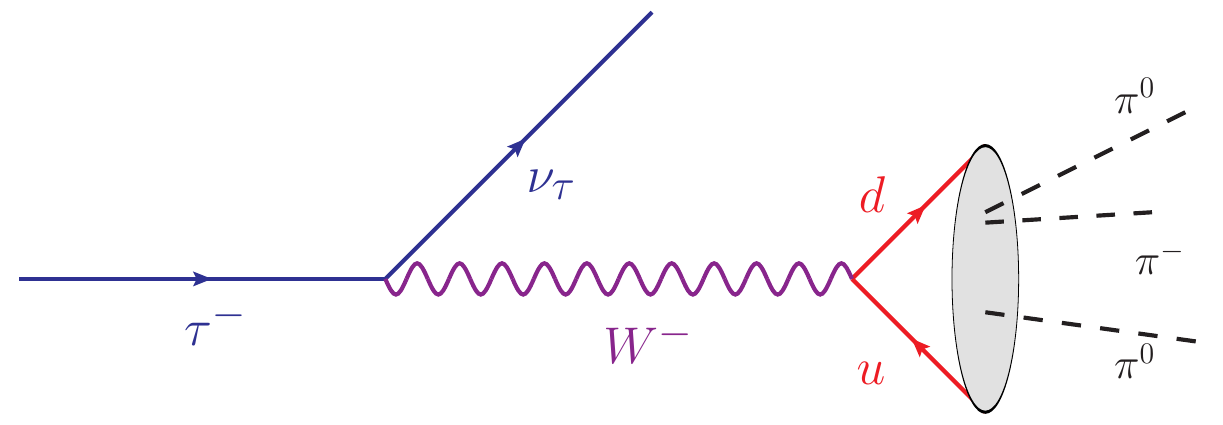}
	\caption{Feynman diagram sketching the hadronic decay of the tau.}
	\label{fig:titlepage}
\end{figure}



\section*{Objectives}
\begin{itemize}
	\item Illustrate the combination of different theoretical techniques to obtain precise predictions in hadronic tau decays.
    \item Highlight how hadronic tau decays provide a unique probe into the transition region between nonperturbative and perturbative QCD.
	\item Explore how these predictions change when potential new physics beyond the Standard Model is introduced.
\end{itemize}

\section{Introduction}\label{intro}

The tau is the heaviest known lepton, with a measured mass of~\cite{HeavyFlavorAveragingGroupHFLAV:2024ctg}
\begin{equation}
m_{\tau}=1.77694(9)\; \mathrm{GeV}\,,
\end{equation}
currently dominated by the recent determination of the Belle-II collaboration~\cite{Belle-II:2023izd}. After being produced, the tau lifetime is~\cite{HeavyFlavorAveragingGroupHFLAV:2024ctg}
\begin{equation}
\tau_\tau=290.3(5)\cdot 10^{-15}\;\mathrm{s}\,,
\end{equation}
making it a very long-lived particle relative to its mass, $\Gamma_{\tau}/m_{\tau}=1/(\tau_\tau m_\tau)\sim 10^{-12}$.

Tau leptons are heavy enough to decay into hadrons, including pseudoscalar nonstrange mesons ($M_{\pi^-}= 0.140 \, \mathrm{GeV}$), strange ones ($M_{K^{-}}=0.494\, \mathrm{GeV}$) and many combinations of them. However, they cannot decay into charmed mesons ($M_{D^{-}}=1.87 \, \mathrm{GeV}$) or baryons ($m_{p}+m_{n}=1.88 \, \mathrm{GeV}$). More than three out of five tau leptons ($64.48(32) \%$)~\cite{HeavyFlavorAveragingGroupHFLAV:2024ctg} decay into hadrons, generating a rich spectrum ranging from the pion threshold up to the tau mass. The most frequent nonstrange and strange modes are shown in table~\ref{tab:tau_decays}. These represent only a small fraction of the numerous hadronic decay modes available to the tau.
\begin{table}[b]
\centering
\begin{tabular}{l c | l c}
\toprule
$\boldsymbol{H^{-}}$ & \textbf{BR (\%)} 
& $\boldsymbol{H^{-}}$ & \textbf{BR (\%)} \\
\midrule
$\pi^- \; \pi^0$ & $25.48(09)$ 
& $\bar{K}^0 \; \pi^-$ & $0.838(14)$ \\
$\pi^-$ & $10.81(05)$
& $K^-$ & $0.696(10)$\\
$\pi^- \; \pi^0 \pi^0$ & $\phantom{0}9.25(10)$
& $K^-  \; \pi^0$ & $0.432(15)$\\
$\pi^- \; \pi^+ \pi^-$ & $\phantom{0}9.01(05)$ 
& $\bar{K}^0 \; \pi^-\pi^0$ & $0.381(13)$  \\
$\pi^- \; \pi^+ \pi^- \pi^0$ & $\phantom{0}4.49(05)$ 
& $K^- \; \pi^+\pi^-$ & $0.293(07)$  \\
$\pi^-\;  \pi^0\pi^0\pi^0$ & $\phantom{0}1.04(07)$ 
& $K^- \; \pi^0\pi^0$ & $0.063(22)$  \\
\bottomrule
\end{tabular}
\caption{Most frequent hadronic nonstrange (left) and strange (right) tau decay modes, $\tau^{-}\to H^{-} \nu_{\tau}$. Branching ratios from Ref.~\cite{HeavyFlavorAveragingGroupHFLAV:2024ctg}. We have chosen to exclude contributions to the channels arising from subsequent weak decays of the $K^0$ (alternative choices can be found in Ref.~\cite{HeavyFlavorAveragingGroupHFLAV:2024ctg}). QCD resonances, such as the $\rho$ or the $a_1$ mesons, cannot typically be separated from the hadronic continuum, and are thus included in their decay products.}
\label{tab:tau_decays}
\end{table}
The experimental determinations are fully dominated by measurements in $e^+e^- \to \tau^+\tau^-$ processes, most notably at LEP (including measurements at ALEPH~\cite{ALEPH:1990ndp}, OPAL~\cite{OPAL:1990yff} and DELPHI~\cite{DELPHI:1990cdc}), CLEO~\cite{CLEO:1991qyy}, BaBar~\cite{BaBar:2001yhh}, Belle~\cite{Belle:2000cnh}, and the corresponding updates. 

In addition to precise branching ratios, collaborations have measured numerous kinematic distributions, providing complementary information. In particular, Belle typically provides the most precise measurements of the hadronic spectral shapes, benefiting from its large statistics and well-separated $\tau$ decay products at lower center-of-mass energy. In contrast, ALEPH achieves greater precision in determining the absolute normalization, owing to its high $\tau^+\tau^-$ selection efficiency, minimal non-$\tau$ background, and finely granular detector—essential for clearly separating the closely spaced decay products at LEP energies~\cite{Davier:2005xq,Davier:2010fmf}.

This large amount of experimental evidence can be confronted with the theoretical state-of-the-art in the field. In this chapter, we provide a detailed introduction to phenomenological methods that allow us to precisely describe and predict a significant part of the experimental information from first principles. From this perspective, the corresponding experimental data becomes a clean probe of the interplay of the different fundamental forces of nature at relatively low energies, where a description of strong interactions in terms of approximately free quarks and gluons does not hold. Sections~\ref{sec:distributions},~\ref{sec:exclusive}, and~\ref{sec:inclusive}, the bulk of this chapter, describe hadronic tau decays within the Standard Model (SM) paradigm, starting from general SM distributions valid for any decay in section~\ref{sec:distributions}. Section~\ref{sec:exclusive} reviews our understanding of the different hadronic channels, including those shown in table~\ref{tab:tau_decays}, with special emphasis on the low-multiplicity modes, which are more accessible from the theoretical point of view. A different approach, summing over various channels to link them to theoretically clean objects—two-point correlation functions of quark currents—is reviewed in section~\ref{sec:inclusive}. Section~\ref{sec:bsm} describes, from an agnostic point of view, what could be different in the presence of physics beyond the Standard Model and what information about it we can extract from experimental data. The hypothetical case of measurable charged lepton flavor violation is also briefly discussed in that section. Conclusions and a brief outline of the theoretical challenges that the field needs to overcome in the future are presented in section~\ref{sec:conclusions}.


\section{Hadronic tau decay distributions}\label{sec:distributions}

The SM provides us with a very precise description of tau decays. Taus are colorless, i.e., they are singlets under SU(3)$_C$, and thus the strong force alone cannot mediate their decay. Local interactions with massless photons do emerge in the SM, $\mathcal{L} \supset -  e  A^{\mu} \bar{\tau}\gamma_{\mu} \tau$, but the number of taus is conserved at each vertex, and thus they cannot trigger their decay either. The same occurs with local interactions involving the $Z$ and the Higgs boson. The only SM mediator left, the one triggering all $\tau$ decays within the SM paradigm, is the $W$ boson. The corresponding SM charged current interaction is
\begin{equation}\label{eq:SMlag}
\mathcal{L}\supset -\frac{g}{\sqrt{2}}[W_\mu^{\dagger}(J^{\mu}_{\tau}+V_{uD}\, J_{q}^{\mu}) + W_{\mu}(J^{\mu,\dagger}_\tau + V_{uD}^*\, J^{\mu,\dagger}_q)  ] \, ,
\end{equation}
where $J_\tau^{\mu}=\bar{\nu}_{\tau,L}\gamma^{\mu}\tau_L$ is the lepton current and $J_{q,D}^{\mu}$ the quark one,
$J_{q,D}^{\mu}=\bar{u}_L\gamma^{\mu}D_L$, $D=d,s$. $g$ is the SU(2)$_{L}$ gauge coupling, $g=\frac{e}{\sin\theta_W}\approx 0.63$. Considering the very large mass of the $W$ boson, $M_W=80.4\; \mathrm{GeV}$, the kinematically allowed decay of the tau requires a second vertex, leading to the topology shown in Fig.~\ref{fig:titlepage} for the hadronic decay, and an analogous one for the leptonic one. This explains the `long' lifetime of the tau. The decay is weak; it is suppressed by a very off-shell $W$ boson propagator, leading to a parametric $m_{\tau}^4/M_{W}^4$ suppression in its width.

Given its pivotal role, let us start by explicitly finding the hadronic tau decay matrix element using the PDG phase convention~\cite{ParticleDataGroup:2024cfk}. In terms of the $S$-matrix,
\begin{equation}
\langle \nu_{\tau}(p_{\nu}) \, H^{-}(p_{H}) | \, S-\mathbb{I} \, | \tau^-(p_{\tau})\rangle = i \, (2\pi)^4 \, \delta^{4}(p_\tau-p_\nu - p_H) \, M_{\tau \to H^{-} \,\nu_\tau} \, ,
\end{equation}
where $p_H$ is the total momentum of the final hadronic state, $H^{-}=\pi^-,\pi^-\pi^0,K^{-}\eta,\dots$ Selecting the term with the two required vertices in the Dyson series of the $S$-matrix, one has
\begin{equation}\label{eq:dermatele}
\begin{aligned}
\langle \nu_\tau \, H^{-} | \,   S-\mathbb{I} \, | \tau^-\rangle &=
\underbrace{i^2}_{\text{Dyson}} \underbrace{\left(-\frac{g}{\sqrt{2}}\right)^2 V_{uD}^*}_{\mathcal{L}} \int d^{4}x \int d^{4}y \; \langle \nu_\tau \, H^-  | T\left( W_{\alpha}(x) J^{\alpha,\dagger}_{q,D}(x) W_{\beta}^{\dagger}(y) J^{\beta}_\tau(y)   \right) | \tau^- \rangle\\
&=\underbrace{(-i)}_{\text{prop}}\frac{-g^2 V_{uD}^*}{2} \int d^{4}x \int d^{4}y \int \frac{d^4 p}{(2\pi)^4} \; e^{-i(x-y)p}\frac{1}{p^{2}-M_W^2} \,  \langle \nu_\tau \, H^-|T(J_{\mu}^{\tau}(y) \, J^{\mu,\dagger}_{q,D}(x)) |\tau^- \rangle \\
&=\frac{i g^2 V_{uD}^*}{2} \, \delta^{4}(p_\tau-p_\nu-p_H) \, \frac{L_{\mu} \; H^\mu}{q^2-M_W^2}  \, ,
\end{aligned}
\end{equation}
where $q^2\equiv (p_\tau-p_\nu)^2$, $L_\mu\equiv\bar{u}_L(p_\nu,s_\nu) \, \gamma_{\mu} \, u_L(p_\tau,s_\tau)$, $H^\mu=\langle H^-| \, J^{\mu,\dagger}_{q,D}(0) \, |0\rangle $ and $2 u_L\equiv 2 \, P_L\,  u = (1-\gamma^5) \, u$. In the second line we have contracted the $W$ propagator\footnote{We have safely neglected, for the sake of simplicity, the very tiny part not proportional to $g_{\mu\nu}$, suppressed by $(m_D \, m_\tau)/M_W^2$.} and in the third we have used translation invariance of the quark current, $J_{q,D}^{\mu}(y)=e^{i\hat{P}y} \, J_{q,D}^{\mu}(0) \, e^{-i\hat{P}y}$. Thus, the matrix element characterizing the hadronic decays of the taus up to tiny EW radiative corrections is
\begin{equation}
M \equiv M_{\tau \to H^{-}\,  \nu_{\tau}}=\frac{g^2 V_{uD}^{*}}{2}\frac{L_{\mu} \, H^{\mu}}{q^2-M_{W}^2} \approx - \frac{g^2 V_{uD}^{*}}{2M_W^2} L_{\mu} \, H^{\mu} \, ,
\end{equation}
where in the last approximation we have used that $q^2<m_{\tau}^2\ll M_{W}^2$. The same result can be directly obtained from the effective Fermi Lagrangian, derived by integrating out the $W$ boson
\begin{equation}\label{eq:fermilag}
\mathcal{L}\supset -2\sqrt{2} \,  G_{F} V_{uD}^* \, J_{q,D}^{\mu,\dagger} \, J_{\mu,\tau} \, ,
\end{equation}
where $G_F=\frac{g^2}{4 \sqrt{2} M_W^2}$. This is an example of the Wilsonian Operator Product Expansion (OPE)~\cite{Wilson:1969zs}. The singularity arising in the first line of Eq.~(\ref{eq:dermatele}) as $y \to x$ is encoded in the $W$ propagator (a c-number) in the second line. $J(y)$ is regular in the neighborhood of $J(x)$. The leading term in the corresponding $y\to x$ expansion leads to the effective local interaction of Eq.~(\ref{eq:fermilag}).

Since the leptonic part of the matrix element is straightforward, and the hadronic part involves nonperturbative strong interactions, it is convenient to obtain general expressions for the hadronic tau decay distributions separating and simplifying the leptonic part as much as possible. The differential decay distribution in the tau rest frame is given by
\begin{equation}
d\Gamma= \frac{(2\pi)^4}{2m_\tau} \, |M|^2 \, d\Phi_n \, .
\end{equation}
Let us simplify $|M|^2$ by summing over the final (approximately massless and experimentally unobservable) neutrino polarizations, using $\sum_{s_{\nu}} u(p_{\nu},s_{\nu}) \bar{u}(p_{\nu},s_{\nu})=\slashed{p}_{\nu}$. Defining $L^{\alpha\beta}=L^{\alpha}L^{\beta,\dagger}$, one has
\begin{equation}
\sum_{s_\nu}|M|^2=8 G_F^2 |V_{uD}|^2   H^{\alpha}H^{\beta,\dagger} \sum_{s_\nu}L_{\alpha\beta}= 8 G_F^2 |V_{uD}|^2    H^{\alpha}H^{\beta,\dagger} \; \; \bar{u}_{L}(p_\tau,s_\tau) \; \gamma_{\beta}\slashed{p}_\nu\gamma_{\alpha}\; u_{L} (p_\tau,s_\tau)  \, .
\end{equation}
If we are interested in simplifying the expression for a tau with polarization $s_{\tau}$, we may use ($\epsilon^{0123}=1$) $\gamma_{\beta} \gamma_{\mu} \gamma_{\alpha} P_L=(g_{\beta\mu}\gamma_{\alpha}+g_{\alpha\mu}\gamma_{\beta}-g_{\beta\alpha}\gamma_{\mu}+i\epsilon_{\delta\beta\mu\alpha}\gamma^{\delta})P_L$ and $\bar{u}(p_\tau,s_\tau) \gamma^{\mu}P_L u(p_\tau,s_\tau)=p_\tau^\mu -m_\tau s_\tau^\mu\equiv L_{\tau}^{\mu}$. One finds\footnote{Equivalently one may instead first impose a fixed $s_{\tau}$ polarization for the initial tau state using the corresponding projector, $P_{s_\tau}=\frac{1+\gamma_5\slashed{s}_{\tau}}{2}$, before `summing' over tau polarizations, namely $P_{s_\tau} u(p_{\tau},s_{\tau}) \bar{u}(p_{\tau},s_{\tau})= P_{s_\tau} \sum_{s'_{\tau}}u(p_{\tau},s'_{\tau}) \bar{u}(p_{\tau},s'_{\tau})=P_{s_\tau}(\slashed{p}_{\tau}+m_{\tau})$. Indeed one trivially has $\sum_{s_\nu}L^{\alpha\beta}=\mathrm{Tr}(\gamma^{\beta}\slashed{p}_{\nu} \gamma^{\alpha} P_L P_{s_\tau}(\slashed{p}_\tau +m_{\tau}))$, leading to the same result.}
\begin{equation}\label{eq:unpol}
\sum_{s_\nu}|M|^2=8 G_F^2 |V_{uD}|^2  \;  \Big(p_{\nu}^{\alpha}L_{\tau}^{\beta}+p_{\nu}^{\beta}L_{\tau}^{\alpha}-g^{\alpha\beta} p_{\nu} \cdot L_{\tau} -i\epsilon^{\alpha\beta\gamma\delta} p_{\nu, \,\gamma} L_{\tau,\,\delta}\Big) \; \langle H^-| \, \bar{D}_L\gamma_{\alpha} u_L \, |0\rangle \, \langle 0| \, \bar{u}_L \gamma_{\beta} D_L \, |H^{+}\rangle \, .
\end{equation}
For unpolarized taus, the polarization average is easily obtained by replacing $L_{\tau}^{\mu}\to p_{\tau}^{\mu}$. 

To simplify the phase space integral, $d\varPhi_n=\delta^{4}(p_\tau-p_\nu-p_H)\frac{d^3p_\nu}{(2\pi)^3 2 E_\nu}\prod_{i=1}^{n} \frac{d^3p_{H,i}}{(2\pi)^3 2 E_{H,i}}$, one may directly start from the recursive decay form~\cite{ParticleDataGroup:2024cfk}
\begin{equation}
d\varPhi_{n}=  \; (2\pi)^3 \, ds \, \delta^{4}(p_{\tau}-q-p_\nu)  \frac{d^3p_\nu}{(2\pi)^3 2 E_\nu} \, \frac{d^3 q}{(2\pi)^3 2 q_0} \, d\phi_{n}  \; ,
\end{equation}
where $q_0$ is constrained to be $q_0=\sqrt{\vec{q}^2+s}$ and $d\phi_{n}$ is the hadronic part of the phase space integral, $d\phi_{n}=\delta^{4}(q-p_{H})\prod_{i=1}^{n} \frac{d^3p_{H,i}}{(2\pi)^3 2 E_{H,i}}$. In the tau rest frame, the Dirac delta simplifies considerably. In particular, by taking into account the constraint on $q_0$ and using the identity $\delta(f(x))=\sum_{x_0}\delta(x-x_0)/|f'(x_0)|$, one obtains
\begin{equation}
\frac{|\vec{q}|^2}{q_0 |\vec{p_{\nu}}|}\,\delta^{4}(p_{\tau}-p_\nu-q)
=\frac{|\vec{q}|}{q_0}\,\delta^{3}(\vec{q}-\vec{p}_{\nu})\,\frac{\delta\left(|\vec{q}|-\frac{m_{\tau}}{2}\left(1-\frac{s}{m_{\tau}^2}\right)\right) \, q_0}{q_0+|\vec{q}|}=\frac{1}{2}\left(1-\frac{s}{m_{\tau}^2}\right)\,\delta^{3}(\vec{q}-\vec{p}_{\nu})\,\delta\left(|\vec{q}|-\frac{m_{\tau}}{2}\left(1-\frac{s}{m_{\tau}^2}\right)\right) \, .
\end{equation}
The phase space integral then reduces to $
d\Phi_n=\frac{1}{8(2\pi)^3}ds \, d\Omega_q\left( 1-\frac{s}{m_{\tau}^2}\right)d\phi_n$ and one finds an expression for the differential distribution in the tau rest frame valid for any hadronic state,
\begin{equation}\label{eq:taurestgen}
d\Gamma= \frac{\pi}{m_\tau} \left( 1-\frac{s}{m_{\tau}^2}\right)\,  G_F^2 |V_{uD}|^2 \, (p_{\nu}^{\alpha}L_{\tau}^{\beta}+p_{\nu}^{\beta}L_{\tau}^{\alpha}-g^{\alpha\beta} p_{\nu} \cdot L_{\tau} -i\epsilon^{\alpha\beta\gamma\delta} p^{\nu}_{\gamma} L^{\tau}_{\delta}) \, ds \, d\Omega_q \;\cdot \; d\phi_n \, \langle H^-| \, \bar{D}_L\gamma_{\alpha} u_L \, |0\rangle \, \langle 0| \, \bar{u}_L \gamma_{\beta} D_L(0) \, |H^{+}\rangle \, .
\end{equation}
We may be interested in integrating over the internal kinematical degrees of freedom of a specific final hadronic state $H^{-}$. In that case, Lorentz invariance restricts the general form of the hadronic tensor to
\begin{equation}\label{eq:lorentHdec}
H^{\alpha \beta}(q)\equiv (2\pi)^3 \, \int d\phi_{n} \langle H^-| \, \bar{D}_L\gamma_{\alpha} u_L \, |0\rangle \, \langle 0| \, \bar{u}_L \gamma_{\beta} D_L(0) \, |H^{+}\rangle=(-g^{\alpha\beta}q^2+q^{\alpha}q^{\beta}) \, H^{(1)}(q^2)+q^{\alpha}q^{\beta} \, H^{(0)}(q^2) \, ,
\end{equation}
from which
\begin{equation}\label{eq:H0andH1}
H^{(0)}(q^2)=\frac{(2\pi)^3}{q^4}\int d\phi_n H_{\alpha}q^{\alpha} H_{\beta}^\dagger q^\beta \qquad \quad , \qquad \quad H^{(1)}(q^2)=-\frac{(2\pi)^3}{3q^2}\int d\phi_n H^{\alpha} \left( g_{\alpha\beta}-\frac{q_{\alpha}q_{\beta}}{q^2}\right) H^{\dagger\beta} \, .
\end{equation}
Plugging Eq.~(\ref{eq:lorentHdec}) into Eq.~(\ref{eq:taurestgen}), one finds the differential distribution for polarized taus,
\begin{equation}
\frac{d\Gamma_{\tau}}{ds\,d\cos\theta}=\frac{m_\tau^3}{8\pi}G_{F}^2 |V_{uD}|^2 \left(1-\frac{s}{m_{\tau}^2} \right)^2\left[ H^{(0)}(s)\,(1+P\cos\theta)+H^{(1)}(s)\left[ \left( 1+2\frac{s}{m_\tau^2} \right) + \left( 1-2\frac{s}{m_\tau^2} \right)P\cos\theta  \right]  \right] \, .
\end{equation}
where $P$ is the degree of polarization and $\theta$ the angle between the tau polarization and $\vec{q}=\sum_i \vec{p}_{H,i}$ in the tau rest frame.\footnote{This agrees with the inclusive version of this expression, given in Ref.~\cite{Braaten:1993ha}.} This expression explicitly shows that one can make universal predictions on the angular distributions independent of the specific hadronic states $H^{(J)}(s)$. For the unpolarized case one finds the invariant mass distribution to be, also adding the short-distance EW correction factor, $S_{\mathrm{EW}}^{\mathrm{had}}$~\cite{Marciano:1988vm,Erler:2002mv},
\begin{equation}\label{eq:invdist}
\frac{d\Gamma_{\tau}}{ds}=\frac{m_\tau^3}{4\pi}G_{F}^2\, S_{\mathrm{EW}}^{\mathrm{had}}\, |V_{uD}|^2 \left(1-\frac{s}{m_{\tau}^2} \right)^2\left[ H^{(0)}(s)+H^{(1)}(s) \left( 1+2\frac{s}{m_\tau^2} \right)    \right] \, .
\end{equation}

\section{Exclusive decays}\label{sec:exclusive}

\subsection{Form factors for hadronic tau decays}\label{sec:formfactors}

In the previous section, the hadronic part of the problem was encoded into local quark currents which, within QCD, interpolate vacuum-to-hadron transitions via non-perturbative interactions. Although, in general, we do not know how to analytically compute these matrix elements, we can use our knowledge of the symmetries preserved non-perturbatively by QCD to restrict their form. Let us review some of the most relevant ones.

\paragraph{Quark equation of motion}
The equations of motion for the flavor-octet quark currents, defined as $\bar{q}\,\lambda_{i}\Gamma \, q$ with $q^T=(u,d,s)$, $\Gamma=\{\gamma^{\mu},\gamma^{\mu}\gamma_5\}$ and $\mathrm{Tr}(\lambda_i)=0$, are preserved non-perturbatively in QCD, yielding (recall that $q=p_H$)
\begin{equation}
\begin{aligned}
q_\mu \langle H^{-}(q) |\bar{D}_L \gamma^{\mu} u_L |0\rangle &= m_D\langle H^{-}(q) |\bar{D}_R  u_L |0\rangle - m_u\langle H^{-}(q) |\bar{D}_L  u_R |0\rangle   \; , \label{eq:eom}
\end{aligned}
\end{equation}
and the same exchanging $L \leftrightarrow R$. As a consequence, we find that $H^{(0)}$ in Eq.~(\ref{eq:H0andH1}) is suppressed by two powers of the light-quark masses for finite $q^2$ values.

\paragraph{Lorentz Invariance}
Further information can be extracted depending on the type of decay. Let us focus on the final states with up to three pseudoscalar mesons, $P_{1},P_{2}, P_3$, which are the relevant asymptotic states of QCD in the isospin limit. Lorentz invariance allows us to parameterize our lack of non-perturbative knowledge using a set of scalar form factors. Given the available independent momenta, the matrix elements can be decomposed as
\begin{equation}\begin{aligned}
\langle P| \bar{D}_L \gamma^{\mu} u_L |0\rangle&=-i \, f_L\, q^\mu \, ,
\\
\langle P_1 P_2| \bar{D}_L \gamma^{\mu} u_L |0\rangle&= (\tilde{p}_{1}-\tilde{p}_2)^\mu \; F_T \, + \, \tilde{F}_L \; q^\mu  \, ,
\\
\langle P_1 P_2 P_3 | \bar{D}_L \gamma^{\mu} u_L |0\rangle&=F_1 \, (\tilde{p}_1-\tilde{p}_{3})^\mu - F_2 \, (\tilde{p}_{2}-\tilde{p}_3)^\mu \, + \, i \, F_3 \,\epsilon^{\mu\alpha\beta\gamma}p_{1\alpha} \, p_{2\beta} \, p_{3\gamma} \,  + F_L \, q^{\mu}  \, ,
&
\end{aligned}\label{eq:ffgen}
\end{equation}
where $\tilde{p}^{i}_{\mu}= (\hat{P}_{T} p_i)_{\mu} \equiv (g_{\mu\nu}-q_{\mu}q_{\nu}/q^2)p_i^{\nu}$. With this definition, the scalar functions $f_L,\tilde{F}_L$ and $F_{L}$ are longitudinal, i.e., they vanish when contracted with the projector $\hat{P}_{T,{\mu}}^{\nu}$,\footnote{Thus, the relation of Eq.~(\ref{eq:eom}) for these modes only concerns $f_L,\tilde{F}_L$ and $F_{L}$.} while the remaining functions are transversal and thus vanish when contracted with $\hat{P}_{L,{\mu}}^{\nu}=q_{\mu}q^{\nu}/q^2$. These scalar functions can depend on the different independent invariants made of the available momenta: none for the first line, one for the second, $q^2=(p_1+p_2)^2$ and three for the third, which for example can be chosen to be $q^2=(p_{1}+p_{2}+p_{3})^2$, $(p_{1}-p_{3})^2$ and $(p_{2}-p_{3})^2$. 

\paragraph{Strangeness}
Since QCD preserves flavor, there are no associated $d \leftrightarrow s$ transitions in the strong sector. Hadronic final states can be classified by strangeness, simply counting the number of kaons minus antikaons, and one can easily see whether they are mediated by $\bar{d}_L\gamma^{\mu}u_L$ or $\bar{s}_L\gamma^{\mu}u_L$. Owing to Eq.~(\ref{eq:taurestgen}), the latter are suppressed with respect to the former by a $|V_{us}|^2/|V_{ud}|^2\sim 0.05$ factor, qualitatively explaining the relative suppression observed in the right panel of table \ref{tab:tau_decays}.

\paragraph{Parity}
The form factors can be separated into vector and axial contributions, e.g. $f_{L}=\frac{f_V-f_A}{2}$ . However, pseudoscalar mesons have negative intrinsic parity,
\begin{align}\label{eq:}
\langle P_1(p_{1,\mu_1}) \, \cdots \, P_n(p_{n,\mu_n}) | &\overset{P}{\rightarrow}  \, (-1)^n \,  \langle P_1(p_1^{\mu_1}) \cdots  P_n(p_n^{\mu_n})| \; ,\\
\bar{D}\gamma^{\mu} u \, \overset{P}{\rightarrow}    \bar{D}\gamma_{\mu} u \qquad &,\qquad \bar{D}\gamma^{\mu}\gamma_5 u \, \overset{P}{\rightarrow} -   \bar{D}\gamma_{\mu} \gamma_5 u \, .
\end{align}
Taking into account that any QCD (or QED) interaction leading to the form factor preserves parity, one immediately obtains
\begin{equation}
\underbrace{F_T,\tilde{F}_L,F_3}_{\text{V}} \quad\qquad , \qquad\quad \underbrace{f_L,F_1,F_2,F_L}_{\text{A}}  \, .
\end{equation}

\paragraph{G-parity}
In the isospin limit ($m_u=m_d$), G-parity is an exact symmetry of QCD. The G-parity transformation is a discrete one obtained by composing charge conjugation and an isospin rotation that conveniently maps both pions and charged light-quark currents into themselves up to a sign, thus giving restrictions to the possible matrix elements. The transformation is defined as $G=C \cdot U_{I_2}^{\pi}$, where $C$ is charge conjugation and $U_{I_2}^{\pi}$ is an isospin rotation of $180$ degrees along the $y$ axis. 
This isospin transformation converts
\begin{equation}
u\overset{U_{I_2}^{\pi}}{\to} d \quad,\quad  d \overset{U_{I_2}^{\pi}}{\to} -u \quad,\quad \pi^+ \overset{U_{I_2}^{\pi}}{\to} -\pi^- \quad ,\quad \pi^- \overset{U_{I_2}^{\pi}}{\to} -\pi^+ \quad ,\quad \pi^0 \overset{U_{I_2}^{\pi}}{\to} -\pi^0 \, .
\end{equation}
Combined with charge conjugation, $\bar{d}\Gamma u \to \eta_{C}^{\Gamma} \, \bar{u}\Gamma d$, one easily finds
\begin{equation}
\bar{d} \Gamma u \overset{G}{\to} \eta_{G}^{\Gamma} \, \bar{d} \Gamma u \qquad \qquad ,\qquad \qquad \pi\overset{G}{\to} -\pi               \, .
\end{equation}
where $\eta_{G}^{\Gamma}=1$ for the vector and the tensor current (so that in the isospin limit cannot mediate transitions into states with odd number of pions but arbitrary etas) and $\eta_{G}^{\Gamma}=-1$ for the axial, pseudoscalar and scalar current (so that in the isospin limit cannot mediate transitions into states with even number of pions plus arbitrary etas). Additional isospin relations allow one to connect form factors appearing in different tau decay channels, or even to those appearing in $e^{+}e^{-}\to \mathrm{hadrons}$, when coupled to the $I=1$ component of the electromagnetic current, $\frac{1}{2}(\bar{u}\gamma^{\mu}u-\bar{d}\gamma^{\mu}d)$. Further estimates can be done by finding relations assuming SU(3)$_{V}$ symmetry, valid in the $m_u=m_d=m_s$ limit. However, given the larger value of the strange mass, $m_{s}$, this approximation is not expected to be precise.

Finally, note that hypothetical stable spin-1 vector or axial mesons can also be described in terms of form factors (see for example~\cite{Tsai:1971vv}),
\begin{equation}
\langle V(A) | \bar{D}\gamma_{\mu}(\gamma_5)u|0 \rangle= M_{V(A)}f_{V(A)} \, \epsilon_{\mu} \, . 
\end{equation}
However, they do not exist in QCD as stable states, making this narrow-width approach a relatively poor approximation. While in the hadronic continuum composed of combinations of pseudoscalar mesons one can see resonance peaks, which one may identify with e.g. the $\rho$ and $a_1$ mesons, they are rather wide, and their separation from the hadronic continuum cannot be performed unambiguously.

\subsection{Decay into one pseudoscalar meson}\label{sec:onemeson}%
The simplest hadronic tau decay involves either a single pion or a single kaon, with kinematics completely determined by energy and momentum conservation. The corresponding form factors are\footnote{Our normalization corresponds to $f_{\pi}\approx 130 \, \mathrm{MeV}$.}
\begin{equation}
\langle \pi^{-}(q) |  \, \bar{d}  \gamma^{\mu}\gamma^{5} u | 0 \rangle = - i\, f_{{\pi}^-} \,q^{\mu} \qquad \, , \qquad \,  \langle K^{-}(q) |  \bar{s}\gamma^{\mu}\gamma^{5} u | 0 \rangle = - i\, f_{{K}^-} \, q^{\mu} \, ,
\end{equation}
Inserting these form factors into Eq.~(\ref{eq:lorentHdec}) immediately yields
\begin{equation}
H^{(0)}\overset{D=d}{=}\frac{f_{\pi^-}^2}{4}\, \delta(q^2-m_{\pi^-}^{2})\,  \qquad \, , \qquad \, H^{(0)}\overset{D=s}{=}\frac{f_{K^-}^2}{4}\, \delta(q^2-m_{K^-}^2)\,  .
\end{equation}
Thus, Eq.~(\ref{eq:invdist}) becomes\footnote{The same procedure for a spin-1 vector/axial meson in the narrow width approximation gives 
\begin{equation}
\Gamma_{\tau \to V(A)\nu_{\tau}}=\frac{m_\tau^3}{16\pi} G_F^2 |V_{uD}|^2 f_{V(A)}^2\left( 1-\frac{M_{V(A)}^2}{m_\tau^2} \right)\left( 1+2\frac{M_{V(A)}^2}{m_\tau^2} \right) \, .
\end{equation}}
\begin{equation}
\Gamma_{\tau^- \to P^- \nu} = \frac{m_\tau^3}{16\pi} G_F^2 |V_{uD}|^2 f_{P^-}^2\left( 1-\frac{m_{P^-}^2}{m_\tau^2} \right)^2 (1+\delta_{\mathrm{RC}}^{(P)})\,,
\quad\text{with }(P,D) = (\pi,d),\,(K,s)\,.
\end{equation}
$\delta_\mathrm{RC}^{(P)}$ refers to (photon-inclusive, see otherwise Ref.~\cite{Guo:2010dv} for estimates on the photon energy spectrum, not yet measured) radiative corrections not accounted in this tree-level evaluation, including the common $S_{\mathrm{EW}}^{\mathrm{had}}$ factor. In the absence of any large logarithm, one would naively expect them to be below the percent level, given that $\alpha/\pi \sim 0.002$. While we do have a large logarithm from the $M_W$ to the $m_\tau$ scale, it can be computed perturbatively, leading to a $\sim 2 \%$ correction \cite{Marciano:1988vm,Braaten:1990ef,Erler:2002mv}. Incorporating estimates of different long-distance corrections one typically finds values clustering around that $2\%$. 

For an approximate numerical assessment let us for example take $\delta_{\mathrm{RC}}^{(\pi)}=0.0194(61)$ and $\delta_{\mathrm{RC}}^{(K)}=0.0204(62)$ from \cite{Cirigliano:2021yto}, based on \cite{Rosner:2015wva,Roig:2019rwf,Arroyo-Urena:2021nil}. Let us also take (GeV units) $f_\pi=0.1302(8)$ and $f_K/f_\pi=1.1934(19)$ from \cite{FlavourLatticeAveragingGroupFLAG:2024oxs}, and $V_{ud}=0.97413(42)$, $V_{us}^2\approx 1-V_{ud}^2$, $m_\pi=0.13957039(18) $, $m_K=0.493677(15)$ , $m_\tau=1.77693(9)$ and $G_F=1.1663788 \cdot 10^{-5}$ from~\cite{ParticleDataGroup:2024cfk}. The SM prediction obtained using these inputs, coming from sectors different than $\tau$ observables, leads to
\begin{equation}
\Gamma_{\tau^- \to \pi^- \nu}^{\mathrm{SM}}=2.460(34) \, \cdot\, 10^{-13}  \qquad , \qquad \Gamma_{\tau^- \to K^- \nu}^{\mathrm{SM}}=1.627(35) \, \cdot \, 10^{-14} \, ,
\end{equation}
in good agreement with the experimental values~\cite{ParticleDataGroup:2024cfk}
\begin{equation}
\Gamma_{\tau^- \to \pi^- \nu}^{\mathrm{exp}}=2.453(12) \,\cdot\, 10^{-13}  \qquad , \qquad \Gamma_{\tau^- \to K^- \nu}^{\mathrm{exp}}=1.578(23) \, \cdot \, 10^{-14} \, .
\end{equation}
Notice that the SM predictive power can be further exploited by considering ratios of decay widths instead, where some of the uncertainties cancel. This is the case of $\Gamma_{\tau^- \to K^- \nu}/\Gamma_{\tau^- \to \pi^- \nu}$ or $\Gamma_{\tau^- \to K^- \nu}/\Gamma_{\pi^- \to \mu^- \nu}$, which can be translated into a precise determination of $V_{us}$, and $\Gamma_{\tau^- \to P^- \nu}/\Gamma_{P^- \to \mu^- \nu}$, which provide powerful tests of lepton flavor universality.

\subsection{Decay into two pseudoscalar mesons}\label{sec:twomesons}%

\subsubsection{Distributions}

Let us re-parameterize Eq.~(\ref{eq:ffgen}) for the two pseudoscalar meson case as
\begin{equation}\label{eq:twohadff}
\langle P_1^{-} P_2^{0}| \bar{D}\gamma^\mu u |0\rangle=C_{P_1P_2}\left[\left(\tilde{p}_{1}-\tilde{p}_2\right)^\mu F_V^{P_1P_2}(q^2) +q^{\mu}\tilde{F}_{S}^{P_1P_2}(q^2) \right]\, ,
\end{equation}
where $(\tilde{p}_{1}-\tilde{p}_2)^{\mu}=\hat{P}_T^{\mu\nu}(p_{1}-p_2)_{\nu}=\left(p_{1}-p_2 - \frac{\Delta_{P_1P_2}}{q^2}q\right)^{\mu}$, with $\Delta_{P_1P_2}=m_{P_1}^2-m_{P_2}^2$. $C_{P_1P_2}$ is typically chosen so that $F_{V}^{PP'}(0)=1$ at leading order in chiral perturbation theory, $C_{\pi\pi}=\sqrt{2}, C_{K\bar{K}}=-1, C_{K\pi}=1/\sqrt{2} , C_{\pi\bar{K}}=-1, C_{K\eta_{8}}=\sqrt{3/2}$ \cite{Pich:2013lsa}. From Eq.~(\ref{eq:eom}) one has
\begin{equation}\label{eq:scalarff}
\tilde{F}_S(q^2)= \frac{m_D-m_u}{q^2}\langle H^{-} H^{'0} | \, \bar{D} u \, |0 \rangle \, .
\end{equation}
Thus, in the chiral limit (isospin for the nonstrange modes), the scalar form factors do not contribute to the distributions. $H^{(0)}$ and $H^{(1)}$ can be readily obtained in terms of these form factors using Eq.~(\ref{eq:H0andH1}). Since the form factors depend only on the scalar invariant $q^2$, the hadronic part can be factored out from the Lorentz-invariant phase space integral, which can most easily be reduced working in the rest frame of the two-hadron system
\begin{equation}
\begin{aligned}\label{eq:phaseint}
 (2\pi)^3\int d\phi_n \, \delta^{4}(q-p_H) &=  (2\pi)^3\int \frac{d^3 p_1}{(2\pi)^3 2 E_1} \frac{d^3 p_2}{(2\pi)^3 2 E_2} \delta^{4}(q-p_1-p_2)=\frac{4\pi}{(2\pi)^3} \int \frac{p_1^2 dp_1}{4 E_1 E_2}\delta\left(\sqrt{s}-E_1(p_1)-E_2(p_1)\right) 
 \\
&=\frac{4\pi}{(2\pi)^3}\int \frac{p_1^2 dp_1}{4E_1 E_2}\frac{E_1 E_2}{p_1(E_1+E_2)} \delta\left(p_1-\sqrt{s}\frac{\lambda_{P_1P_2}^{1/2}}{2}\right)=\frac{\lambda_{P_1P_2}^{1/2}}{16\pi^2 }\, .
\end{aligned}
\end{equation}
where $\lambda_{P_1P_2}=\lambda(s,m_{P_1}^2,m_{P_2}^2)/s^2$, and $\lambda(x,y,z)\equiv x^2+y^2+z^2-2xy-2xz-2yz$. One thus has
\begin{equation}
\begin{aligned}
H^{(0)}(s)&=\frac{C_{P_1P_2}^2 |\tilde{F}_S|^2}{4}\frac{\lambda_{P_1P_2}^{1/2}}{16\pi^2 } \, \qquad , \qquad
H^{(1)}(s)&=\frac{\lambda_{P_1P_2} C_{P_1P_2}^2 |F_V|^2}{12} \frac{\lambda_{P_1P_2}^{1/2}}{16\pi^2 }  \, ,
\end{aligned}
\end{equation}
which, combined with Eq.~(\ref{eq:invdist}), leads to
\begin{equation}\label{eq:disttwo}
\frac{d\Gamma_{\tau}}{ds}=\frac{G_{F}^2 \, |V_{uD}|^2 \, m_\tau^3}{768\pi^3} \, S_{\mathrm{EW}}^{\mathrm{had}}\,|C_{P_1P_2}|^2 \,  \left(1-\frac{s}{m_{\tau}^2} \right)^2\left[ 3 |\tilde{F}_S(s)|^2\lambda_{P_1P_2}^{1/2}+|F_V(s)|^2\lambda_{P_1P_2}^{3/2} \left( 1+2\frac{s}{m_\tau^2} \right)    \right] \, ,
\end{equation}
in agreement with the literature (see e.g.~\cite{Pich:2013lsa}). One may prefer not to integrate over the internal degrees of freedom and instead retain the explicit dependence on the angle formed by the tau and the charged pion momenta in the hadronic rest frame. Taking $p_2=q-p_1$ and $p_\tau=q+p_\nu$, the needed scalar products between leptonic and hadronic momenta in Eq.~(\ref{eq:taurestgen}) (see Eq.~(\ref{eq:twohadff})) are
\begin{equation}
\begin{aligned}
p_{\nu}\cdot p_1 &= E_\nu (E(|\vec{p}_1|)-|\vec{p}_1| \cos\theta) =\frac{m_\tau^2 - s}{4\,s}\Bigl[s + m_{P_1}^2 - m_{P_2}^2 - \lambda_{P_1P_2}^{\frac{1}{2}}\,s\,\cos\theta\Bigr]  \, , \\
&q\cdot p_\nu = \frac{m_\tau^2 - q^2}{2},\qquad q\cdot p_{1} = \frac{q^2 + m_{P_1}^2 - m_{P_2}^2}{2}\,.
\end{aligned}
\end{equation}
This trivially gives us
\begin{equation}
\begin{aligned}
\frac{d\Gamma}{ds \, d\cos\theta}&=
\frac{G_{F}^2 |V_{uD}|^2 m_{\tau}^3}{512\pi^3}\lambda^{1/2}_{P_1P_2}  \, S_{\mathrm{EW}}^{\mathrm{had}}\, |C_{P_1P_2}|^2\left(1-\frac{s}{m_{\tau}^2}\right)^2 (\mathcal{P}_{0}(s)+\mathcal{P}_{1}(s)\cos\theta + \mathcal{P}_2(s) \cos^2\theta) \, ,\\
\mathcal{P}_0(s)&=|\tilde{F}_S|^2  + |F_{V}|^2 \lambda_{P_1P_2} \frac{s}{m_{\tau}^2} \quad ,\quad
\mathcal{P}_{1}(s)=-2\,\mathrm{Re}(F_V^* \tilde{F}_S)\,\lambda^{1/2}_{P_1P_2} \quad , \quad \mathcal{P}_{2}(s)= \left(1-\frac{s}{m_{\tau}^2}\right) |F_V|^2\lambda_{P_1P_2} \quad , 
\end{aligned}
\end{equation}
again in agreement with the literature, e.g. see \cite{Aguilar:2024ybr}. Knowledge of the distributions in different angles (or integral moments) can thus give back $|F_V|^2$, $|\tilde{F}_S|^2$ and $\mathrm{Re}(F_V^* \tilde{F}_S)$ plus certain predictions independent of the specific form factors. Further angular distributions can also be obtained if the polarization of the tau is also considered~\cite{Kuhn:1992nz}.
 Integrating over $d \cos\theta$, one recovers Eq.~(\ref{eq:disttwo}).

\subsubsection{The two-pion mode}

More than a $25\%$ of tau leptons decay into one charged and one neutral pion. As argued above, parity guarantees its decay to be mediated, within the SM, by the vector current,
\begin{equation}\label{eq:twopionff}
\langle \pi^{-} \pi^{0}| \bar{d}\gamma^\mu u |0\rangle=\sqrt{2} \, \left[\left(p_{-}-p_0 - \frac{\Delta_{\pi^- \pi^0}}{q^2}q\right)^\mu \, F_V^{\pi^-\pi^0}(q^2) +q^{\mu}\tilde{F}_{S}^{\pi^- \pi^0}(q^2) \right] \approx \sqrt{2} (p^{-}-p^0)^\mu\, F_V^{\pi^-\pi^0}(q^2) \, .
\end{equation}
In the last approximation we have used the fact that $\Delta_{\pi^- \pi^0}$ is very small and that, using Eq.~(\ref{eq:scalarff}), $\tilde{F}_{S}^{\pi^-\pi^0}(q^2)=\frac{m_d-m_u}{q^2}\langle \pi^{-} \pi^0 | \bar{d}u | 0\rangle$ plus the fact that $\langle \pi^{-} \pi^0 | \bar{d}u | 0\rangle$ itself is zero in the isospin limit owing to G-parity, as argued in the previous section. Using the expressions above, one straightforwardly obtains
\begin{equation}
\frac{d\Gamma}{ds \, d\cos\theta}=
\frac{G_{F}^2 |V_{uD}|^2 m_{\tau}^3}{256\pi^3}\lambda^{3/2}_{\pi^- \pi^0} \, S_{\mathrm{EW}}^{\mathrm{had}}\, |F_{V}^{\pi^- \pi^0}|^2 \left(1-\frac{s}{m_{\tau}^2} \right)^2 \Bigg(   \frac{s}{m_{\tau}^2}
+\cos^2 \theta \left(1-\frac{s}{m_{\tau}^2}\right) \Bigg)  \, ,
\end{equation}
which integrating over angle gives,
\begin{equation}
\frac{d\Gamma}{ds}=\frac{G_{F}^2 \, |V_{uD}|^2 \, m_\tau^3}{384\pi^3} \lambda_{\pi^{-}\pi^{0}}^{3/2} \, S_{\mathrm{EW}}^{\mathrm{had}}\, |F_V^{\pi^{-}\pi^{0}}(s)|^2  \left(1-\frac{s}{m_{\tau}^2} \right)^2 \left( 1+2\frac{s}{m_\tau^2} \right)     \, .
\end{equation}

\begin{figure}[tb]
	\centering
	\includegraphics[width=0.7\textwidth]{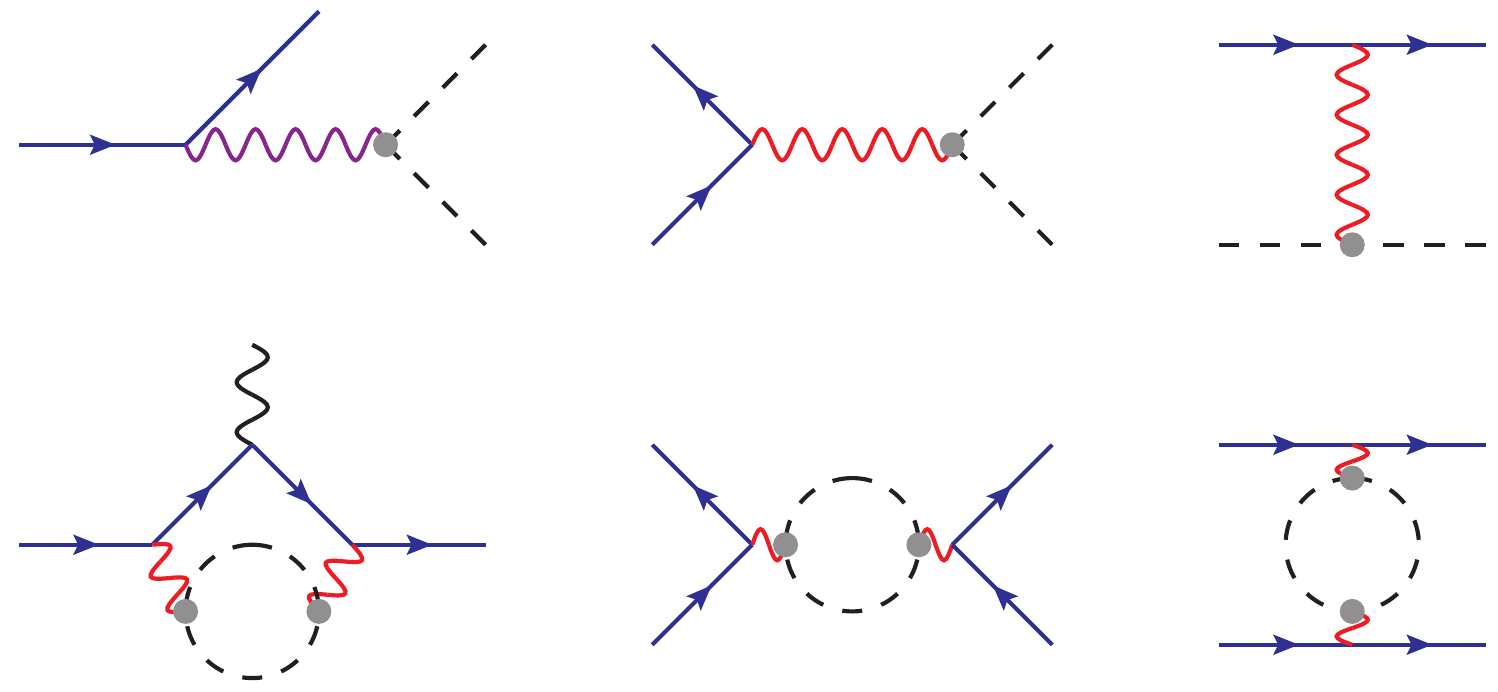}
	\caption{Processes involving the vector form factor of the pion. Pions, leptons and bosons are represented, respectively, with dashed, continuous and wiggled lines (photons in red, $W$ in purple).}
	\label{fig:vff}
\end{figure}

The study of the corresponding invariant mass distribution of the tau is thus the study of the modulus of the vector form factor in the allowed kinematic regime. A substantial amount of work has been dedicated to the study of the form factor. It not only appears in hadronic tau decays, but an isospin rotation relates it (up to isospin breaking corrections) to the hadronization of an off-shell photon,
\begin{equation}
F_{V}(q^2) (p_+-p_-)_{\mu}=\frac{1}{2}\langle \pi^{+}\pi^-|\bar{u}\gamma_{\mu }u -\bar{d}\gamma_{\mu}d |0\rangle =\langle \pi^+ \pi^- | j_{\mu}^{\mathrm{em}}|0\rangle \, ,
\end{equation}
which is traditionally known as the Conserved Vector Current (CVC) test. In the last equality we have used once again $G$-parity (see section \ref{sec:formfactors}) to discard any contribution from the $I=0$ part, $\frac{1}{6}(\bar{u}\gamma^{\mu}u+\bar{d}\gamma^{\mu}d-2\bar{s}\gamma^{\mu}s)$. Thus the same form factor naturally appears in many processes in particle physics, such as electron-positron annihilation into hadrons, in pion-lepton scattering, in pion beta decay or in the contribution of the hadronic vacuum polarization to precision observables such as the anomalous magnetic moment of the muon or the running of the electromagnetic coupling. This is illustrated in Fig.~\ref{fig:vff}.

\begin{figure}[tbh]
	\centering
	\includegraphics[width=0.35\textwidth]{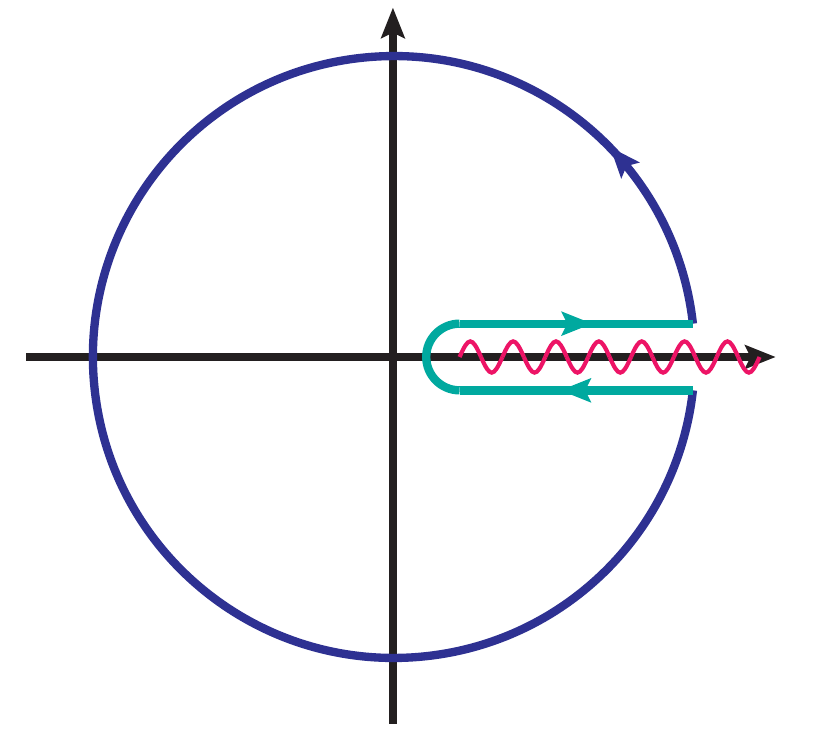}
	\caption{Circuit of integration used to obtain the usual dispersion relations for the vector form factor of the pion.}
	\label{fig:circuit}
\end{figure}

While we do not know how to precisely compute the vector form factor from first principles, let us list some properties we know about it.
\paragraph{Asymptotic behaviour}
For large momentum transfer, one can factorize the amplitude into a hard-scattering part—calculable in perturbative QCD—and a soft part, namely the pion’s parton distribution function. Within this framework, one finds that the pion form factor behaves asymptotically as 
\begin{equation}
F_{V}(Q^2)\to \frac{8 f_{\pi}^2 \alpha_s(Q^2)}{Q^2} \,.
\end{equation}
The corresponding perturbative series appear to be, however, extremely poorly behaved. See for example~\cite{Lepage:1980fj,Melic:1998qr,Simula:2023ujs,RuizArriola:2024gwb} and references therein.
\paragraph{Analyticity}
The form factor is analytic in the whole complex plane except for the physical hadronic cut, which starts at $s=4m_\pi^2$. Considering this analytic structure, one can write dispersion relations either for $F_V=|F_V|\, e^{i\delta_{V}}$ or for $\ln F_V=\ln |F_V| + i \delta_V$, using the integration kernel $\frac{1}{s^n(s-q^2-i\epsilon)}$. Integrating them along the circuit of Fig.~\ref{fig:circuit}, sending the radius to infinity, one finds
\begin{align}
F_{V}(s)&=\sum_{m=0}^{n-1}\frac{s^{m}}{m!}F_{V}^{(m)}(0)+\frac{s^{n}}{\pi} \int^{\infty}_{4m_{\pi}^2} \frac{ds'}{s'^n}\frac{\mathrm{Im}\, F_V(s')}{s'-s-i\epsilon} \, ,\\
F_{V}(s)&=\mathrm{exp}\left(\sum_{m=0}^{n-1}\frac{s^{m}}{m!}(\ln F_V)^{(m)}(0)\right) \, \mathrm{exp}\left(\int^{\infty}_{4m_\pi^2}\frac{ds'}{s'^{n}}\frac{\delta_V(s')}{s-s'-i\epsilon} \right) \, ,
\end{align}
where minimal values for $n$ can be obtained from the asymptotic behavior above. The last equation, expressed as an exponential of a dispersion integral of the scattering phase, is known as the Omnès representation of the vector form factor~\cite{Omnes:1958hv}. Knowledge of either $\mathrm{Im}\,F_V$ or $\delta_V$ up to sufficiently higher energies (lower with more subtractions/higher n) is enough to extract $F_{V}(s)$, except for a polynomial subtraction (detailed knowledge of $F_V$ at zero energies), which is lower at lower $n$.

\paragraph{Unitarity}
In the elastic region, $s<16m_{\pi}^2$, unitarity of the $S$-matrix implies
\begin{equation}
\mathrm{Im}\, F_{V}(s)=\left(1-\frac{4m_{\pi}^2}{s} \right)^{1/2} F_{V}(s) \, T_{1}^{1}(s) \, ,
\end{equation}
where $T_{1}^{1}(s)$ is the partial wave amplitude for $\pi\pi$ with angular momenta and isospin equal to one. Thus ($\mathrm{Im}\, F_{V}$ is obviously real),
$\delta_V$ is fully determined by the phase of the isospin $I=1$ pion scattering amplitude, $\delta_{V}(s)=\delta_{I}^{1}(s)$~\cite{Watson:1954uc}. With enough subtractions, the elastic region should dominate and $\delta_I^1(s)$ can determine $F_V(s)$ up to a polynomial ambiguity. Generalizations beyond the elastic regime also exist (see for example Ref.~\cite{Oller:2025leg} and references therein).

\begin{figure}[tb]
	\centering
	\includegraphics[width=0.6\textwidth]{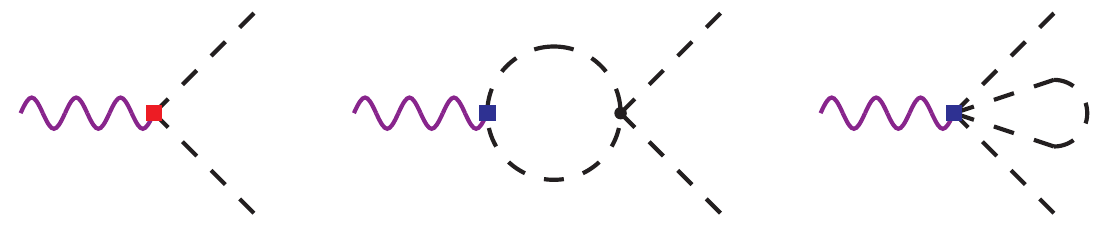}
	\caption{Topologies contributing to the vector form factor at the NLO in the chiral counting, not including wave-function renormalization. At this order, the imaginary part is fully determined by the second topology}
	\label{fig:vffchpt}
\end{figure}

\paragraph{Low-energy behavior}
The effective field theory describing QCD at low energies, where the pseudoscalars are the only dynamical degrees of freedom, is chiral perturbation theory \cite{Gasser:1983yg,Gasser:1984ux}, leading to an expansion in powers of $\frac{p^2}{(4\pi F)^2}\sim \frac{m_{P}^2}{(4\pi F)^2}\sim  \frac{m_{q} |\langle\bar{q} q \rangle | }{(4\pi)^2F^4}$, where $F\approx f_{\pi}/\sqrt{2}$. Within it, one can work out the form of the vector current order by order in the chiral expansion, to be recombined from vertices emerging from the Dyson series of the EFT
\begin{equation}
\langle \pi^{-}\pi^{0}|\, J_{\mu} \, |0\rangle= \langle \pi^{-}\pi^{0}|\, (J_{\mu}^{\chi,\mathrm{LO}}[\Phi]+  J_{\mu}^{\chi,\mathrm{NLO}}[\Phi]+\cdots )\, e^{iS_{\mathrm{int}}[\Phi]} \, |0\rangle \, .
\end{equation}
In the isospin limit, one has at NLO, apart from the wave function renormalization factor, the topologies of Fig.~\ref{fig:vffchpt}, leading to\footnote{Obviously, chiral perturbation theory recovers $F_{V}(0)=1$ in the isospin limit, as required by the definition of electric charge. In fact, it can be shown that this also holds up to second order in isospin breaking \cite{Gasser:1984ux,Ademollo:1964sr}.}
\begin{equation}
F_V(s) \, =\, 1\, +\,\frac{2 L_9^r(\mu)}{F^2}\, s \, -\,
\frac{s}{96\pi^2 F^2}\,
\left[ A\left(\frac{M_\pi^2}{s},\frac{M_\pi^2}{\mu^2}\right) +
{1\over 2}\, A\left(\frac{M_K^2}{s},\frac{M_K^2}{\mu^2}\right)  \right]\, .
\end{equation}
where $L_9^{r}(\mu)$ is a low-energy constant,
\begin{equation}
A\left(\frac{M_P^2}{s},\frac{M_P^2}{\mu^2}\right) \, =\,
\log{\left( \frac{M^2_P}{\mu^2}\right)} + {8 M^2_P\over s} -
\frac{5}{3}  + 
\sigma_P^3 \,\log{\left(\frac{\sigma_P+1}{\sigma_P-1}\right)}\, ,
\end{equation}
with $\sigma_P \equiv \sqrt{1 - 4 M_P^2/s}$. This satisfies the requirements of unitarity and dispersion relations, but only up to the first unaccounted chiral order. For example the imaginary part just above the pion threshold is trivially given by $\mathrm{Im}\,F_{V}(s)=\frac{s\sigma_{\pi}^3}{96\pi F_{\pi}^2}$, as required by unitarity and the corresponding chiral perturbation theory result for pion scattering. Additionally, the chiral perturbation result contains some extra information on the quark mass dependence.

\paragraph{$\rho$ meson dominance and resonance chiral theory} At leading order in large-$N_C$ there are no chiral loops, as $F\to\infty$, while vector meson resonances are expected to remain in the spectra with finite mass (e.g. see Ref.~\cite{Nieves:2011gb}). The vector form factor is expected to go as
\begin{equation}
F_V(s)\approx \sum_{V} \frac{c_V\, M_V^2}{M_{V}^2-s} \approx \frac{M_\rho^2}{M_{\rho}^2-s} \, .
\end{equation}
In the last approximation we have truncated the series at the first vector resonance, the $\rho$. This is called $\rho$-meson dominance approximation, which gives a surprisingly good approximation. Obviously, the form factor in the strict large-$N_c$ limit cannot be used at the resonance peak, since it yields a divergent result (the $\rho$ never decays into two pions). A minimal solution often found in the literature is adding a small Breit-Wigner-like width to the denominator. As one decreases $N_C$, $(4\pi F)$ becomes relatively close to $M_{\rho}$, leading to the usual chiral counting. Instead, one may invoke large $N_C$ to integrate out the $\rho$, $M_\rho \ll 4\pi F$ and work out the corresponding form factor. This is implemented in resonance chiral theory~\cite{Ecker:1988te,Ecker:1989yg} and the corresponding form factor can be found in the literature~\cite{Rosell:2004mn,Pich:2008jm}.

\begin{figure}[tb]
	\centering
	\includegraphics[width=0.35\textwidth]{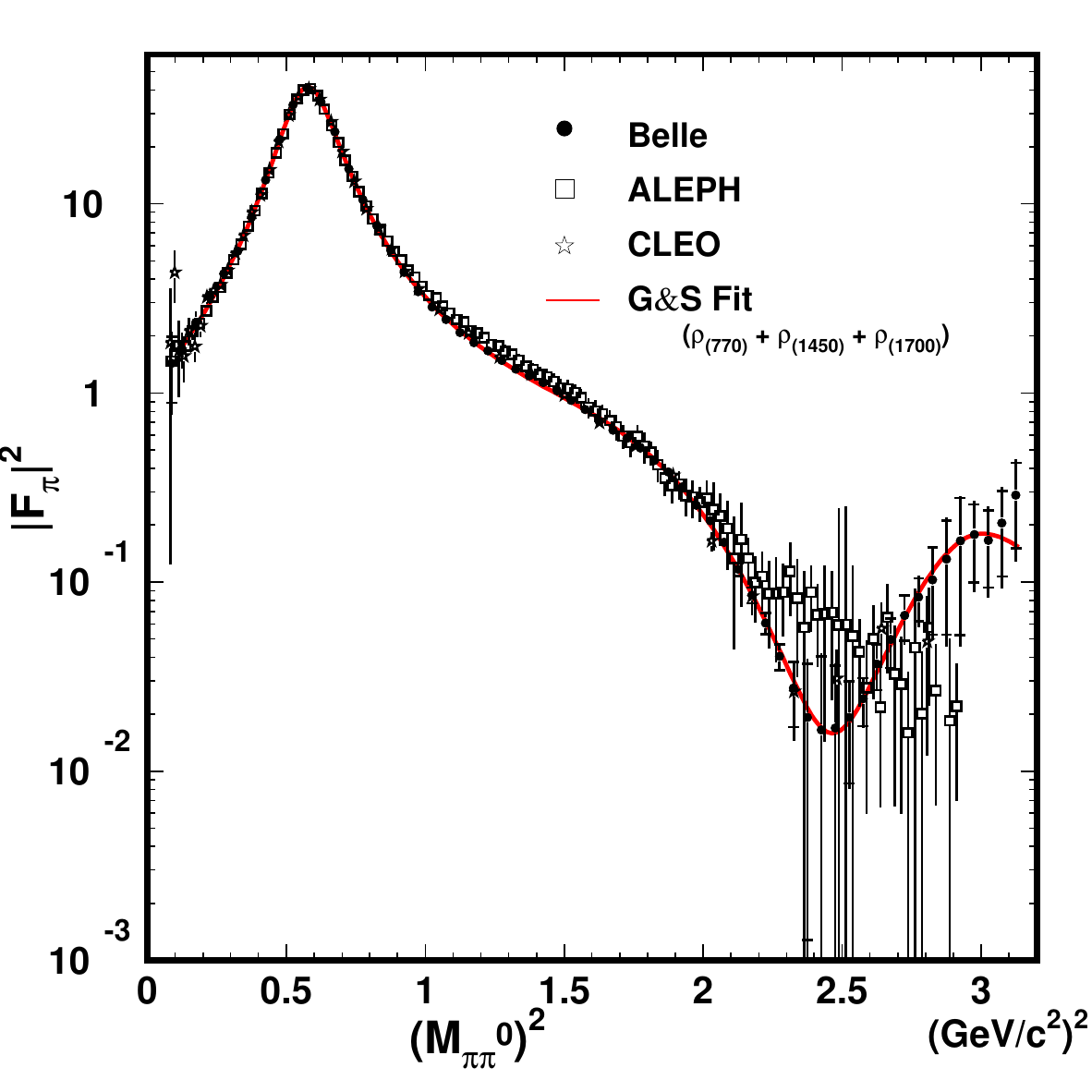}    \includegraphics[width=0.55\linewidth]{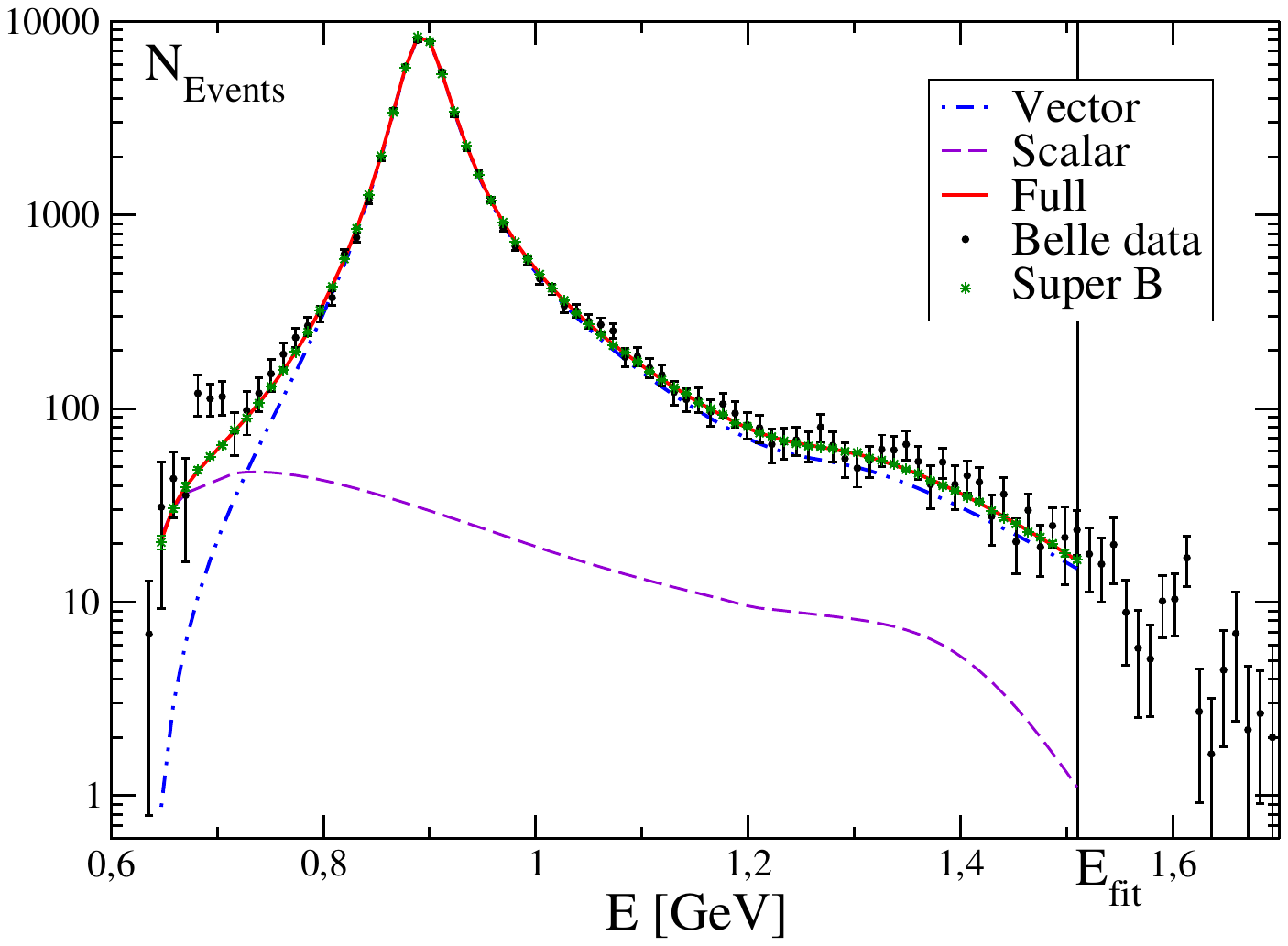}
	\caption{On the left, vector form factor of the pion as extracted from $\tau^-\to\pi^{-}\pi^0\nu_{\tau}$ data. Figure taken from.~\cite{Belle:2008xpe}. On the right, invariant mass distribution for the $\tau \to K_{S}\pi^-\nu$ channel from~\cite{Belle:2007goc}. Figure taken from Ref.~\cite{Antonelli:2013usa}.}
	\label{fig:vffbelle}
\end{figure}

Many different phenomenological works (see for example~\cite{Guerrero:1997ku,Colangelo:2018mtw,RuizArriola:2024gwb,Kirk:2024oyl} and references therein) combine (improved versions of) all these elements in different ways trying to minimize the amount of assumptions or parameters needed to fit the experimental data at the existing precision. Currently the most precise experimental shape for the (tau) vector form factor comes from Belle~\cite{Belle:2008xpe}. Their result, compared with previous ones in that reference, is displayed in Fig.~\ref{fig:vffbelle}. It is clear that the two-pion channel contribution is strongly dominated by the (wide) region near the $\rho$ peak. 

\subsubsection{Other two-meson modes}
The same kind of tools have been extensively used to study the other two-pseudoscalar modes. Let us briefly review them.

\paragraph{$K\pi$}
This is the most frequent Cabibbo-suppressed mode, as can be seen in  table~\ref{tab:tau_decays}. In the isospin limit one has $\mathrm{Br}(\bar{K}^{0}\pi^-)=2\, \mathrm{Br}(K^-\pi^0)$, which is a good first approximation. The leading (percent level) isospin breaking effects, predominantly arising from $m_{u} \neq m_d$, have been addressed in Ref.~\cite{Antonelli:2013usa}. The most precise experimental determination of the invariant mass distribution comes from Belle~\cite{Belle:2007goc} and is displayed in Figure~\ref{fig:vffbelle}, together with the fit of Ref.~\cite{Antonelli:2013usa}.

Once again, the contribution of the scalar form factor is suppressed with respect to the one of the vector form factor. However, the suppression is not so strong, since its contribution only vanishes in the $m_s=m_u$ limit, owing to Eq.~(\ref{eq:scalarff}). The scalar form factor is not well described by naive saturation of narrow-width resonances. It has been studied in detail for example in Refs.~\cite{Jamin:2001zq,Jamin:2006tj}, exploiting dispersion relations plus $K\pi$ scattering data for the needed phase.

The leading contribution comes from the vector form factor. Due to the flavour-changing phenomena involved, there are not so many physical processes involving the same vector form factor compared to the pion one. It does however appear in $K\to\pi\ell\nu$ decays, in a different kinematic regime. The corresponding form factor is dominated by contributions near two resonance peaks, the $K^{*}(892)$ and the $K^*(1410)$ vector mesons. They have been studied in detail in Refs.~\cite{Jamin:2006tk,Jamin:2008qg,Boito:2008fq,Boito:2010me}.

This channel is expected to provide the first observable signal of CP violation in lepton decays. Direct CP violation in the SM is too small to be observed in this kind of decay and, considering the very strong bounds from precise CP-odd probes such as EDMs, it is also hard to conceive any BSM scenario with large direct CP violation in this channel~\cite{Cirigliano:2017tqn,Rendon:2019awg}. One still expects to observe CP violation via $K^{0}-\bar{K^0}$ mixing, with an asymmetry value largely independent of the tau decay dynamics \cite{Bigi:2005ts}
\begin{equation}
\mathcal{A}_\tau\equiv \frac{\Gamma(\tau^{+} \to \, K_S \, \pi^+\,  \bar{\nu}_\tau)\,-\,\Gamma(\tau^{-} \to \, K_S \, \pi^-\,  \nu_\tau)}{\Gamma(\tau^{+} \to \, K_S \, \pi^+\,  \bar{\nu}_\tau)\,+\,\Gamma(\tau^{-} \to \, K_S \, \pi^-\,  \nu_\tau)}=\frac{|\langle K_S | K_0 \rangle M_{\tau^+\to K^0}|^2-|\langle K_S | \bar{K}_0 \rangle M_{\tau^-\to \bar{K}^0}|^2  }{|\langle K_S | K_0 \rangle M_{\tau^+\to K^0}|^2+|\langle K_S | \bar{K}_0 \rangle M_{\tau^-\to \bar{K}^0}|^2  }\approx \frac{|\langle K_S|K^0\rangle|^2-|\langle K_S|\bar{K}^0\rangle|^2}{|\langle K_S|K^0\rangle|^2+|\langle K_S|\bar{K}^0\rangle|^2}\sim 3\cdot 10^{-3} \, ,
\end{equation}
which is in some tension with the BaBar measurement~\cite{BaBar:2011pij}, still compatible with zero but with preferred opposite sign. Some technical aspects regarding the interpretation of the experimental measurement can be found in Ref.~\cite{Grossman:2011zk}

\paragraph{$\eta \pi$}
Having one pion in the final state, this decay is, within the SM, forbidden in the isospin limit due to G-parity. Both the scalar (due to the suppression from Eq.~$(\ref{eq:eom})$) and the vector form factor enter the amplitude at NLO in isospin breaking. 

A dispersive treatment of the vector form factor is more involved, considering there is no elastic regime. Instead, near the $\eta\pi$ threshold, the $\pi^-\pi^0\to \eta\pi$ re-scattering is expected to dominate the phase shift~\cite{Descotes-Genon:2014tla}. The different phenomenological estimates of the vector form factor contribution (e.g. see~\cite{Escribano:2016ntp,Moussallam:2021flg} and references therein) cluster around $\mathrm{BR_V}(\tau \to \eta \pi \nu)\sim 2\cdot 10^{-6}$. The scalar form factor instead should have a small elastic regime, considering that, as argued above, the scalar form factor of the two-pion channel vanishes in the isospin limit. The different analyses also consider other intermediate states such as $K^-K^{0}$ in the unitarity relations, but they strongly differ in the results, giving $\mathrm{BR_S}(\tau \to \eta \pi \nu)\sim (1-10)\,\mathrm{BR_V}(\tau \to \eta \pi \nu)$. The possible decay into $\eta '\pi$ has also been analyzed in~\cite{Escribano:2016ntp}. 

On the experimental side, this decay mode has not yet been observed. Current bounds on the branching ratios are still above the $10^{-5}$ level, being dominated by Belle~\cite{Hayasaka:2009zz}.

\paragraph{$K\bar{K}$ and $K\eta$}

\begin{figure}
    \centering
\includegraphics[width=0.4\linewidth]{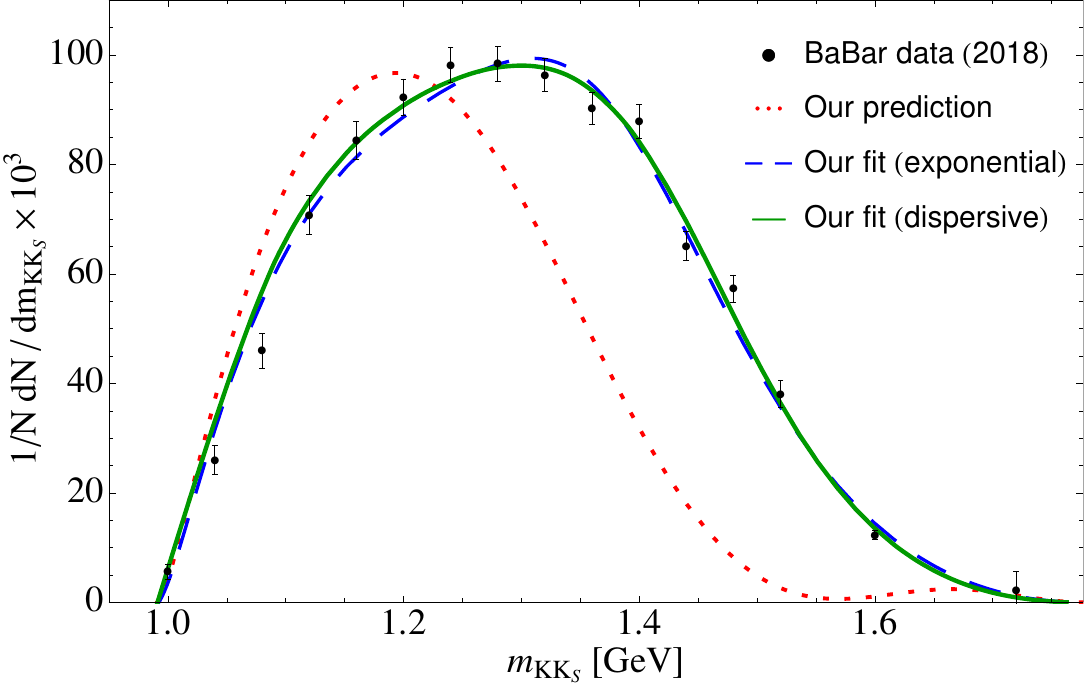}
\includegraphics[width=0.4\linewidth]{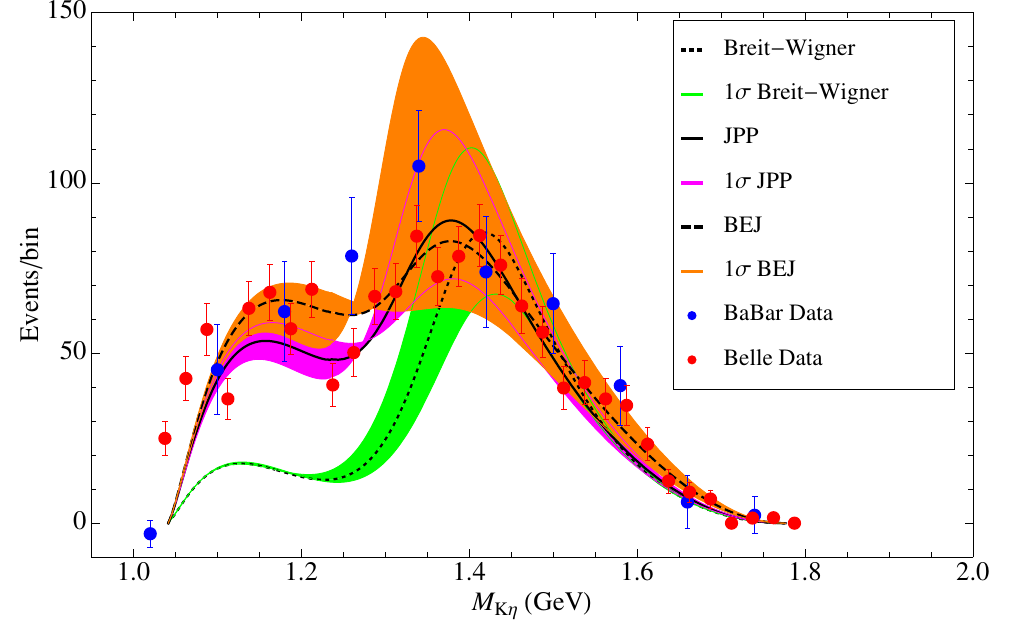}
    \caption{Phenomenological fits to the experimental distributions of Refs.~\cite{Gonzalez-Solis:2019iod,Escribano:2013bca}. Figures taken from the corresponding references.}
    \label{fig:KetaKK}
\end{figure}

The $K\bar{K}$ and the $K\eta$ channels have, respectively, the same quantum numbers as the $\pi\pi$ and $K\pi$ and are thus affected by the same resonances. However, their kinematic thresholds lie above the masses of the $\rho$ and $K^*$ resonances, qualitatively explaining the dramatic SU(3)$_V$ breaking observed in their corresponding branching ratios. Phenomenological studies can be found in Refs.~\cite{Escribano:2013bca,Gonzalez-Solis:2019iod}. In Fig.~\ref{fig:KetaKK} we show fits to the experimental distributions made in those references, using data from Belle and BaBar~\cite{Belle:2008jjb,BaBar:2010bul,BaBar:2018qry}.

\subsection{Three-pseudoscalar modes}\label{sec:threemesons}%

Following section \ref{sec:formfactors}, the hadronization of the SM quark currents for three pseudoscalars may be parameterized as
\begin{equation}\label{eq:threepion}
    J^\mu(p_1, p_2, p_3) = \langle H_1(p_1) H_2(p_2) H_3(p_3)  | J_V^\mu -J_A^\mu | 0 \rangle  =  [(\tilde{p}_{1}-\tilde{p}_3)F_1^A + (\tilde{p}_{2}-\tilde{p}_{3})F_2^A]^\mu + i \epsilon^{\mu \alpha \beta \gamma} p_{1\alpha} p_{2\beta} p_{3\gamma} \, F_3^V + q^\mu \tilde{F}_4^P \, ,
\end{equation}
where, once again $p_i^\mu\equiv(\hat{P}_T p_i)^\mu$, $q=p_1+p_2+p_3$,
and $\tilde{F}_{4}^P=\frac{m_D+m_u}{s}\langle P_1 P_2 P_3 | \bar{D}\gamma_5 u |0\rangle$, following from Eq.~(\ref{eq:eom}). Thus, the $\tilde{F}_4$ contribution vanishes in the chiral limit. A very large number of angular distributions can be built in a four-body decay, encoding precision tests of the SM independent of the actual form factor values. The general structure of the angular distributions in terms of the form factors can be found in Ref.~\cite{Kuhn:1992nz}. In this context, powerful software tools are essential to analyze the corresponding multi-dimensional distributions, including simulations adapted to the different experimental set-ups. TAUOLA is currently the most used Monte Carlo event generator for hadronic tau decays~\cite{Jadach:1993hs}.

If one is only interested in the hadronic invariant mass distributions, one can directly perform the Lorentz contractions in~Eq.~(\ref{eq:H0andH1}) to leave the distributions as a function of the three independent invariants
\begin{equation}
\begin{aligned}
W_{SA}&\equiv \frac{4}{s}(q \cdot H) (q \cdot H^{\dagger} )=s |\tilde{F}_4^{P}|^2 \, , \\
W_{A+B}&\equiv -4 \hat{P}_{T,\beta}^{\alpha} H_{\alpha}H^{\beta,\dagger} =-  \hat{P}^{(T)}{}^\mu_{\nu} [(p_{1}-p_3)F_1^A + (p_{2}-p_3)F_2^A]^\nu  [(p_{1}-p_3)F_1^{A,*} + (p_{2}-p_3)F_2^{A,*}]_\mu + \mathrm{Gram} \, |F_{3}^{V}|^2\, ,
\end{aligned}
\end{equation}
where $\mathrm{Gram}$ is the Gram determinant, defined as $\mathrm{Gram}=-\epsilon^{\mu p_1p_2p_3}\epsilon_{\mu p_1p_2p_3}$. The different contractions can be written as functions of $s_{13}=(p_1-p_3)^2,s_{23}=(p_2-p_3)^2$ and $s=q^2$. The differential phase space can then be reduced to these three variables, integrating the rest. For example, in the hadronic rest frame, one can use the three-dimensional Dirac delta to integrate over $\vec{p}_{3}$, then choose $\vec{p}_1$ along the $z$ axis and place $\vec{p}_2$ in the $x$–$z$ plane, so that one can integrate over the solid angle of the first hadron and the azimuthal part of the second. The remaining Dirac delta can then be used to integrate over $\cos\theta_{12}$. After changing variables one finds
\begin{equation}
(2\pi)^3\int d\phi_{3} \, \delta^{4}(p_1+p_2+p_3)=
(2\pi)^3 \int \frac{1}{(2\pi)^9 2 E_{1} 2E_{2} 2E_{3}} p_1^2 dp_1 p_2^2 dp_2 \frac{8\pi^2 E_3}{p_1p_2}=\frac{\pi^2}{(2\pi)^6}\int dE_1 dE_2=\int \frac{ds_{13}\, ds_{23}}{256\pi^4 s} \, .
\end{equation}
Thus one has,
\begin{equation}
H^{(1)}=\int \frac{ds_{13}\, ds_{23}}{256\pi^4 s} \, \frac{W_{A+B}}{3\cdot 4 s}  \, \quad \quad , \, \quad \quad H^{(0)}=\int \frac{ds_{13} \, ds_{23}}{256\pi^4 s} \frac{W_{SA}}{4 s}   \,    ,
\end{equation}
leading to the well-known result,
\begin{equation}
\frac{d\Gamma_{\tau\to \nu_\tau H_1 H_2 H_3}}{ds}=\frac{m_\tau^3}{4096\pi^5 s^2}G_{F}^2 |V_{uD}|^2 \, S_{\mathrm{EW}}^{\mathrm{had}}\,\left(1-\frac{s}{m_{\tau}^2} \right)^2 \int ds_{13} \, ds_{23} \left[ W_{SA}+\frac{1}{3} W_{A+B} \left( 1+2\frac{s}{m_\tau^2} \right)    \right] \, .
\end{equation}

\begin{figure}[tb]
    \centering
\includegraphics[width=0.6\linewidth]{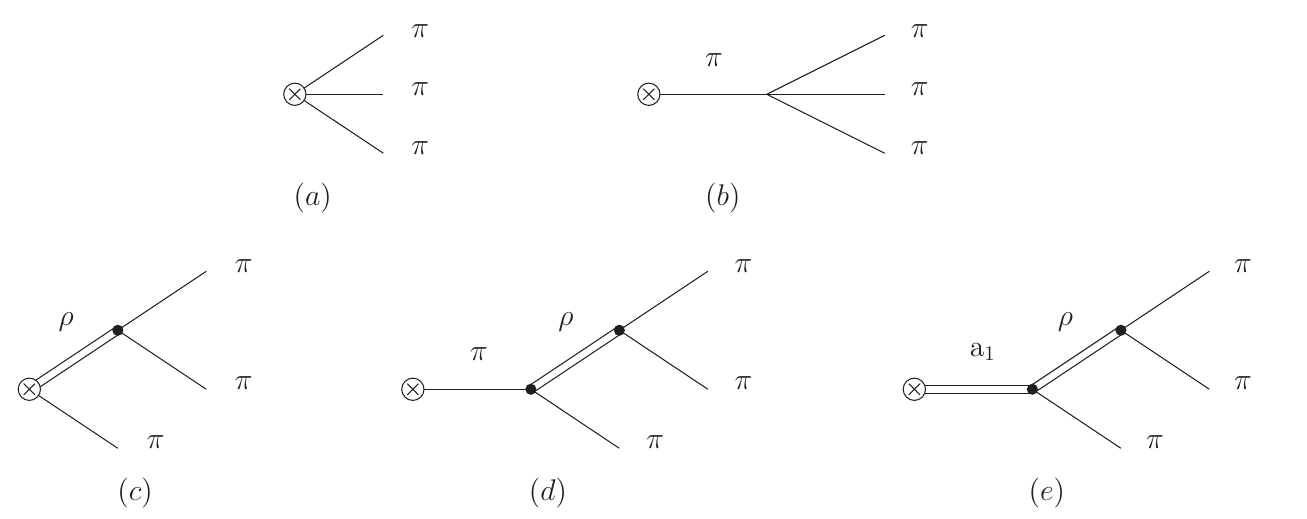}
\includegraphics[width=0.25\linewidth]{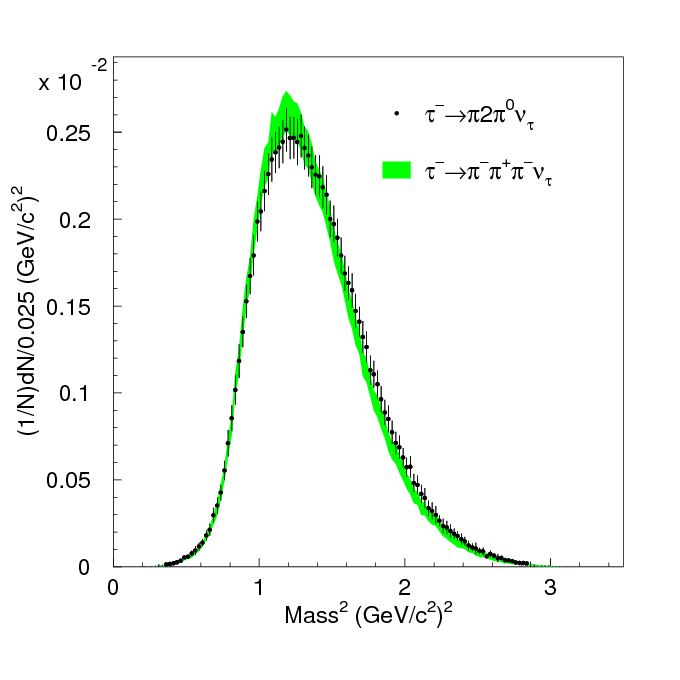}
    \caption{Left. Leading $\tau\to\pi\pi\pi\nu_{\tau}$ contributions involving the $\rho$ and the $a_1$ resonances within the resonance chiral theory approach. Right. Differential $\tau\to 3\pi\nu_{\tau}$ distributions from ALEPH data. Figures taken from Ref.~\cite{GomezDumm:2003ku} and \cite{ALEPH:2005qgp}.}
    \label{fig:threepions}
\end{figure}

The increase in both the number of form factors and kinematical variables makes the theoretical assessment of the distributions particularly challenging and, for the moment we do not know how to precisely predict the invariant mass distributions from first principles in general. The dominant three-pseudoscalar decay mode of the $\tau$ is into $3\pi$, which is the first mode whose quantum numbers allow for an intermediate $a_1$ resonance. In the isospin limit, it can only be mediated by the axial current, owing to G-parity (see section~\ref{sec:formfactors}), and thus $F_3^V=0$ in that limit. Different attempts of describing data can be found in the literature, including early attempts to incorporate resonances~\cite{Kuhn:1990ad}, the description near the threshold using chiral perturbation theory~\cite{Colangelo:1996hs}, and attempts to accommodate the large-$N_c$ expansion within the resonance chiral theory approach~\cite{GomezDumm:2003ku,Dumm:2009va}. Within the last framework, the leading contributions involving the $\rho$ and the $a_{1}$ resonances are shown in Fig.~\ref{fig:threepions}. Within the assumptions of that work, one finds $\mathrm{\frac{d\Gamma}{ds}}(\tau^- \to \pi^-\pi^-\pi^+\nu_{\tau})\approx \frac{d\Gamma}{ds}(\tau^- \to \pi^0\pi^0\pi^-\nu_{\tau})$, which agrees with the experimental result, as can be seen in the ALEPH result of Fig.~\ref{fig:threepions}. This is, however, not simply a consequence of isospin symmetry. The isospin relations between $J_{\pi^+\pi^-\pi^-}$ and $J_{\pi^0\pi^0\pi^-}$ in Eq.~(\ref{eq:threepion}) also involve exchange of momenta~\cite{Colangelo:1996hs,Girlanda:1999fu} and, as a consequence, the corresponding BR cannot be inferred from each other from isospin symmetry alone. Examples of potential contributions different for both modes (even in the chiral limit) can be found in Ref.~\cite{Sanz-Cillero:2017fvr}.

Phenomenological analyses also exist for $\pi\pi\eta$~\cite{GomezDumm:2012dpx} and $K\bar{K}\pi$~\cite{Dumm:2009kj}. Due to G-parity, the former can only be mediated, in the isospin limit, by $F_3^V$. In that limit, it can be related to $e^{+}e^{-}$ data. Current tau data~\cite{Belle:2008jjb} show some tension \cite{Arteaga:2022xxy}, which would imply large isospin-breaking effects. For the $K\bar{K}\pi$ channel, $e^+e^-$ data alone are insufficient since both vector and axial-vector currents contribute to the amplitude.

Higher-multiplicity modes have been observed in various experiments, but they become increasingly challenging to predict theoretically. An attempt to the $4\pi$ channel can be seen in~\cite{Ecker:2002cw}. Their corresponding invariant mass distributions encode, however, very fundamental information when added together, as we review in the next section.

\section{Inclusive decays}\label{sec:inclusive}

As we move to higher energies and multiplicities, the growing complexity in terms of hadrons and form factors makes precise predictions nearly impossible. However, summing over hadronic channels reveals a transition from a resonance-dominated ‘jungle’ into smoother, more uniform distributions. This can be understood by linking the corresponding hadronic sums to two-point correlation functions of quark currents which, for sufficiently large momenta, can be described in terms of approximately free quarks and gluons. From this perspective, hadronic tau decays serve as an ideal probe to explore the transition from QCD confinement to asymptotic freedom. In this section we illustrate in some detail how a precise understanding of certain inclusive observables can be achieved.

\subsection{The Källén-Lehmann spectral representation of two-point functions}

Hadronic distributions are closely connected to the Källén–Lehmann spectral representation~\cite{Kallen:1952zz,Lehmann:1954xi} of two-point correlation functions of quark currents. We sketch this following similar lines as in Refs.~\cite{Weinberg:1995mt,deRafael:1997ea,Zwicky:2016lka,Gonzalez-Alonso:2010vnm}. In section~\ref{sec:distributions}, we found the hadronic part of the invariant mass distributions to be of the form\footnote{For simplicity, we perform the derivation explicitly for scalar currents, though the argument generalizes to other current types.} 
\begin{equation}
H_{J, H}(q^2)=(2\pi)^{3} \, \int d\phi_n \, \delta^{4}(q-p_n) \, \langle 0 | J(0) | H^+(p_n) \rangle \, \langle H^-(p_n) | J^\dagger(0)|0\rangle  \, ,
\end{equation}
which, considering that $q=p_n$, is defined for $q^0>0$ and $q^2>q_{\mathrm{th}}^2$. We may define the inclusive spectral function by summing over all possible hadronic states $H$
\begin{equation}\label{eq:spectral}
\rho_{J}(q^2)\equiv \sum_{H} H_{J,H}(q^2)=(2\pi)^3 \sum_n d\phi_n \, \delta^{4}(q-p_n) \, \langle 0 | J(0) | H^+(p_n) \rangle \, \langle H^-(p_n) | J^{\dagger}(0) |0\rangle \, ,
\end{equation}
once again for $q^2>0$ and $q_0>0$. We may generalize it for arbitrary arguments by multiplying on the lhs by $
\int_{t_{\mathrm{th}}}^{\infty}   dt \, \theta(q^0)\,\delta(t-q^2)=1$. To express this result as a two-point function, we use translation invariance on the right-hand side, noting that $J(0)=e^{-i\hat{P}x}\, J(x)\, e^{i\hat{P}x}$, and finally integrate in both sides of the equality by $\int \frac{d^{4}q}{(2\pi)^3} \, e^{-iqx}$. One finds
\begin{equation}\label{eq:prod}
F_{J}(x)\equiv\int \frac{d^{4}q}{(2\pi)^3} \int_{t_{\mathrm{th}}}^{\infty}   dt \, e^{-iqx}\, \theta(q^0)\, \delta(t-q^2) \,\rho_{J}(q^2)=\,   \langle 0|J^{}(x)  J^{\dagger}(0) |0\rangle  \, .
\end{equation}
Exchanging the integration variables, the lhs becomes $F_{J}(x)=\int_{t_{\mathrm{th}}}^{\infty} dt \,\Delta(x,t)\, \rho_{J}(t) $, with $\Delta(x,t)=\int \frac{d^{4}q}{(2\pi)^3}e^{-iqx}\,\theta(q^0)\,\delta(q^2-t^2)$. Recalling that
the usual Feynman propagator is $\Delta_{F}(x,t)=i\, [\theta(x^0)\Delta(x,t) +\theta(-x^0)\Delta(-x,t)]=\int \frac{d^{4}p}{(2\pi)^4}\frac{e^{-ipx}}{t-p^2-i\epsilon}$,
one has
\begin{equation}
\int d^4x \, e^{iqx} \, i\, [\theta(x^0) F_{J}(x)+\theta(-x^0)F_{J}(-x)]=\int^{\infty}_{t_{th}} dt \, \frac{\rho_{J}(t)}{t-q^2-i\epsilon} \, .
\end{equation}
We cannot straightforwardly identify it with $i \int \frac{d^4q}{(2\pi)^4} e^{iqx} \langle 0|T(J(x)J^{\dagger}(0))|0\rangle$ yet using Eq.~(\ref{eq:prod}), unless $F_{J}(-x)=F_{J^{\dagger}}(-x)$. But this is actually the case.\footnote{Causality requires $\langle 0| [J(0),J^{\dagger}(-x)] |0 \rangle=F_{J}(x)-F_{J^{\dagger}}(-x)=\int^{\infty}_{0}\,dt \, (\Delta(x,t)\,\rho_{J}(t)-\Delta(-x,t)\,\rho_{J^{\dagger}}(t))$ to vanish for space-like separations ($x^2<0$), where $\Delta(x,t)=\Delta(-x,t)$ holds automatically. This is sufficient to guarantee that $\rho_{J}(t)=\rho_{J^\dagger}(t)$ and then $F_{J}(-x)=F_{J^{\dagger}}(-x)$.} With these considerations, one straightforwardly obtains the standard relation,\footnote{Here, we implicitly assume that the integral converges sufficiently fast as $t\to \infty$. Otherwise, one must introduce suitable subtractions for the real part.}
\begin{equation}
\Pi_{J}(q^2)\equiv i\int d^{4}x \, e^{iqx} \langle 0 |T(J(x)J^{\dagger}(0))|0\rangle =\int^{\infty}_{t_{th}} dt \, \frac{\rho_{J}(t)}{t-q^2-i\epsilon} \to \rho_{J}(q^2)=\frac{1}{\pi}\,\mathrm{Im}\,\Pi_{J}(q^2) \, .
\end{equation}
where we have taken $\mathrm{Im}\,\frac{1}{x-i\epsilon}=\pi \,\delta(x)$.

\subsection{Vector plus axial distributions}
We can now recall Eq.~(\ref{eq:invdist}),
\begin{equation}  
\frac{d\Gamma_{\tau}}{ds}=\frac{m_\tau^3}{4\pi}G_{F}^2 |V_{uD}|^2 \, S_{\mathrm{EW}}^{\mathrm{had}}\,\left(1-\frac{s}{m_{\tau}^2} \right)^2\left[ H^{(0)}(s)+H^{(1)}(s) \left( 1+2\frac{s}{m_\tau^2} \right)    \right] \, ,
\end{equation}
and sum over all possible (strange or nonstrange) hadronic channels to replace $H^{(J)}\to \frac{1}{\pi}\mathrm{Im} \,\Pi^{{(J)}}_{H^{\dagger}_\beta,H^{\dagger}_{\alpha}}$. Let us give it in a more standard form by expanding in vector and axial currents and considering that $\Pi_{VA,AV}=0$ due to parity. Normalizing by the $\tau \to e \nu_{e}\bar{\nu}_{\tau}$ width, $\Gamma_{\tau\to e}\approx \frac{G_{F}^2 m_{\tau}^5}{192 \pi^3}\,S_{\mathrm{EW}}^{\mathrm{lep}}$, one finds
\begin{equation}\label{eq:Rtau1}
\frac{dR_{\tau}}{ds}\equiv\frac{d\Gamma_{\tau}}{\Gamma_{\tau\to e}ds}=\frac{12\pi}{m_{\tau}^2}\, \mathrm{S}_{\mathrm{EW}}  \,\left(1-\frac{s}{m_{\tau}^2} \right)^2\left[ \mathrm{Im}\,\Pi^{(1+0)}(s) \left( 1+2\frac{s}{m_\tau^2}\right) -2\frac{s}{m_{\tau}^2}\mathrm{Im}\,\Pi^{(0)}(s)    \right] \, .
\end{equation}
where, defining,
\begin{equation}\label{eq:twopoint}
\Pi^{\mu\nu}_{uD,\Gamma}(q)\equiv i \int d^{4}x\, e^{iqx} \langle 0 | T(\bar{u}(x)\Gamma^{\mu}D(x) \, \bar{D}(0)\Gamma^{\nu}u(0) ) |0\rangle =(-g^{\mu\nu}q^2-q^{\mu}q^{\nu})\, \Pi_{uD,\Gamma}^{(1)}(q^2) + q^{\mu}q^{\nu} \, \Pi^{(0)}_{uD,\Gamma} \, ,
\end{equation}
we have
\begin{equation}
\Pi^{(J)}\equiv |V_{ud}|^2 (  \Pi_{ud,V}^{(J)} + \Pi_{ud,A}^{(J)}) + |V_{us}|^2 (  \Pi_{us,V}^{(J)} + \Pi_{us,A}^{(J)}) \, \qquad \quad \Pi^{(1+0)}\equiv\Pi^{(1)}+\Pi^{(0)} \, .
\end{equation}
Notice how, looking at Eq.~(\ref{eq:twopoint}), $\Pi^{\mu\nu}$ being regular as $q\to 0$ does not imply that $\Pi^{(1)}(s)$ cannot have a kinematic pole at $q^2=0$, as far as $\Pi^{(0)}(s)$ cancels it in the $\Pi^{(1+0)}(s)$ combination. Of course one can separate into $R_{\tau,d}$ and $R_{\tau,s}$ by counting the number of kaons in the final state and, for the nonstrange channel, one can approximately separate $V$ and $A$, mostly using G-parity, as explained in section~\ref{sec:formfactors}.

Regarding the correlators, we can easily compute them in the partonic massless approximation. For example
\begin{equation}
\Pi_{ud,V}^{\mu\nu}(q)=-i N_{C} \int d^{4}x\,  e^{iqx}\, \mathrm{Tr}[\gamma^{\mu}iS_{F}(x)\gamma^{\nu}iS_{F}(-x)]=i\, N_C \int \frac{d^d k}{(2\pi)^d}\frac{\mathrm{Tr}(\gamma^{\mu}\slashed{k}\gamma^{\nu}(\slashed{k}-\slashed{q}))}{k^2 (k-q)^2}=-\frac{N_C}{12\pi^2}[\log(-q^2)+C](-g^{\mu\nu}q^2+q^{\mu}q^{\nu}) \, ,
\end{equation}
\begin{figure}[tb]
    \centering
\includegraphics[width=0.25\linewidth]{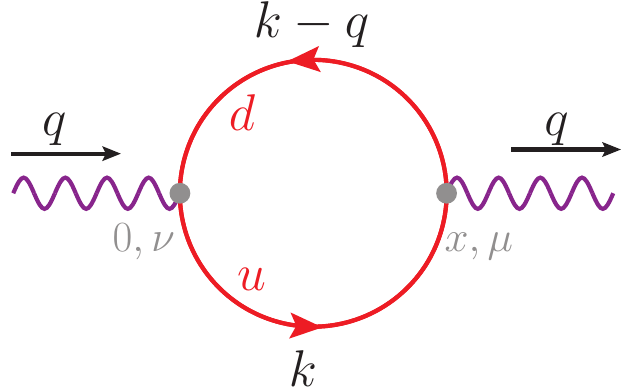}
    \caption{Two point function in the partonic approximation.}
    \label{fig:partonic}
\end{figure}
which we represent in Figure~\ref{fig:partonic}, and analogously for $\Pi^{\mu\nu}_{ud,A}$ and $\Pi_{uD,\Gamma}^{\mu\nu}$. $iS_{F}(x)$ is the usual Dirac propagator. Thus, in this approximation $\mathrm{Im}\,\Pi^{(1+0)}_{ud,V}(s)=\frac{N_C}{12\pi}, \mathrm{Im}\,\Pi^{(0)}=0$. Inserting it into Eq.~(\ref{eq:Rtau1}), one has 
\begin{equation}
R_{\tau,d}^{\mathrm{partonic}}=N_C |V_{ud}|^2\approx 2.86  \qquad \quad , \quad\qquad  R_{\tau,s}^{\mathrm{partonic}}=N_C |V_{us}|^2\approx 0.14 \quad ,
\end{equation}
which, compared to the experimental result~\cite{HeavyFlavorAveragingGroupHFLAV:2024ctg}
\begin{equation}\label{eq:Rtaud}
R_{\tau,d}=3.470\, \pm \, 0.008  \qquad \qquad , \qquad \qquad R_{\tau,s}=0.1632\, \pm \, 0.0027 \quad ,
\end{equation}
gives a reasonable first approximation. One may wonder why. The logarithmic cut predicted by the (massless) quark description in Eq.~(\ref{eq:twopoint}) has nothing to do with the physical cut of the correlator, starting with the two-pion channel and dominated by resonance contributions, such as the $\rho$ and the $a_1$. Indeed, as shown in Fig.~\ref{fig:specv+a}, where we compare the partonic prediction with the experimental tau one from ALEPH~\cite{Davier:2013sfa}, one observes a resonant spectrum, not a constant one. Of course, this failure at low energies is not a surprise when one considers perturbative QCD corrections to the free-quark limit, $\mathrm{Im}\,\Pi^{\mathrm{partonic}}$, as they scale as $\alpha_s(s)$. For large $s$ values, incorporating these corrections appears to improve the agreement with data, as shown in Fig.~\ref{fig:specv+a}, but at low $s$ values $\alpha_{s}(s)$ diverges and thus $\mathrm{Im}\,\Pi^{\mathrm{partonic}}$ cannot provide a first approximation to the observed spectrum.
\begin{figure}[tb]
    \centering
\includegraphics[width=0.5\linewidth]{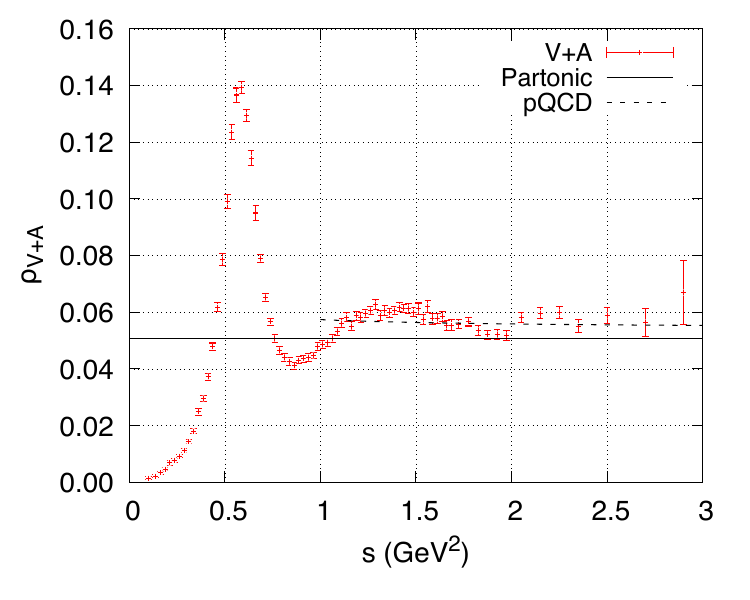}
    \caption{Nonstrange spectral function, $\rho_{V+A}(s)=\frac{1}{\pi}\mathrm{Im}\,\Pi^{(1+0)}_{ud,V+A}(s)$ in the partonic approximation, $\mathrm{Im}\,\Pi^{\mathrm{partonic}}_{ud,V+A}=\frac{N_C}{6\pi}$, including pQCD corrections (dashed line), and using experimental ALEPH data~\cite{Davier:2013sfa}.}
    \label{fig:specv+a}
\end{figure}

A much better understanding on why the partonic approximation works reasonably well for $R_{\tau}$ emerges as we invoke analyticity~\cite{Braaten:1991qm}. The correlators are analytic in the whole complex plane except for the physical cut.\footnote{We have incorporated the pion and kaon contributions into the continuum integrals.} If we integrate along the circuit of Fig.~\ref{fig:circuit} any of the correlators, $\Pi=\Pi_{uD,\Gamma}^{(1+0)}$,\footnote{The extra contribution to $R_{\tau}$ from the longitudinal part is dominated by the pion pole contribution and is suppressed by $\mathcal{O}(\frac{f_{\pi}^2\, m_{\pi}^2}{m_{\tau}^4})$. Normally one directly works with $\Pi=\Pi^{(1+0)}$, subtracting that tiny piece on the experimental size.} times an arbitrary monomial function $\left(\frac{s}{s_0}\right)^n$, where $s_0$ is the integration radius, one easily finds
\begin{equation}
A^{(n)}_{\Gamma} (s_0)\,\equiv\,\pi
\int_{s_{\mathrm{th}}}^{s_0} \frac{ds}{s_0}\; \left(\frac{s}{s_0}\right)^n\; \rho_{\Gamma} (s)=\;
\frac{i}{2}\; \oint_{|s|=s_0} \frac{ds}{s_0}\; \left(\frac{s}{s_0}\right)^n \;\Pi_{\Gamma}(s) \;,  \; R_{\tau,d}=R_{\tau,d}(m_{\tau}^2)=12|V_{ud}|^2 S_{\mathrm{EW}}\pi \, (A_{V+A}^{(0)}(m_{\tau}^2)-3 A_{V+A}^{(2)}(m_{\tau}^2)+2 A_{V+A}^{(3)}(m_{\tau}^2)) \, . 
\end{equation}
One can thus evaluate $R_{\tau}$ without needing to go below $|s|=m_{\tau}^2$, where perturbative QCD, which is the leading term of the Operator Product Expansion (OPE) in the QCD vacuum~\cite{Shifman:1978bx}, can be used. Additionally, one is away from the physical cut, where the OPE is expected not to work so well, when performing the integral and the region near it has a double kinematic zero, $\propto \left(1-s/s_0\right)^2$, for $R_{\tau}(s_0)$, further suppressing it.

\begin{figure}[tb]
    \centering
\includegraphics[width=0.45\linewidth]{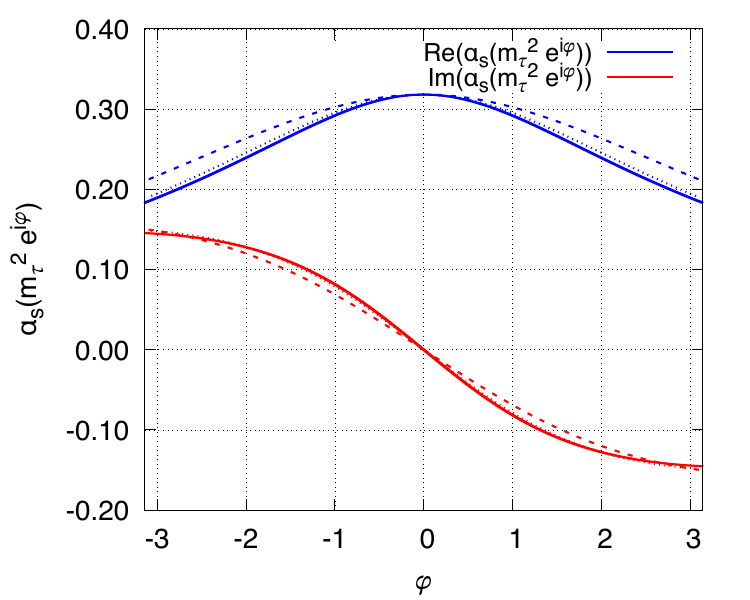}
    \caption{Analytic continuation of the QCD coupling for $\alpha_s(m_{\tau})=0.318$. Dashed and dotted lines display the solutions using only $\beta_1$, or both $\beta_1$ and $\beta_2$, instead of the full set ($\beta_1$ to $\beta_5$).}
    \label{fig:alpha_re_im}
\end{figure}

\begin{figure}[tb]
    \centering
\includegraphics[width=0.45\linewidth]{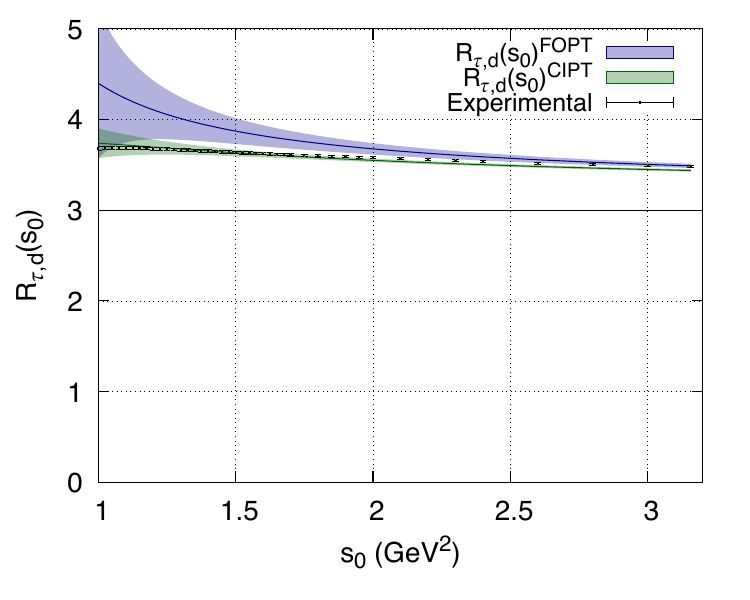}
\includegraphics[width=0.45\linewidth]{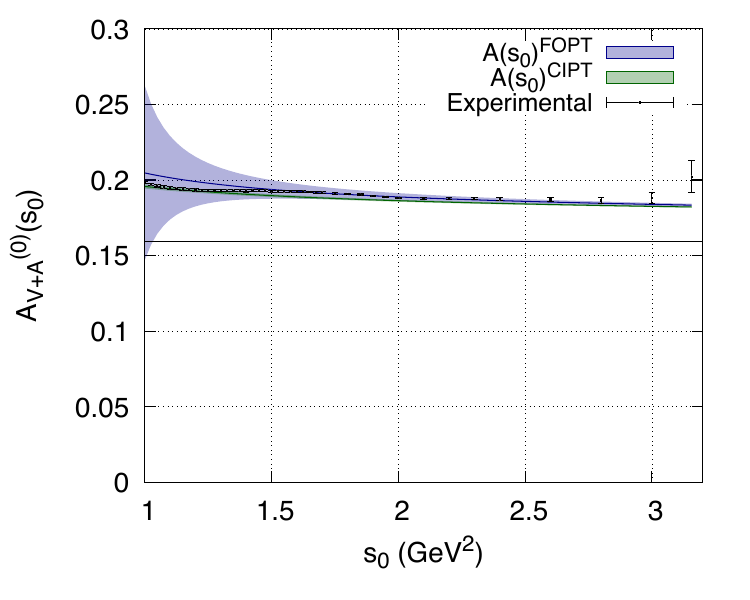}
    \caption{Comparison between the purely perturbative prediction with the CIPT and FOPT prescriptions for $\alpha_s=0.1185$ with experimental data for $R_{\tau,d}(s_0)$ and $A^{(0)}_{V+A}(s_0)$. Perturbative uncertainties have been naively estimated by modifying the first unknown coefficient $K_5$ in a generous interval, $K_5=(-125, 675)$, and changing the renormalization scale $\mu^2=\xi s_0$, with $\xi=(1/2,2)$.}
    \label{fig:RtauA}
\end{figure}

In order to evaluate $R_{\tau}$ using the OPE, one can use that the perturbative expansion of the Adler function, the logarithmic derivative of the correlator, (free from subtraction ambiguities) is known up to $5$ loops~\cite{Baikov:2008jh}. For Euclidean momenta ($q^2=-Q^2$) one has for $\Pi=\Pi_{uD,\Gamma}$, $\Gamma=V,A$\footnote{The fact that $\Pi_{uD,V}=\Pi_{uD,A}=\Pi_{uD,V+A}/2$ at all orders in massless pQCD is a consequence of the conservation of chiral symmetry at the Lagrangian level. One can show this by simply decomposing the corresponding quark currents into chiral ones ($\gamma_{\mu}=\gamma_\mu(P_L+P_R))$) and using that there are no vertices to convert $q_{L}$ into $q_{R}$ and vice-versa. The motivated practitioner is however warned that there are some technicalities involved to recover this exact perturbative result if one uses dimensional regularization. See for example~\cite{Trueman:1979en}.}
\begin{equation}
D(Q^2)=-Q^2\frac{d\Pi(Q^2)}{dQ^2} =\sum_n K_n \, \left(\frac{\alpha_{s}(Q)}{\pi}\right)^n\, .
\end{equation}
with $K_0 = K_1 = 1$, while for $n_f=3$ quark flavors $K_2=1.63982$, $K_3=6.37101$ and $K_4=49.0757$ ($\overline{\mathrm{MS}}$ scheme). For the integral, one can perform the analytic continuation to find\footnote{We follow the notation of Ref.~\cite{Pich:2022tca}.}
\begin{align}\label{eq:pert}
A^{(n)}_{\mathrm{pert}}(s_0)&=\frac{1}{8\pi^2(n+1)}\sum_m K_m \int_{-\pi}^{\pi} d\varphi \;\left( 1-(-1)^{n+1}e^{i\varphi (n+1)} \right)  a_{s}^{m}\left( s_{0}e^{i\varphi}\right) \, ,
\end{align}
with $a_s\equiv \frac{\alpha_s}{\pi}$, thus requiring the analytic continuation of the $\beta$ function,
\begin{equation}
Q\frac{d\alpha_s(Q)}{dQ}=\alpha_s \sum_{n=1}\beta_{n} \, a_s^n  \, \, ,
\end{equation}
starting with $\beta_1=-9/2$. Taking as reference value $\alpha_s(M_Z)=0.1185$, or $\alpha_s(m_\tau)=0.318$, approximately corresponding to the current most precise lattice determinations (e.g. see~\cite{DallaBrida:2022eua,Brida:2025gii}), one finds Fig.~\ref{fig:alpha_re_im} when numerically solving up to the last known perturbative order, which is qualitatively similar to Fig.~18 of Ref.~\cite{Davier:2005xq}.
Two common prescriptions for the perturbative series are: (i) expressing $A_{\mathrm{pert}}^{(n)}(s_0)$ as a truncated polynomial in $a(s_0)$ (Fixed Order Perturbation Theory, FOPT), or (ii) employing the exact solution of the $\beta$ function in the integral of Eq.~(\ref{eq:pert}) up to the last known order (Contour Improved Perturbation Theory, CIPT)~\cite{LeDiberder:1992jjr}. An extensive literature can be found about the suitability of one or the other prescription and how to implement them together with Borel re-summations and renormalons methods aiming to account for the asymptotic higher-order perturbative tail, which, up to the last known order, is relatively well behaved. One finds, using that input for $\alpha_s$,
\begin{equation}
R_{\tau,d}^{\mathrm{pert, FOPT}}(m_{\tau}^2)\approx 3.49  \qquad , \qquad R_{\tau,d}^{\mathrm{pert, CIPT}}(m_{\tau}^2) \approx 3.44 \, ,
\end{equation}
which nicely agrees with the result of Eq.~(\ref{eq:Rtaud}). Perturbative QCD works quite well to describe data  at $s_0\approx m_{\tau}^2$, as further illustrated by plotting the $s_0$-dependence of $R_{\tau,d}(s_0)$ and $A^{(0)}(s_0)$ in Fig.~\ref{fig:RtauA}, where we use ALEPH data~\cite{Davier:2013sfa}. The corresponding experimental spectral functions are provided with covariance matrices, which are essential to propagate uncertainties. Unfortunately, nonperturbative corrections are observed to be larger for the separate semi-inclusive $V$ and $A$ channels and perturbation theory does not provide an accurate description at low $s_0$ for the individual $V$ and $A$ channels. Even at $s_0 = m_{\tau}^2$, one finds $R_{\tau,d}^{V-A}\neq 0$, signaling nonzero power corrections. We will further explore the intriguing physics of the $V-A$ spectrum in the next subsection.

\begin{figure}[t]
    \centering
    \includegraphics[width=0.8\linewidth]{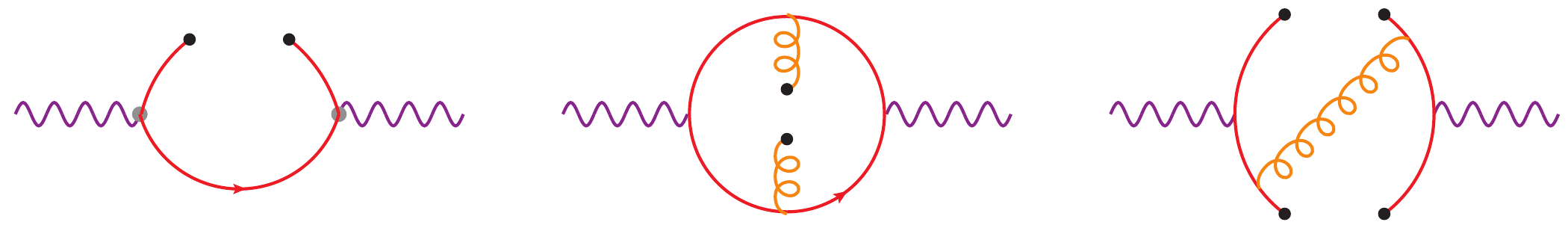}
    \caption{Topologies contributing to the power corrections of two-point correlation functions~\cite{Shifman:1978bx}. From left to right, contributions proportional to quark, gluon and four-quark condensate.}
    \label{fig:ope}
\end{figure}

Perturbation theory is only the leading term of an Operator Product Expansion in the QCD vacuum~\cite{Shifman:1978bx}. In the same way that quark currents or four-quark operators can mediate flavor transitions, provided they have compatible quantum numbers, higher-dimensional operators with the same quantum numbers as the QCD vacuum give corrections to the perturbative series. As usual, the corresponding Wilson coefficients can be computed perturbatively, order by order in $\alpha_s(\mu)$, while the vacuum hadronic matrix elements are generally not well known. Example of diagrams involved in the calculation are shown in Fig.~\ref{fig:ope}.\footnote{A pedagogical step-by-step introduction can be found in Ref.~\cite{Pascual:1984zb}.} At NLO in $\alpha_s$ they have the form\footnote{Power corrections are in general different for different correlators. We omit the corresponding indices to lighten the notation.}
\begin{equation}\label{eq:opeeuc}
\left.\Pi^{\mathrm{OPE}}(s)\right|_{D>0}\, =\,\sum_{D>0}\,\frac{\mathcal{O}_{D}(\mu)+\mathcal{P}_{D}\,\ln{(-s/\mu^2)} }{(-s)^{D/2}} \, .
\end{equation}
The factors $\mathcal{P}_{D}$, suppressed by $\alpha_s$, determine the QCD running of the coefficients $\mathcal{O}_{D}(\mu)$. Up to tiny (and well known) quark mass corrections, one has $\mathcal{O}_2(\mu)=\mathcal{P}_2=\mathcal{P}_4=0$. For the tau observables one finds
\begin{equation}\label{eq:suleadingPowers}
\left. A^{(n)}(s_0)\right|_{D>0}\, =\, -\pi\,\sum_{p=2} \frac{d_{p}^{(n)}}{(-s_0)^p}\, \qquad ,\qquad
d_{p}^{(n)}\, =\, \left\{\begin{array}{lr}
\mathcal{O}_{2p}(s_0), & \quad\text{if } p= n+ 1 \, ,\\[4pt]
\frac{\displaystyle\mathcal{P}_{2p}}{\displaystyle n-p+1}, &\quad\text{if } p\neq n+ 1 \, .\\
\end{array}\right.
\end{equation}
It is thus clear why $R_{\tau,d}(m_{\tau}^2)$ and $A^{(0)}(m_{\tau}^2)$ are particularly clean observables from the perspective of power corrections, as they are parametrically suppressed by $\frac{\langle\mathcal{O}_6 \rangle}{m_{\tau}^6}$ and $\alpha_s\frac{\langle\mathcal{O}_6 \rangle}{m_{\tau}^6}$.\footnote{A naive order of magnitude assessment for the former can be made by using large-$N_c$ factorization estimates of the dimension $6$ condensate, where one may take $\mathcal{O}_{6,V+A}\sim\mathcal{O}_{6,V+A}^{N_C} \approx 0.001 \cdot \, \mathrm{GeV}^6$, to find $\Delta R_{\tau,d}(s_0)\sim 0.01\left(\frac{m_{\tau}^2}{s_0}\right)^3 $, i.e. power corrections are expected to have a very minor role at the tau mass and rapidly become relevant at lower $s_0$ values.}

Beyond the OPE, there are further effects, known as quark-hadron duality violations, enhanced in the analytic continuation of the correlator. Using again analyticity, the  size of these effects can be expressed in the form \cite{Chibisov:1996wf,Cata:2008ye,Gonzalez-Alonso:2010kpl,Boito:2017cnp}
\begin{equation}
\label{eq:DV_A}
\Delta A^\omega (s_0)\,\equiv\, \frac{i}{2}\; \oint_{|s|=s_0} \frac{ds}{s_0}\; \omega (s) \left\{\Pi(s)-\Pi^{\mathrm{OPE}}(s)\right\}
\, =\, -\pi\int_{s_0}^\infty \frac{ds}{s_0}\;\omega(s)\;\Delta\rho^{\mathrm{DV}}(s)
\, , \qquad  \, \Delta\rho_\mathcal{J}^{\mathrm{DV}}(s)\,\equiv\, \rho(s) - \rho^{\mathrm{OPE}}(s) \, ,
\end{equation}
where the weight function $\omega(s)$ refers to any combination of monomial functions. The existence of quark-hadron duality violations becomes obvious when one observes Fig.~\ref{fig:specv+a}. For large-enough values of $s$, the OPE provides the correct average value of $\rho(s)$, but it cannot reproduce a hadronic resonance structure that generates oscillations around this mean value. These local differences diminish rapidly with increasing $s$, suggesting an exponential suppression of duality violations at higher energies, typically modeled as $\Delta\rho^{\mathrm{DV}}(s)\sim e^{-\gamma s}$. In that case, the DV correction on the right-hand-side of Eq.~(\ref{eq:DV_A}) is completely dominated by the region of $s$ values just slightly above $s_0$. In fact, the relatively large oscillations of the spectral function at $s_0\lesssim m_{\tau}^2$ already have a very minor numerical role in the total integrals $A^\omega (s_0)$, especially for weights with zeros at $s=s_0$, such as $R_{\tau}(s_0)$.\footnote{This can be quantitatively checked. Notice, see Fig.~\ref{fig:specv+a}, that this is largely independent of the $\alpha_s$ value.}

Overall, the comparison between experimental data and the OPE allows for an extraction of $\alpha_s(m_{\tau}^2)$ with a typical uncertainty of $\sim 5\, \%$, which translates into an extraction of $\alpha_s(M_Z^2)$ close to the percent level~\cite{ParticleDataGroup:2024cfk,Baikov:2008jh,Davier:2013sfa,Pich:2016bdg,Boito:2020xli,Ayala:2022cxo}. Such precision measurements have provided a crucial test of asymptotic freedom, despite initial doubts,
see for example Ref.~\cite{Shifman:1995qj}.

A similar OPE analysis can be performed for the strange V+A channel; however, due to the lack of precise experimental data on the strange spectral functions, one usually relies on $R_{\tau,s}$ to extract $V_{us}$ by comparing it with the nonstrange spectrum,
\begin{equation}
\frac{R_{\tau,d}}{|V_{ud}|^2}-\frac{R_{\tau,s}}{|V_{us}|^2}=\delta R_{th} \, .
\end{equation}
$\delta R_{th}$ parametrizes the effect of SU(3)$_V$ ($m_d \neq m_s$) breaking and may be estimated using the OPE, see e.g.~\cite{Gamiz:2006xx} and references therein. Using current experimental data \cite{HeavyFlavorAveragingGroupHFLAV:2024ctg} this leads to a too small $V_{us}$ value compared to other determinations.

A major novelty in the inclusive field is the possibility of computing $R_{\tau,D}$ from first principles using lattice QCD instead of the OPE. This has emerged as a side application of progress originally made in predicting the (smeared) $e^{+}e^{-}\to \mathrm{hadrons}$ spectrum in the context of the muon g-2 theory program. The first results~\cite{Evangelista:2023fmt,ExtendedTwistedMass:2024myu} appear to approximately agree with the OPE expectations.

\subsection{Vector minus axial nonstrange distributions}

\begin{figure}[tb]
    \centering
\includegraphics[width=0.5\linewidth]{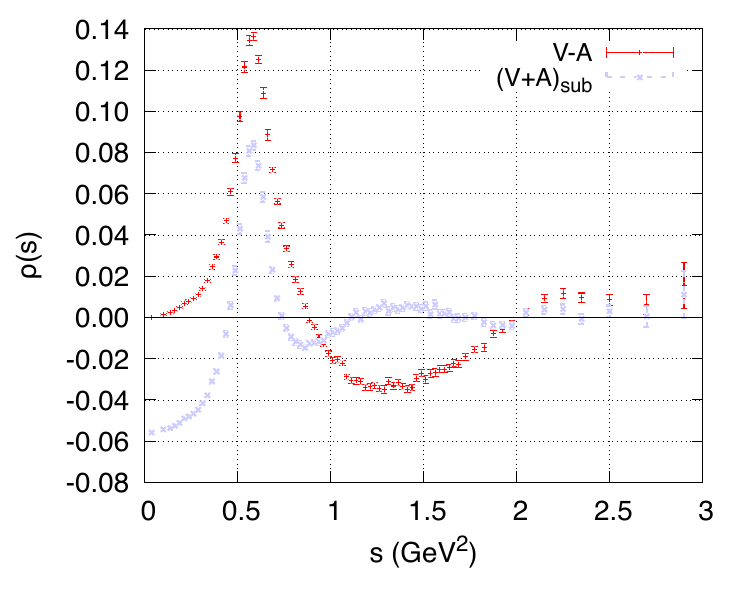}
    \caption{Nonstrange spectral function, $\rho_{V-A}(s)=\frac{1}{\pi}\mathrm{Im}\,\Pi^{(1+0)}_{ud,V-A}(s)$ using experimental ALEPH data \cite{Davier:2013sfa}. For comparison we display in faint blue the $V+A$ subtracting $\mathrm{\rho}^{\mathrm{pert}}(2 \, \mathrm{GeV}^2)$, which gives a very good approximation of the perturbative spectral function in the regime where it is well defined.}
    \label{fig:specv-a}
\end{figure}

\begin{figure}[tb]
    \centering
\includegraphics[width=0.49\linewidth]{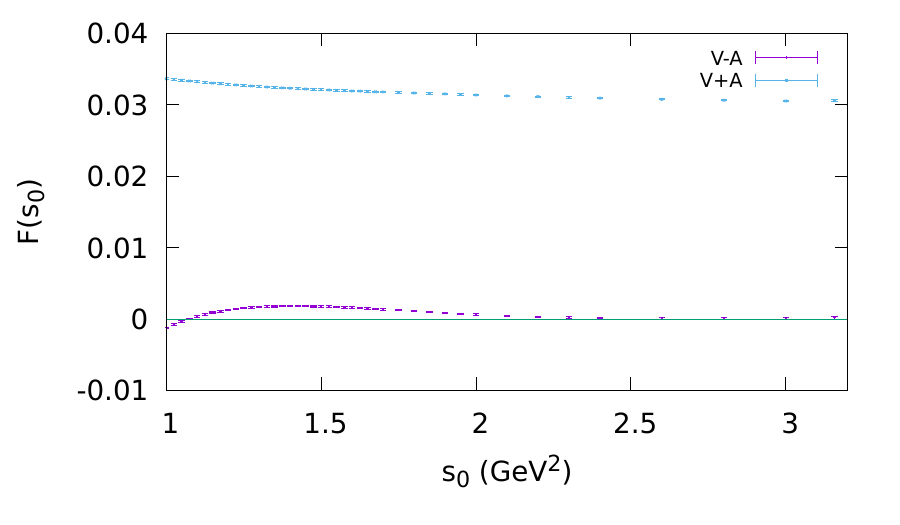}
\includegraphics[width=0.49\linewidth]{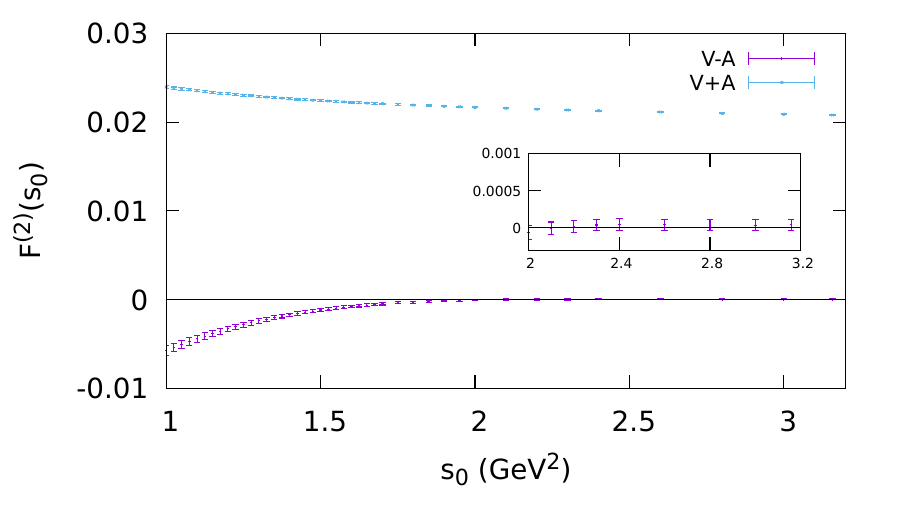}
    \caption{Weinberg-like finite energy sum rules, corresponding to Eq~(\ref{eq:WSR}). Figure taken from Ref.~\cite{Pich:2021yll}. The corresponding $V+A$ integral is displayed to showcase the size of the cancellation.}
    \label{fig:WSR}
\end{figure}

Given the large uncertainties in the perturbative series, one may question the significance of nonperturbative contributions beyond the perturbative noise of the asymptotic $\alpha_s$ expansion. One may even consider whether the nonperturbative information may be fully encoded in the coefficients of the perturbative series. A clear illustration that this is not the case is provided by the $VV-AA$ correlator. As discussed in section~\ref{sec:exclusive}, the dominant nonstrange channels can be separated into vector and axial components using parity and G-parity. We have thus experimental access to the spectral function corresponding to ($J_{\mu}^{L(R)}=\bar{u}\gamma_{\mu}P_{L(R)} d$)
\begin{equation}
\Pi^{\mu\nu}_{V-A}(q)=i\int d^{4}x \,  e^{iqx}\, \langle 0 | T(J_{V}^{\mu}(x)J_{V}^{\dagger\nu}(0)-J_{A}^{\mu}(x)J_{A}^{\dagger\nu}(0)) |0 \rangle=4 i\int d^{4}x \, e^{iqx} \langle 0 |T(J^{\mu}_L(x) J^{\nu}_R(0))  |0\rangle \, ,
\end{equation}
where we have used parity. In massless perturbative QCD there are no vertices converting $q_{L}\leftrightarrow q_R$, considering that chiral symmetry is conserved. This correlator vanishes, in the chiral limit, at all orders in perturbation theory. However, it does receive power corrections. Apart from the very tiny and well controlled (determined by the well-known quark masses and quark condensate) light-quark mass corrections, the OPE of the corresponding $\Pi^{(1+0)}$ correlator starts at dimension $D=6$ via a four-quark operator. Thus $\lim_{Q^2\to \infty}Q^4\Pi_{V-A}= 0$.
One can use this scaling to recover, in the chiral limit, the so-called Weinberg sum rules~\cite{Weinberg:1967kj}, simply integrating along the circuit of Fig.~\ref{fig:circuit} (now including the pion pole at $q^2=0$). one obtains
\begin{equation}
\int^{\infty}_0 dq^2 \rho_{V-A}(q^2)=0 \qquad , \qquad \int^{\infty}_0 dq^2 \, q^2 \rho_{V-A}(q^2)=0 \, .
\end{equation}
In the large-$N_C$ limit, the QCD spectrum consists of free, stable mesons~\cite{tHooft:1973alw,Witten:1979kh}, which for the correlators in question implies
\begin{equation}
\rho_{V-A}^{N_C\to \infty}(q^2)=-f_{\pi}^2\, \delta(q^2)+ \sum_{V}f_{V}^2\,\delta(q^2-M_{V}^2) - \sum_A f_{A}^2\, \delta(q^2-M_A^2) \; ,
\end{equation}
No se te that, at the strictly local level, quark-hadron duality never arises in the exact large-$N_C$ limit. In the real world, this is a decent approximation below the GeV, where we actually observe the $\rho$ and the $a_{1}$ peak. However, the real-world spectrum does exhibit a transition into quark-hadron duality at higher energies, where chiral symmetry is effectively restored, and thus $\rho_V\approx \rho_A$. This leads to the original Weinberg result 
\begin{align}
\rho_{V-A}(q^2)&= f_{\rho}^2\,\delta(M_{\rho}^2-q^2)-f_{\pi}^2\,\delta(-q^2)- f_{a_1}^2\,\delta(M_{a_1}^2-q^2) \to f_{a_1}^2=f_{\rho}^2-f_{\pi}^2  \quad , \quad f_{a_1}^2M_{a_1}^2=f_{\rho}^2 \, M_{\rho}^2 \, .
\end{align}
Currently, we can do much better due to improved knowledge of the OPE and more precise experimental data from tau decays. In Fig.~\ref{fig:specv-a}, we show the corresponding ALEPH spectral function \cite{Davier:2013sfa}. Quark-hadron duality (predicting $\rho_{V-A}\approx 0$) is observed to be a very poor approximation below the $\tau$ mass. 
%
%
%
This nonzero structure is signature of the spontaneous chiral symmetry breaking of massless QCD, $\mathrm{SU}(3)_{\mathrm{L}}\times \mathrm{SU}(3)_{\mathrm{R}} \to \mathrm{SU}(3)_\mathrm{V}$. Using the OPE, one can define finite energy sum rules that also incorporate quark mass and power corrections. Once again, in order to reduce both experimental uncertainties and duality violations (see Eq.~(\ref{eq:DV_A})), now more dramatic, it is convenient to use weights with zeros at $s=s_0$. For example, Ref.~\cite{Pich:2021yll} defines
\begin{equation}
F_{V\pm A}(s_{0})\,\equiv\, \int^{s_{0}}_{s_{\mathrm{th}}}\frac{ds}{s_{0}}\; \left(1-\frac{s}{s_{0}} \right)\frac{1}{\pi}\operatorname{Im}\Pi_{V\pm A}(s)\quad
, \quad  F^{(2)}_{V\pm A}(s_{0})\equiv \int^{s_{0}}_{s_{\mathrm{th}}}\frac{ds}{s_{0}}  \left(1-\frac{s}{s_{0}} \right)^2\frac{1}{\pi}\operatorname{Im}\Pi_{V\pm A}(s)
 -\frac{\langle\mathcal{O}_{6,V\pm A}^d(s_{0})\rangle}{s_{0}^{3}} \, .
\label{eq:WSR}
\end{equation}
which, for the V-A channel, are zero up to tiny duality violations. $\langle\mathcal{O}_{6,V- A}^d(s_{0})\rangle$ plays a very minor numerical role at $s_0=m_{\tau}^2$. It is related to pseudo-goldstone matrix elements~\cite{Donoghue:1999ku} that can be extracted from the lattice, agreeing with the large-$N_c$ estimates. The corresponding experimental curves obtained in Ref.~\cite{Pich:2021yll} are displayed in Fig.~\ref{fig:WSR}. Despite having a local spectrum as distinctly nonzero as the $V+A$ channel and exhibiting much more pronounced local duality violations, the appropriate integrals exhibit rapid convergence to the quark-hadron duality prediction (zero) below the tau mass.

Poles in the weight functions, such as $1/s^n$, can also be introduced when integrating along the circuit of Fig.~\ref{fig:circuit}. These poles yield information on $\Pi_{LR}(s)$ and its derivatives evaluated at zero energy. There, one can use chiral perturbation theory and translate the comparison into a precise determination of a low-energy constant, $L_{10}$. See for example~\cite{Boito:2015fra,Gonzalez-Alonso:2016ndl} and references therein.

\section{Hadronic tau decays beyond the standard model}\label{sec:bsm}

\subsection{Effective Field Theory generalization}

We have seen how in the SM one can describe hadronic tau decays with a Lagrangian made out of charged-current interactions, obtained after integrating out the $W$ boson. In general, all low-energy interactions of the SM are described by an effective Lagrangian in which the heavy degrees of freedom, such as $W$, $Z$, $t$ are integrated out. Its presence gets encoded into the Wilson coefficients of operators of dimension larger than four, such as that of Eq.~(\ref{eq:fermilag}). In the presence of extra heavy particles beyond the standard model, one can write down the most general Lagrangian invariant under $\mathrm{U}(1)_{\mathrm{em}}\times \mathrm{SU}(3)_{\mathrm{C}}$ that can mediate the hadronic tau decays. One obtains~\cite{Cirigliano:2009wk}
\begin{eqnarray}
\label{eq:leff1} 
{\cal L}_{\rm eff} 
&=& - \frac{G_\mu V_{uD}}{\sqrt{2}}  \Bigg[
\Big(1 + \epsilon_L^{ D\ell}  \Big) \bar{\ell}  \gamma_\mu  (1 - \gamma_5)   \nu_{\ell} \cdot \bar{u}   \gamma^\mu (1 - \gamma_5 ) D
+  \epsilon_R^{D\ell}  \   \bar{\ell} \gamma_\mu (1 - \gamma_5)  \nu_\ell    \cdot \bar{u} \gamma^\mu (1 + \gamma_5) D
\nonumber\\
&&+~ \bar{\ell}  (1 - \gamma_5) \nu_{\ell} \cdot \bar{u}  \Big[  \epsilon_S^{D\ell}  -   \epsilon_P^{D\ell} \gamma_5 \Big]  D
+{1 \over 4} \hat \epsilon_T^{D\ell} \,   \bar{\ell}   \sigma_{\mu \nu} (1 - \gamma_5) \nu_{\ell}    \cdot  \bar{u}   \sigma^{\mu \nu} (1 - \gamma_5) D
\Bigg]+{\rm h.c.}, 
\end{eqnarray}
The full information on possible BSM gets encoded in $10$ couplings for the tau sector, $\epsilon_{\Gamma}^{D\tau}$. In the SM limit, they vanish and, thus, one recovers the Lagrangian of Eq.~(\ref{eq:fermilag}).

If additionally one assumes new physics to be above the EW scale, one can match with the SMEFT, building the most general $D=6$ extension of the SM Lagrangian compatible with $\mathrm{SU}(3)\times  \mathrm{SU}(2)_\mathrm{L} \times \mathrm{U}(1)_\mathrm{Y}$. One can classify the tree-level contributions in two types.

\begin{figure}[tb]
    \centering
\includegraphics[width=0.69\linewidth]{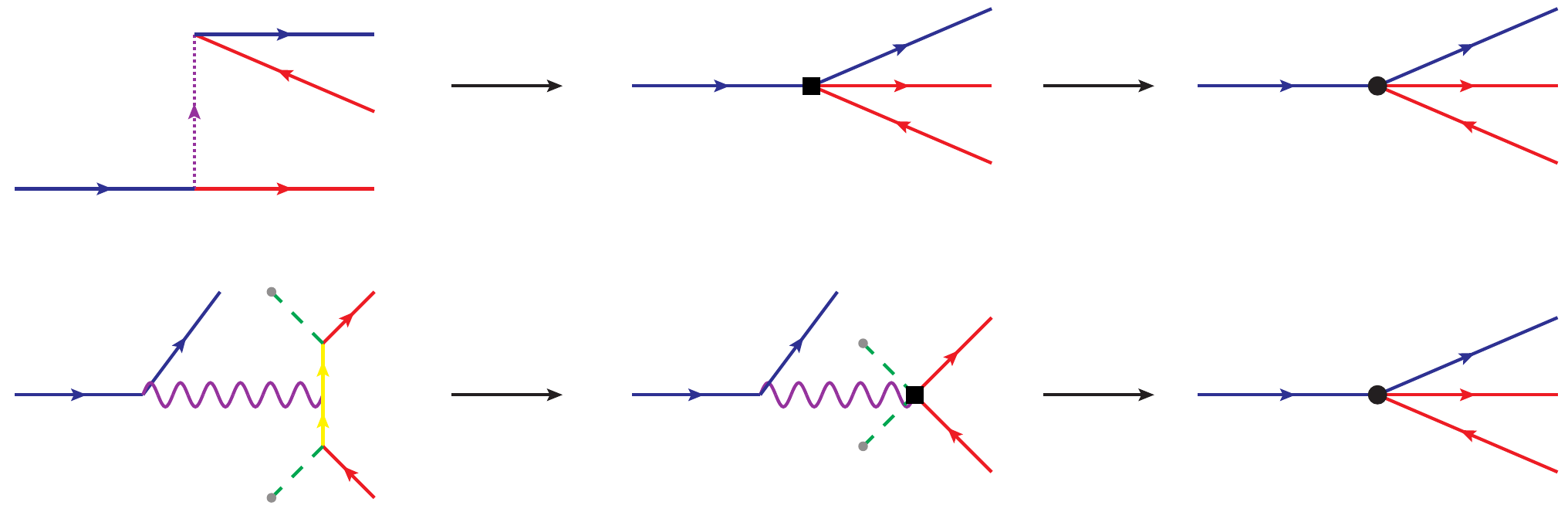}
    \caption{Example of BSM inducing four-fermion operators (leptoquark) and operators modifying vertices (vector-like quark) potentially leading to modification of charged current interactions.}
    \label{fig:bsm}
\end{figure}

\paragraph{Four-fermion operators}
These contact terms directly contain the charged current operators of the Lagrangian above. They are
\begin{equation}
Q_{lq}^{(3)}=\bar{l}_{p}\gamma_{\mu}\tau^{I}l_{r} 
\, \bar{q}_{s}\gamma^{\mu}\tau^{I}q_{t} 
\quad , \quad 
Q_{ledq}=\bar{l}_{p}^{j}e_{r} \, \bar{d}_{s}q_{t}^{j}
\quad , \quad 
Q_{lequ}^{(1)}=\bar{l}_{p}^{j} \, e_{r}\epsilon_{jk} \, \bar{q}_{s}^{k}u_{t} 
\quad , \quad 
Q_{lequ}^{(3)}=\bar{l}_{p}^{j}\sigma_{\mu\nu}e_{r} \, \epsilon_{jk} \, \bar{q}_{s}^{k}\sigma^{\mu\nu}u_{t} \, ,
\end{equation}
where $p,r,s,t$ are family indices (omitted in the operators) and $j,k, I$ are $\mathrm{SU}(2)_\mathrm{L}$ ones, being $l=(\nu_L,e_L)^T$ and $q=(u_L,d_L)^T$ the corresponding doublets. They could be induced at tree level for instance by leptoquarks, which are hypothetical colored bosons that can couple to quarks and leptons in the same vertex.
\paragraph{Operators modifying vertices}
They contain one fermion current, a covariant derivative and two Higgs doublets. They are
 \begin{equation}
  \mathcal{Q}_{\varphi l}^{(3)}=\varphi^{\dagger}i\overset{\leftrightarrow}{D}\left.\hspace{-0.1cm}^{I}_{\mu}\right.\varphi \,\bar{l}_{p}\gamma^{\mu}l_{r}    \quad , \quad
 \mathcal{Q}_{\varphi q}^{(3)}=\varphi^{\dagger}i\overset{\leftrightarrow}{D}\left.\hspace{-0.1cm}_{\mu}^{I}\right.\varphi \,\bar{q}_{p}\gamma^{\mu}q_{r} \quad , \quad
 \mathcal{Q}_{\varphi ud}=i(\tilde{\varphi}^{\dagger}D_{\mu}\varphi)(\bar{u}_{p}\gamma^{\mu}d_{r}) \, ,
\end{equation}
where $\varphi$ is the Higgs doublet and ${D}_{\mu}^{I}$ contains the covariant derivative (see e.g. Ref.~\cite{Grzadkowski:2010es} for its precise definition). For instance, they could be generated at tree level by vector-like fermions. After electroweak symmetry breaking, one has $\varphi \supset 1/\sqrt{2} \, (0,v)^T$, which for these operators adds an anomalous contribution (right-handed for the last one) to the $W$ vertex, parametrically suppressed by $v^2/{\Lambda_{\mathrm{BSM}}^2}$, before integrating it out. Once the additional \( M_W^2 \sim v^2 \) factor in the denominator is included, both operator classes have the same parametric suppression. A simple tree-level example of the different steps for both cases is shown in Fig.~\ref{fig:bsm}.
The suppression of the matrix elements with respect to the SM for both classes is only $\epsilon_{\Gamma}^{D\tau}\sim v^2/\Lambda_{\mathrm{BSM}}^2$. A process mediated by weak currents, suppressed in the SM by powers of the masses of the corresponding bosons, is more sensitive to potential new physics and knowledge of combinations of the $\epsilon_{\Gamma}^{D\tau}$ couplings beyond the percent level naturally probes new physics particles with masses well above the TeV scale.

When searching for new physics, a change in strategy is required. In this context, we are not interested in trying to qualitatively grasp all the different pieces entering the extremely rich and complex hadronic dynamics emerging from QCD. Instead we have to deal with them as obstacles to making precise predictions, not bothering ourselves with theoretically inaccessible observables within the current state-of-the-art, such as the branching ratio of $\tau\to K\pi\pi\nu$ or the inclusive spectral function at exactly $1\, \mathrm{GeV}$. In order to obtain bounds of the $\epsilon_{\Gamma}^{D\tau}$ at or beyond percent level, one should focus on either
\begin{enumerate}
\item Observables where one has a very good control on the SM prediction and very precise experimental value.
\item Observables that come with an extra SM suppression.
\end{enumerate}
Note that extremely high precision in new physics distributions, or $\epsilon_{\Gamma}^{D\tau}$ prefactors, is unnecessary, given that no new physics has yet been observed (whether a new physics bound is $|\epsilon_{\Gamma}^{D\tau}|<10^{-3}$ or $1.1|\epsilon_{\Gamma}^{D\tau}|<10^{-3}$ is largely irrelevant). In our case, the most obvious observable in the first category is the $\tau\to P \nu_{\tau}$ channel, with $P=K,\pi$. One can easily generalize the form factors in the presence of BSM currents (see e.g. Ref.~\cite{Cirigliano:2021yto} for details) to find\footnote{One needs to be particularly careful when taking the inputs for the SM part and incorporate any possible BSM contamination on them. For instance, the definition of the Lagrangian of Eq.~(\ref{eq:leff1}) has been chosen such that $G_{\mu}$ is the muon decay parameter, as obtained from $\mu \to e^-\nu_\mu \bar{\nu}_e$, which only in the SM limit corresponds to the $G_{F}$ input, trivially related to the $M_{W}$ mass. This needs to be taken into account when matching to the SMEFT. On the other hand, $\hat{V}_{uD}$ is the input obtained from superallowed $\beta$ decays, mediated by vector current. The corresponding BSM shift, $\epsilon_{L}^{De}+\epsilon_{R}^{De}$, is incorporated into $\delta_{\mathrm{BSM}}^{(P)}$. Finally, $f_{P}$ may be taken for example from lattice QCD simulations where only experimental inputs largely insensitive to new physics, such as $M_{\Omega}$, are used.}
\begin{equation}
\Gamma(\tau \to P \nu_\tau) 
= \frac{m_\tau^3 f_P^2 G_\mu^2 |\hat{V}_{uD}|^2}{16\pi} \left( 1- \frac{m_P^2}{m_\tau^2} \right)^2 (1 + \delta_{\rm RC}^{(P)} ) \left( 1+ 2\,\delta_{\rm BSM}^{(P)} \right)~\quad ,\quad
\delta_{\rm BSM}^{(P)} ~= ~\epsilon_L^{D\tau} - \epsilon_L^{De} - \epsilon_R^{D\tau} - \epsilon_R^{De} - \frac{m_{P}^2}{m_\tau \, (m_u+m_D)} \epsilon_P^{D\tau}~\, .
\end{equation}
In principle, the same kind of generalization can be done for the two-pseudoscalar mode, finding a result of the form (see e.g.~\cite{Garces:2017jpz,Miranda:2018cpf,Cirigliano:2021yto})
\begin{equation}
\label{PP_spectrum}
\frac{d\Gamma^{}}{d s}   = \left[ \frac{d \hat \Gamma^{}}{d s} \right]_{\rm SM}  \bigg( 1 + 2 (\epsilon_L^{D\tau} + \epsilon _R^{D\tau} -  \epsilon_L^{De} - \epsilon _R^{De}) 
 +  a_S (s) \, \epsilon_S^{D\tau}  + 
a_T (s) \, \hat{\epsilon}_T^{D\tau}   +  {\cal O}(\epsilon^2)  \bigg) \, .
\end{equation}
Notice that new physics contributions, involving new form factors encoded in $a_{S}(s)$ and $a_{T}(s)$, can modify both the shape and the normalization. While precise enough experimental data exist for many modes, the challenge is how to precisely determine $d\hat{\Gamma}_{\mathrm{SM}}/ds$, considering they depend on generally unknown form factors.\footnote{One needs to be particularly careful when performing fits assuming a given parametrization for them, since it may become unclear whether any deviation from data may come from new physics or from an incomplete parametrization of the SM form factors.} As argued in Ref.~\cite{Cirigliano:2018dyk}, a possible exception is the $\tau^-\to \pi^-\pi\, \nu_{\tau}$ channel, which in the standard model only depends on one form factor, which can be taken for example from $e^{+}e^{+}\to\pi^{-}\pi^{0}$ data, after considering isospin breaking corrections.\footnote{Potential heavy new physics contamination in that process is negligible because the leading contribution is electromagnetic, thus not suppressed within the SM. Within the uncertainties one cannot even observe the effect of the SM $Z$ boson at that energy.}

The last type of observable in the first category are the precise SM tests with inclusive distributions. Generalizing the Källén-Lehmann spectral representation for the case in which beyond the standard model quark currents are present and, using G-parity, to re-interpret the SM nonstrange $VV$ vs $AA$ separation as a (at linear order in new physics) $(VV,VT)$ vs ($AA,AP,AS$) one, eventually finding (see Ref.~\cite{Cirigliano:2021yto} for details)
\begin{equation}
\rho_{V}^{\rm exp}(s)
\approx\left(1+2\epsilon_{L+R}^{d\tau}-2\epsilon_{L+R}^{de}\right)\frac{1}{\pi}\operatorname{Im}\Pi_{VV}^{(1+0)}(s)+6 \,\hat \epsilon^{d\tau}_{T}\left(1+\frac{2s}{m_{\tau}^{2}}\right)^{-1}\frac{\operatorname{Im}\Pi_{VT}}{\pi\,m_{\tau} }(s) \; ,\; 
\rho_{A}^{\rm exp}(s>4\,m_{\pi}^2)
\approx\left(1+2\epsilon_{L+R}^{d\tau}-2\epsilon_{L+R}^{de}-4\epsilon_{R}^{d\tau}\right)\frac{1}{\pi}\operatorname{Im}\Pi_{AA}^{(1+0)}(s)  \, ,
\label{eq:rhoAexp}
\end{equation}
which using the OPE (and now also lattice QCD) can be used to set bounds on potential new physics. The same, currently limited by the experimental situation, applies for the strange sector.

Regarding the second category, i.e. observables with an extra SM suppression, the most obvious example is the $\tau\to \pi\eta \nu_{\tau}$, whose SM suppression was discussed in section~\ref{sec:exclusive}. This suppression is not present if BSM currents, particularly the scalar one, are present. The corresponding extraction of new physics bounds was studied in detail in Ref.~\cite{Garces:2017jpz}.

Overall, as shown in Ref.~\cite{Cirigliano:2021yto}, the state-of-the-art of the field provides explicit confirmation of the SM-like (V-A) $\times$ (V-A) nature of charged current interactions involving the heaviest lepton and the lightest quarks at the percent level. A better understanding of long-distance radiative and isospin-breaking corrections is required to further improve these bounds. Given their nonperturbative nature, input from the lattice QCD community appears to be essential.

\subsection{Charged Lepton Flavor Violation}
Lepton flavor violation has not been observed for charged leptons. Even if we trivially extend the standard model to accommodate the tiny neutrino mass term ($\nu$SM), thus technically allowing for some charged lepton flavor violation, the corresponding suppression in the amplitudes, scaling as $(m_\nu/M_W)^2$ due to the unitarity of the PMNS matrix,  is too small to be observed in charged lepton decays. Nevertheless, one may entertain the possibility that beyond the standard model physics could trigger these decays. 

\begin{figure}[tbh]
    \centering
\includegraphics[width=0.59\linewidth]{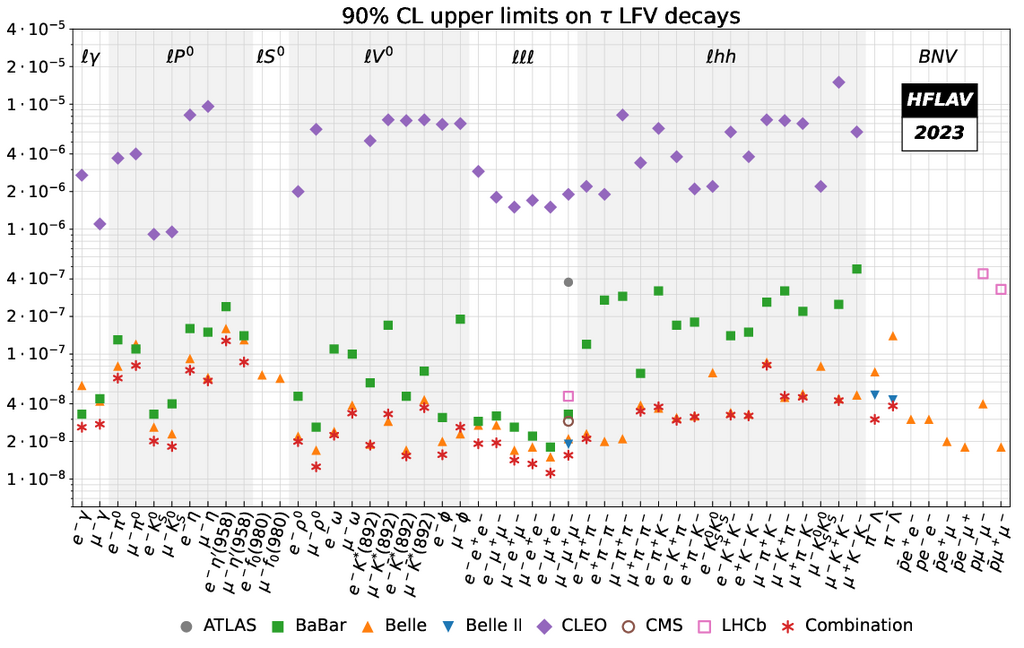}
    \caption{Current experimental limits on cLFV tau decays branching ratios. Table taken from Ref.~\cite{HeavyFlavorAveragingGroupHFLAV:2024ctg}.}
    \label{fig:taulfv}
\end{figure}

Assuming lepton flavor conservation is not an approximate symmetry for beyond the standard model particles, it appears plausible to observe it first in charged lepton flavor violation observables, as they naturally probe masses up to $10^4-10^5$ TeV for $\mu\to e$ transitions~\cite{Davidson:2022jai} and beyond $10 \, \mathrm{TeV}$ for $\tau\to e$ and $\tau\to \mu$ transitions.\footnote{Obviously some symmetry needs to be invoked to make the tau probes competitive. Typically if a model allows both $\tau\to e$ and $\tau\to \mu$ at tree level, one still expects bounds on $\mu\to e$ from loop corrections to dominate~\cite{Ardu:2022pzk}.} The current experimental status of lepton flavor violation in (mostly) hadronic $\tau$ decays is summarized in Figure~\ref{fig:taulfv}, taken from Ref.~\cite{HeavyFlavorAveragingGroupHFLAV:2024ctg}.

If charged lepton flavor conservation is violated by particles above the tau scale, one can study it with an EFT containing only the light SM degrees of freedom and symmetries, i.e. $\mathrm{SU}(3)_\mathrm{C} \times \mathrm{U}(1)_{\mathrm{em}}$.\footnote{This EFT is often called the LEFT~\cite{Jenkins:2017jig}, which also contains the Lagrangian studied in the previous subsection.} At dimension $D=6$ potential semileptonic operators emerge violating lepton flavor number~\cite{Celis:2014asa}\footnote{Dipole operators, starting at $D=5$, are also technically relevant for these decays, but usually better constrained by $\tau\to \ell \gamma$ searches.}
\begin{align}  \label{taulag2} 
\mathcal{L}_{eff}^{(\ell q)} &\supset   - \frac{1}{\Lambda^2}  \,  \sum_{q=u,d,s}  \Bigl\{    \left( \mathrm{C^{q}_{VR}} \, \bar \ell \,  \gamma^{\rho} \,  P_R \, \tau  \; + \mathrm{C^{q}_{VL}}  \, \bar \ell  \, \gamma^{\rho} \, P_L \, \tau  \;\right)    \bar q \, \gamma_{\rho} \,  q   \nonumber + \left( \mathrm{C^{q}_{AR}} \,   \bar \ell \, \gamma^{\rho} \,  P_R \, \tau  + \mathrm{C^{q}_{AL}} \,   \bar \ell \, \gamma^{\rho} \, P_L \, \tau \right)   \bar q \, \gamma_{\rho} \gamma_{5} \, q \nonumber \\ 
&+  \,   \left(  \mathrm{\tilde{C}^{q}_{SR}} \, \bar \ell \, P_L \, \tau     + \mathrm{\tilde{C}^{q}_{SL}} \bar \ell  \, P_R \, \tau  \right)   \bar q \,q \nonumber +  \,  \left(   \mathrm{\tilde{C}^{q}_{PR}}  \, \bar \ell \, P_L \, \tau   + \mathrm{\tilde{C}^{q}_{PL}}  \,  \bar \ell \, P_R \, \tau  \right)   \bar q \, \gamma_{5}\, q \\
&+  \,   \left( \mathrm{\tilde{C}^{q}_{TR}}  \,   \bar \ell \, \sigma^{\rho \nu}  P_L \, \tau  + \mathrm{\tilde{C}^{q}_{TL}}  \,  \bar \ell \, \sigma^{\rho \nu} \,  P_R\,  \tau     \right)  \bar q \, \sigma_{\rho \nu}\,  \, q  
+ \mathrm{h.c.} \Bigr\}  \,,
\end{align}
for $\ell=\mu,e$ and similar for $\bar{s}d$ operators if one also allows for flavor changing neutral currents in the quark sector.

From here, the logic is not very different compared to the standard decays. Hadronic matrix elements can be compiled into form factors, now mediated by neutral currents, and one can compute the branching ratios as functions of them. The form factors need to be estimated with the same type of nonperturbative techniques already described, without much motivation for reaching very high precision, as estimating the order of magnitude is often sufficient to get an idea of the potential reach of each decay mode in exploring new physics scenarios.

In the next step of this EFT ladder, one can assume that the corresponding BSM is heavier than the EW scale and one usually matches to the $D=6$ SMEFT. If it is much heavier than the EW scale, one should then consider the SMEFT running~\cite{Jenkins:2013wua,Jenkins:2017jig,Alonso:2013hga}. Examples of works incorporating these steps and analyzing the complementarity with different sectors are Refs.~\cite{Celis:2013xja,Husek:2020fru,Cirigliano:2021img,Plakias:2023esq}.


\section{Conclusions}
\label{sec:conclusions}
Hadronic tau decays constitute a powerful test of the interactions among quarks and leptons at relatively low energies. There are a large number of decay modes and experimentally observed distributions measured with very high precision, typically going beyond what we can currently predict from first principles. Unfortunately we cannot yet confirm that the numbers of table~\ref{tab:tau_decays} and the corresponding distributions match exactly those predicted by the Standard Model, even taking the Lagrangian inputs from other sectors, mostly due to hadronic uncertainties. New approaches to handle multi-hadron distributions in the resonance regime are needed.

Nevertheless, the strong efforts of the community have allowed for very precise predictions for many exclusive and inclusive modes within the standard model. They constitute fundamental tests directly probing the transition between the confinement regime and the asymptotic freedom in strong interactions, mostly matching experiments at the percent-level accuracy, confirming the success of the electroweak theory in one more sector. To achieve precision significantly beyond the percent level for the cleanest observables, improved experimental data and, particularly, a better understanding of long-distance radiative corrections are required.

\begin{ack}[Acknowledgments]%
I would like to thank Nils Hermansson-Truedsson, Bogdan Malaescu, Antonio Pich and Pablo Roig for useful comments on the manuscript. This work has been supported by the Generalitat Valenciana (Spain) through the plan GenT program (CIDEIG/2023/12) and by the Spanish Government (Agencia Estatal de Investigación MCIN/AEI/10.13039/501100011033) Grants No. PID2020–114473GB-I00 and No. PID2023-146220NB-I00.
\end{ack}


\bibliographystyle{Numbered-Style} 
\bibliography{reference}

\begin{thebibliography*}{100}
\providecommand{\bibtype}[1]{}
\providecommand{\url}[1]{{\tt #1}}
\providecommand{\urlprefix}{URL }
\expandafter\ifx\csname urlstyle\endcsname\relax
  \providecommand{\doi}[1]{doi:\discretionary{}{}{}#1}\else
  \providecommand{\doi}{doi:\discretionary{}{}{}\begingroup \urlstyle{rm}\Url}\fi
\providecommand{\bibinfo}[2]{#2}
\providecommand{\eprint}[2][]{\url{#2}}
\makeatletter\def\@biblabel#1{\bibinfo{label}{[#1]}}\makeatother

\bibtype{Article}%
\bibitem{HeavyFlavorAveragingGroupHFLAV:2024ctg}
\bibinfo{author}{Swagato Banerjee}, et al. (\bibinfo{collaboration}{Heavy Flavor Averaging Group (HFLAV)}), \bibinfo{title}{{Averages of $b$-hadron, $c$-hadron, and $\tau$-lepton properties as of 2023}}  (\bibinfo{year}{2024}), \eprint{2411.18639}.

\bibtype{Article}%
\bibitem{Belle-II:2023izd}
\bibinfo{author}{I. Adachi}, et al. (\bibinfo{collaboration}{Belle-II}), \bibinfo{title}{{Measurement of the \ensuremath{\tau}-lepton mass with the Belle II experiment}}, \bibinfo{journal}{Phys. Rev. D} \bibinfo{volume}{108} (\bibinfo{number}{3}) (\bibinfo{year}{2023}) \bibinfo{pages}{032006}, \bibinfo{doi}{\doi{10.1103/PhysRevD.108.032006}}, \eprint{2305.19116}.

\bibtype{Article}%
\bibitem{ALEPH:1990ndp}
\bibinfo{author}{D. Decamp}, et al. (\bibinfo{collaboration}{ALEPH}), \bibinfo{title}{{ALEPH: A detector for electron-positron annihilations at LEP}}, \bibinfo{journal}{Nucl. Instrum. Meth. A} \bibinfo{volume}{294} (\bibinfo{year}{1990}) \bibinfo{pages}{121--178}, \bibinfo{doi}{\doi{10.1016/0168-9002(90)91831-U}}, \bibinfo{note}{[Erratum: Nucl.Instrum.Meth.A 303, 393 (1991)]}.

\bibtype{Article}%
\bibitem{OPAL:1990yff}
\bibinfo{author}{K. Ahmet}, et al. (\bibinfo{collaboration}{OPAL}), \bibinfo{title}{{The OPAL detector at LEP}}, \bibinfo{journal}{Nucl. Instrum. Meth. A} \bibinfo{volume}{305} (\bibinfo{year}{1991}) \bibinfo{pages}{275--319}, \bibinfo{doi}{\doi{10.1016/0168-9002(91)90547-4}}.

\bibtype{Article}%
\bibitem{DELPHI:1990cdc}
\bibinfo{author}{P.~A. Aarnio}, et al. (\bibinfo{collaboration}{DELPHI}), \bibinfo{title}{{The DELPHI detector at LEP}}, \bibinfo{journal}{Nucl. Instrum. Meth. A} \bibinfo{volume}{303} (\bibinfo{year}{1991}) \bibinfo{pages}{233--276}, \bibinfo{doi}{\doi{10.1016/0168-9002(91)90793-P}}.

\bibtype{Article}%
\bibitem{CLEO:1991qyy}
\bibinfo{author}{Y. Kubota}, et al. (\bibinfo{collaboration}{CLEO}), \bibinfo{title}{{The CLEO-II detector}}, \bibinfo{journal}{Nucl. Instrum. Meth. A} \bibinfo{volume}{320} (\bibinfo{year}{1992}) \bibinfo{pages}{66--113}, \bibinfo{doi}{\doi{10.1016/0168-9002(92)90770-5}}.

\bibtype{Article}%
\bibitem{BaBar:2001yhh}
\bibinfo{author}{Bernard Aubert}, et al. (\bibinfo{collaboration}{BaBar}), \bibinfo{title}{{The BaBar detector}}, \bibinfo{journal}{Nucl. Instrum. Meth. A} \bibinfo{volume}{479} (\bibinfo{year}{2002}) \bibinfo{pages}{1--116}, \bibinfo{doi}{\doi{10.1016/S0168-9002(01)02012-5}}, \eprint{hep-ex/0105044}.

\bibtype{Article}%
\bibitem{Belle:2000cnh}
\bibinfo{author}{A. Abashian}, et al. (\bibinfo{collaboration}{Belle}), \bibinfo{title}{{The Belle Detector}}, \bibinfo{journal}{Nucl. Instrum. Meth. A} \bibinfo{volume}{479} (\bibinfo{year}{2002}) \bibinfo{pages}{117--232}, \bibinfo{doi}{\doi{10.1016/S0168-9002(01)02013-7}}.

\bibtype{Article}%
\bibitem{Davier:2005xq}
\bibinfo{author}{Michel Davier}, \bibinfo{author}{Andreas Hocker}, \bibinfo{author}{Zhiqing Zhang}, \bibinfo{title}{{The Physics of Hadronic Tau Decays}}, \bibinfo{journal}{Rev. Mod. Phys.} \bibinfo{volume}{78} (\bibinfo{year}{2006}) \bibinfo{pages}{1043--1109}, \bibinfo{doi}{\doi{10.1103/RevModPhys.78.1043}}, \eprint{hep-ph/0507078}.

\bibtype{Article}%
\bibitem{Davier:2010fmf}
\bibinfo{author}{M. Davier}, \bibinfo{author}{A. Hoecker}, \bibinfo{author}{G. Lopez~Castro}, \bibinfo{author}{B. Malaescu}, \bibinfo{author}{X.~H. Mo}, \bibinfo{author}{G. Toledo~Sanchez}, \bibinfo{author}{P. Wang}, \bibinfo{author}{C.~Z. Yuan}, \bibinfo{author}{Z. Zhang}, \bibinfo{title}{{The Discrepancy Between tau and e+e- Spectral Functions Revisited and the Consequences for the Muon Magnetic Anomaly}}, \bibinfo{journal}{Eur. Phys. J. C} \bibinfo{volume}{66} (\bibinfo{year}{2010}) \bibinfo{pages}{127--136}, \bibinfo{doi}{\doi{10.1140/epjc/s10052-009-1219-4}}, \eprint{0906.5443}.

\bibtype{Article}%
\bibitem{ParticleDataGroup:2024cfk}
\bibinfo{author}{S. Navas}, et al. (\bibinfo{collaboration}{Particle Data Group}), \bibinfo{title}{{Review of particle physics}}, \bibinfo{journal}{Phys. Rev. D} \bibinfo{volume}{110} (\bibinfo{number}{3}) (\bibinfo{year}{2024}) \bibinfo{pages}{030001}, \bibinfo{doi}{\doi{10.1103/PhysRevD.110.030001}}.

\bibtype{Article}%
\bibitem{Wilson:1969zs}
\bibinfo{author}{Kenneth~G. Wilson}, \bibinfo{title}{{Nonlagrangian models of current algebra}}, \bibinfo{journal}{Phys. Rev.} \bibinfo{volume}{179} (\bibinfo{year}{1969}) \bibinfo{pages}{1499--1512}, \bibinfo{doi}{\doi{10.1103/PhysRev.179.1499}}.

\bibtype{Article}%
\bibitem{Braaten:1993ha}
\bibinfo{author}{Eric Braaten}, \bibinfo{title}{{Tau polarimetry with inclusive decays}}, \bibinfo{journal}{Phys. Rev. Lett.} \bibinfo{volume}{71} (\bibinfo{year}{1993}) \bibinfo{pages}{1316--1319}, \bibinfo{doi}{\doi{10.1103/PhysRevLett.71.1316}}, \eprint{hep-ph/9304227}.

\bibtype{Article}%
\bibitem{Marciano:1988vm}
\bibinfo{author}{W.~J. Marciano}, \bibinfo{author}{A. Sirlin}, \bibinfo{title}{{Electroweak Radiative Corrections to tau Decay}}, \bibinfo{journal}{Phys. Rev. Lett.} \bibinfo{volume}{61} (\bibinfo{year}{1988}) \bibinfo{pages}{1815--1818}, \bibinfo{doi}{\doi{10.1103/PhysRevLett.61.1815}}.

\bibtype{Article}%
\bibitem{Erler:2002mv}
\bibinfo{author}{Jens Erler}, \bibinfo{title}{{Electroweak radiative corrections to semileptonic tau decays}}, \bibinfo{journal}{Rev. Mex. Fis.} \bibinfo{volume}{50} (\bibinfo{year}{2004}) \bibinfo{pages}{200--202}, \eprint{hep-ph/0211345}.

\bibtype{Article}%
\bibitem{Tsai:1971vv}
\bibinfo{author}{Yung-Su Tsai}, \bibinfo{title}{{Decay Correlations of Heavy Leptons in e+ e- ---\ensuremath{>} Lepton+ Lepton-}}, \bibinfo{journal}{Phys. Rev. D} \bibinfo{volume}{4} (\bibinfo{year}{1971}) \bibinfo{pages}{2821}, \bibinfo{doi}{\doi{10.1103/PhysRevD.13.771}}, \bibinfo{note}{[Erratum: Phys.Rev.D 13, 771 (1976)]}.

\bibtype{Article}%
\bibitem{Guo:2010dv}
\bibinfo{author}{Zhi-Hui Guo}, \bibinfo{author}{Pablo Roig}, \bibinfo{title}{{One meson radiative tau decays}}, \bibinfo{journal}{Phys. Rev. D} \bibinfo{volume}{82} (\bibinfo{year}{2010}) \bibinfo{pages}{113016}, \bibinfo{doi}{\doi{10.1103/PhysRevD.82.113016}}, \eprint{1009.2542}.

\bibtype{Article}%
\bibitem{Braaten:1990ef}
\bibinfo{author}{Eric Braaten}, \bibinfo{author}{Chong-Sheng Li}, \bibinfo{title}{{Electroweak radiative corrections to the semihadronic decay rate of the tau lepton}}, \bibinfo{journal}{Phys. Rev. D} \bibinfo{volume}{42} (\bibinfo{year}{1990}) \bibinfo{pages}{3888--3891}, \bibinfo{doi}{\doi{10.1103/PhysRevD.42.3888}}.

\bibtype{Article}%
\bibitem{Cirigliano:2021yto}
\bibinfo{author}{Vincenzo Cirigliano}, \bibinfo{author}{David D\'\i{}az-Calder\'on}, \bibinfo{author}{Adam Falkowski}, \bibinfo{author}{Mart\'\i{}n Gonz\'alez-Alonso}, \bibinfo{author}{Antonio Rodr\'\i{}guez-S\'anchez}, \bibinfo{title}{{Semileptonic tau decays beyond the Standard Model}}, \bibinfo{journal}{JHEP} \bibinfo{volume}{04} (\bibinfo{year}{2022}) \bibinfo{pages}{152}, \bibinfo{doi}{\doi{10.1007/JHEP04(2022)152}}, \eprint{2112.02087}.

\bibtype{Article}%
\bibitem{Rosner:2015wva}
\bibinfo{author}{Jonathan~L. Rosner}, \bibinfo{author}{Sheldon Stone}, \bibinfo{author}{Ruth~S. Van~de Water}, \bibinfo{title}{{Leptonic Decays of Charged Pseudoscalar Mesons - 2015}}  (\bibinfo{year}{2015}), \eprint{1509.02220}.

\bibtype{Article}%
\bibitem{Roig:2019rwf}
\bibinfo{author}{Pablo Roig}, \bibinfo{title}{{Semileptonic $\tau$ decays: powerful probes of non-standard charged current weak interactions}}, \bibinfo{journal}{EPJ Web Conf.} \bibinfo{volume}{212} (\bibinfo{year}{2019}) \bibinfo{pages}{08002}, \bibinfo{doi}{\doi{10.1051/epjconf/201921208002}}, \eprint{1903.02682}.

\bibtype{Article}%
\bibitem{Arroyo-Urena:2021nil}
\bibinfo{author}{M.~A. Arroyo-Ure\~na}, \bibinfo{author}{G. Hern\'andez-Tom\'e}, \bibinfo{author}{G. L\'opez-Castro}, \bibinfo{author}{P. Roig}, \bibinfo{author}{I. Rosell}, \bibinfo{title}{{Radiative corrections to \ensuremath{\tau}\textrightarrow{}\ensuremath{\pi}(K)\ensuremath{\nu}\ensuremath{\tau}[\ensuremath{\gamma}]: A reliable new physics test}}, \bibinfo{journal}{Phys. Rev. D} \bibinfo{volume}{104} (\bibinfo{number}{9}) (\bibinfo{year}{2021}) \bibinfo{pages}{L091502}, \bibinfo{doi}{\doi{10.1103/PhysRevD.104.L091502}}, \eprint{2107.04603}.

\bibtype{Article}%
\bibitem{FlavourLatticeAveragingGroupFLAG:2024oxs}
\bibinfo{author}{Y. Aoki}, et al. (\bibinfo{collaboration}{Flavour Lattice Averaging Group (FLAG)}), \bibinfo{title}{{FLAG Review 2024}}  (\bibinfo{year}{2024}), \eprint{2411.04268}.

\bibtype{Article}%
\bibitem{Pich:2013lsa}
\bibinfo{author}{Antonio Pich}, \bibinfo{title}{{Precision Tau Physics}}, \bibinfo{journal}{Prog. Part. Nucl. Phys.} \bibinfo{volume}{75} (\bibinfo{year}{2014}) \bibinfo{pages}{41--85}, \bibinfo{doi}{\doi{10.1016/j.ppnp.2013.11.002}}, \eprint{1310.7922}.

\bibtype{Article}%
\bibitem{Aguilar:2024ybr}
\bibinfo{author}{Daniel A.~L\'opez Aguilar}, \bibinfo{author}{Javier Rend\'on}, \bibinfo{author}{Pablo Roig}, \bibinfo{title}{{CP violation in two-meson Tau decays induced by heavy new physics}}, \bibinfo{journal}{JHEP} \bibinfo{volume}{01} (\bibinfo{year}{2025}) \bibinfo{pages}{105}, \bibinfo{doi}{\doi{10.1007/JHEP01(2025)105}}, \eprint{2409.05588}.

\bibtype{Article}%
\bibitem{Kuhn:1992nz}
\bibinfo{author}{Johann~H. Kuhn}, \bibinfo{author}{E. Mirkes}, \bibinfo{title}{{Structure functions in tau decays}}, \bibinfo{journal}{Z. Phys. C} \bibinfo{volume}{56} (\bibinfo{year}{1992}) \bibinfo{pages}{661--672}, \bibinfo{doi}{\doi{10.1007/BF01474741}}, \bibinfo{note}{[Erratum: Z.Phys.C 67, 364 (1995)]}.

\bibtype{Article}%
\bibitem{Lepage:1980fj}
\bibinfo{author}{G.~Peter Lepage}, \bibinfo{author}{Stanley~J. Brodsky}, \bibinfo{title}{{Exclusive Processes in Perturbative Quantum Chromodynamics}}, \bibinfo{journal}{Phys. Rev. D} \bibinfo{volume}{22} (\bibinfo{year}{1980}) \bibinfo{pages}{2157}, \bibinfo{doi}{\doi{10.1103/PhysRevD.22.2157}}.

\bibtype{Article}%
\bibitem{Melic:1998qr}
\bibinfo{author}{B. Melic}, \bibinfo{author}{B. Nizic}, \bibinfo{author}{K. Passek}, \bibinfo{title}{{Complete next-to-leading order perturbative QCD prediction for the pion form-factor}}, \bibinfo{journal}{Phys. Rev. D} \bibinfo{volume}{60} (\bibinfo{year}{1999}) \bibinfo{pages}{074004}, \bibinfo{doi}{\doi{10.1103/PhysRevD.60.074004}}, \eprint{hep-ph/9802204}.

\bibtype{Article}%
\bibitem{Simula:2023ujs}
\bibinfo{author}{Silvano Simula}, \bibinfo{author}{Ludovico Vittorio}, \bibinfo{title}{{Dispersive analysis of the experimental data on the electromagnetic form factor of charged pions at spacelike momenta}}, \bibinfo{journal}{Phys. Rev. D} \bibinfo{volume}{108} (\bibinfo{number}{9}) (\bibinfo{year}{2023}) \bibinfo{pages}{094013}, \bibinfo{doi}{\doi{10.1103/PhysRevD.108.094013}}, \eprint{2309.02135}.

\bibtype{Article}%
\bibitem{RuizArriola:2024gwb}
\bibinfo{author}{Enrique Ruiz~Arriola}, \bibinfo{author}{Pablo Sanchez-Puertas}, \bibinfo{title}{{Phase of the electromagnetic form factor of the pion}}, \bibinfo{journal}{Phys. Rev. D} \bibinfo{volume}{110} (\bibinfo{number}{5}) (\bibinfo{year}{2024}) \bibinfo{pages}{054003}, \bibinfo{doi}{\doi{10.1103/PhysRevD.110.054003}}, \eprint{2403.07121}.

\bibtype{Article}%
\bibitem{Omnes:1958hv}
\bibinfo{author}{R. Omnes}, \bibinfo{title}{{On the Solution of certain singular integral equations of quantum field theory}}, \bibinfo{journal}{Nuovo Cim.} \bibinfo{volume}{8} (\bibinfo{year}{1958}) \bibinfo{pages}{316--326}, \bibinfo{doi}{\doi{10.1007/BF02747746}}.

\bibtype{Article}%
\bibitem{Watson:1954uc}
\bibinfo{author}{Kenneth~M. Watson}, \bibinfo{title}{{Some general relations between the photoproduction and scattering of pi mesons}}, \bibinfo{journal}{Phys. Rev.} \bibinfo{volume}{95} (\bibinfo{year}{1954}) \bibinfo{pages}{228--236}, \bibinfo{doi}{\doi{10.1103/PhysRev.95.228}}.

\bibtype{Article}%
\bibitem{Oller:2025leg}
\bibinfo{author}{Jos\'e~Antonio Oller}, \bibinfo{title}{{Coupled-channel formalism}}  (\bibinfo{year}{2025}), \eprint{2501.10000}.

\bibtype{Article}%
\bibitem{Gasser:1983yg}
\bibinfo{author}{J. Gasser}, \bibinfo{author}{H. Leutwyler}, \bibinfo{title}{{Chiral Perturbation Theory to One Loop}}, \bibinfo{journal}{Annals Phys.} \bibinfo{volume}{158} (\bibinfo{year}{1984}) \bibinfo{pages}{142}, \bibinfo{doi}{\doi{10.1016/0003-4916(84)90242-2}}.

\bibtype{Article}%
\bibitem{Gasser:1984ux}
\bibinfo{author}{J. Gasser}, \bibinfo{author}{H. Leutwyler}, \bibinfo{title}{{Low-Energy Expansion of Meson Form-Factors}}, \bibinfo{journal}{Nucl. Phys. B} \bibinfo{volume}{250} (\bibinfo{year}{1985}) \bibinfo{pages}{517--538}, \bibinfo{doi}{\doi{10.1016/0550-3213(85)90493-6}}.

\bibtype{Article}%
\bibitem{Ademollo:1964sr}
\bibinfo{author}{M. Ademollo}, \bibinfo{author}{Raoul Gatto}, \bibinfo{title}{{Nonrenormalization Theorem for the Strangeness Violating Vector Currents}}, \bibinfo{journal}{Phys. Rev. Lett.} \bibinfo{volume}{13} (\bibinfo{year}{1964}) \bibinfo{pages}{264--265}, \bibinfo{doi}{\doi{10.1103/PhysRevLett.13.264}}.

\bibtype{Article}%
\bibitem{Nieves:2011gb}
\bibinfo{author}{J. Nieves}, \bibinfo{author}{A. Pich}, \bibinfo{author}{E. Ruiz~Arriola}, \bibinfo{title}{{Large-Nc Properties of the rho and f0(600) Mesons from Unitary Resonance Chiral Dynamics}}, \bibinfo{journal}{Phys. Rev. D} \bibinfo{volume}{84} (\bibinfo{year}{2011}) \bibinfo{pages}{096002}, \bibinfo{doi}{\doi{10.1103/PhysRevD.84.096002}}, \eprint{1107.3247}.

\bibtype{Article}%
\bibitem{Ecker:1988te}
\bibinfo{author}{G. Ecker}, \bibinfo{author}{J. Gasser}, \bibinfo{author}{A. Pich}, \bibinfo{author}{E. de Rafael}, \bibinfo{title}{{The Role of Resonances in Chiral Perturbation Theory}}, \bibinfo{journal}{Nucl. Phys. B} \bibinfo{volume}{321} (\bibinfo{year}{1989}) \bibinfo{pages}{311--342}, \bibinfo{doi}{\doi{10.1016/0550-3213(89)90346-5}}.

\bibtype{Article}%
\bibitem{Ecker:1989yg}
\bibinfo{author}{G. Ecker}, \bibinfo{author}{J. Gasser}, \bibinfo{author}{H. Leutwyler}, \bibinfo{author}{A. Pich}, \bibinfo{author}{E. de Rafael}, \bibinfo{title}{{Chiral Lagrangians for Massive Spin 1 Fields}}, \bibinfo{journal}{Phys. Lett. B} \bibinfo{volume}{223} (\bibinfo{year}{1989}) \bibinfo{pages}{425--432}, \bibinfo{doi}{\doi{10.1016/0370-2693(89)91627-4}}.

\bibtype{Article}%
\bibitem{Rosell:2004mn}
\bibinfo{author}{I. Rosell}, \bibinfo{author}{J.~J. Sanz-Cillero}, \bibinfo{author}{A. Pich}, \bibinfo{title}{{Quantum loops in the resonance chiral theory: The Vector form-factor}}, \bibinfo{journal}{JHEP} \bibinfo{volume}{08} (\bibinfo{year}{2004}) \bibinfo{pages}{042}, \bibinfo{doi}{\doi{10.1088/1126-6708/2004/08/042}}, \eprint{hep-ph/0407240}.

\bibtype{Article}%
\bibitem{Pich:2008jm}
\bibinfo{author}{A. Pich}, \bibinfo{author}{I. Rosell}, \bibinfo{author}{J.~J. Sanz-Cillero}, \bibinfo{title}{{Form-factors and current correlators: Chiral couplings L(10)mu) **r(mu) and C(87)**r(mu) at NLO in 1/N(C)}}, \bibinfo{journal}{JHEP} \bibinfo{volume}{07} (\bibinfo{year}{2008}) \bibinfo{pages}{014}, \bibinfo{doi}{\doi{10.1088/1126-6708/2008/07/014}}, \eprint{0803.1567}.

\bibtype{Article}%
\bibitem{Guerrero:1997ku}
\bibinfo{author}{Francisco Guerrero}, \bibinfo{author}{Antonio Pich}, \bibinfo{title}{{Effective field theory description of the pion form-factor}}, \bibinfo{journal}{Phys. Lett. B} \bibinfo{volume}{412} (\bibinfo{year}{1997}) \bibinfo{pages}{382--388}, \bibinfo{doi}{\doi{10.1016/S0370-2693(97)01070-8}}, \eprint{hep-ph/9707347}.

\bibtype{Article}%
\bibitem{Colangelo:2018mtw}
\bibinfo{author}{Gilberto Colangelo}, \bibinfo{author}{Martin Hoferichter}, \bibinfo{author}{Peter Stoffer}, \bibinfo{title}{{Two-pion contribution to hadronic vacuum polarization}}, \bibinfo{journal}{JHEP} \bibinfo{volume}{02} (\bibinfo{year}{2019}) \bibinfo{pages}{006}, \bibinfo{doi}{\doi{10.1007/JHEP02(2019)006}}, \eprint{1810.00007}.

\bibtype{Article}%
\bibitem{Kirk:2024oyl}
\bibinfo{author}{Matthew Kirk}, \bibinfo{author}{Bastian Kubis}, \bibinfo{author}{M\'eril Reboud}, \bibinfo{author}{Danny van Dyk}, \bibinfo{title}{{A simple parametrisation of the pion form factor}}, \bibinfo{journal}{Phys. Lett. B} \bibinfo{volume}{861} (\bibinfo{year}{2025}) \bibinfo{pages}{139266}, \bibinfo{doi}{\doi{10.1016/j.physletb.2025.139266}}, \eprint{2410.13764}.

\bibtype{Article}%
\bibitem{Belle:2008xpe}
\bibinfo{author}{M. Fujikawa}, et al. (\bibinfo{collaboration}{Belle}), \bibinfo{title}{{High-Statistics Study of the tau- ---\ensuremath{>} pi- pi0 nu(tau) Decay}}, \bibinfo{journal}{Phys. Rev. D} \bibinfo{volume}{78} (\bibinfo{year}{2008}) \bibinfo{pages}{072006}, \bibinfo{doi}{\doi{10.1103/PhysRevD.78.072006}}, \eprint{0805.3773}.

\bibtype{Article}%
\bibitem{Belle:2007goc}
\bibinfo{author}{D. Epifanov}, et al. (\bibinfo{collaboration}{Belle}), \bibinfo{title}{{Study of tau- ---\ensuremath{>} K(S) pi- nu(tau) decay at Belle}}, \bibinfo{journal}{Phys. Lett. B} \bibinfo{volume}{654} (\bibinfo{year}{2007}) \bibinfo{pages}{65--73}, \bibinfo{doi}{\doi{10.1016/j.physletb.2007.08.045}}, \eprint{0706.2231}.

\bibtype{Article}%
\bibitem{Antonelli:2013usa}
\bibinfo{author}{Mario Antonelli}, \bibinfo{author}{Vincenzo Cirigliano}, \bibinfo{author}{Alberto Lusiani}, \bibinfo{author}{Emilie Passemar}, \bibinfo{title}{{Predicting the $\tau$ strange branching ratios and implications for $V_{us}$}}, \bibinfo{journal}{JHEP} \bibinfo{volume}{10} (\bibinfo{year}{2013}) \bibinfo{pages}{070}, \bibinfo{doi}{\doi{10.1007/JHEP10(2013)070}}, \eprint{1304.8134}.

\bibtype{Article}%
\bibitem{Jamin:2001zq}
\bibinfo{author}{Matthias Jamin}, \bibinfo{author}{Jose~Antonio Oller}, \bibinfo{author}{Antonio Pich}, \bibinfo{title}{{Strangeness changing scalar form-factors}}, \bibinfo{journal}{Nucl. Phys. B} \bibinfo{volume}{622} (\bibinfo{year}{2002}) \bibinfo{pages}{279--308}, \bibinfo{doi}{\doi{10.1016/S0550-3213(01)00605-8}}, \eprint{hep-ph/0110193}.

\bibtype{Article}%
\bibitem{Jamin:2006tj}
\bibinfo{author}{M. Jamin}, \bibinfo{author}{J.~A. Oller}, \bibinfo{author}{A. Pich}, \bibinfo{title}{{Scalar K pi form factor and light quark masses}}, \bibinfo{journal}{Phys. Rev. D} \bibinfo{volume}{74} (\bibinfo{year}{2006}) \bibinfo{pages}{074009}, \bibinfo{doi}{\doi{10.1103/PhysRevD.74.074009}}, \eprint{hep-ph/0605095}.

\bibtype{Article}%
\bibitem{Jamin:2006tk}
\bibinfo{author}{Matthias Jamin}, \bibinfo{author}{Antonio Pich}, \bibinfo{author}{Jorge Portoles}, \bibinfo{title}{{Spectral distribution for the decay tau ---\ensuremath{>} nu(tau) K pi}}, \bibinfo{journal}{Phys. Lett. B} \bibinfo{volume}{640} (\bibinfo{year}{2006}) \bibinfo{pages}{176--181}, \bibinfo{doi}{\doi{10.1016/j.physletb.2006.06.058}}, \eprint{hep-ph/0605096}.

\bibtype{Article}%
\bibitem{Jamin:2008qg}
\bibinfo{author}{Matthias Jamin}, \bibinfo{author}{Antonio Pich}, \bibinfo{author}{Jorge Portoles}, \bibinfo{title}{{What can be learned from the Belle spectrum for the decay - tau- ---\ensuremath{>} nu(tau) K(S) pi-}}, \bibinfo{journal}{Phys. Lett. B} \bibinfo{volume}{664} (\bibinfo{year}{2008}) \bibinfo{pages}{78--83}, \bibinfo{doi}{\doi{10.1016/j.physletb.2008.04.049}}, \eprint{0803.1786}.

\bibtype{Article}%
\bibitem{Boito:2008fq}
\bibinfo{author}{Diogo~R. Boito}, \bibinfo{author}{Rafel Escribano}, \bibinfo{author}{Matthias Jamin}, \bibinfo{title}{{K pi vector form-factor, dispersive constraints and tau ---\ensuremath{>} nu(tau) K pi decays}}, \bibinfo{journal}{Eur. Phys. J. C} \bibinfo{volume}{59} (\bibinfo{year}{2009}) \bibinfo{pages}{821--829}, \bibinfo{doi}{\doi{10.1140/epjc/s10052-008-0834-9}}, \eprint{0807.4883}.

\bibtype{Article}%
\bibitem{Boito:2010me}
\bibinfo{author}{D.~R. Boito}, \bibinfo{author}{R. Escribano}, \bibinfo{author}{M. Jamin}, \bibinfo{title}{{K $\pi$ vector form factor constrained by $\tau -> K\ pi \nu_\tau$ and $K_{l3}$ decays}}, \bibinfo{journal}{JHEP} \bibinfo{volume}{09} (\bibinfo{year}{2010}) \bibinfo{pages}{031}, \bibinfo{doi}{\doi{10.1007/JHEP09(2010)031}}, \eprint{1007.1858}.

\bibtype{Article}%
\bibitem{Cirigliano:2017tqn}
\bibinfo{author}{Vincenzo Cirigliano}, \bibinfo{author}{Andreas Crivellin}, \bibinfo{author}{Martin Hoferichter}, \bibinfo{title}{{No-go theorem for nonstandard explanations of the $\tau\to K_S\pi\nu_\tau$ CP asymmetry}}, \bibinfo{journal}{Phys. Rev. Lett.} \bibinfo{volume}{120} (\bibinfo{number}{14}) (\bibinfo{year}{2018}) \bibinfo{pages}{141803}, \bibinfo{doi}{\doi{10.1103/PhysRevLett.120.141803}}, \eprint{1712.06595}.

\bibtype{Article}%
\bibitem{Rendon:2019awg}
\bibinfo{author}{Javier Rend\'on}, \bibinfo{author}{Pablo Roig}, \bibinfo{author}{Genaro Toledo~S\'anchez}, \bibinfo{title}{{Effective-field theory analysis of the $\tau^{-}\rightarrow (K \pi)^{-}\nu_{\tau}$ decays}}, \bibinfo{journal}{Phys. Rev. D} \bibinfo{volume}{99} (\bibinfo{number}{9}) (\bibinfo{year}{2019}) \bibinfo{pages}{093005}, \bibinfo{doi}{\doi{10.1103/PhysRevD.99.093005}}, \eprint{1902.08143}.

\bibtype{Article}%
\bibitem{Bigi:2005ts}
\bibinfo{author}{I.~I. Bigi}, \bibinfo{author}{A.~I. Sanda}, \bibinfo{title}{{A 'Known' CP asymmetry in tau decays}}, \bibinfo{journal}{Phys. Lett. B} \bibinfo{volume}{625} (\bibinfo{year}{2005}) \bibinfo{pages}{47--52}, \bibinfo{doi}{\doi{10.1016/j.physletb.2005.08.033}}, \eprint{hep-ph/0506037}.

\bibtype{Article}%
\bibitem{BaBar:2011pij}
\bibinfo{author}{J.~P. Lees}, et al. (\bibinfo{collaboration}{BaBar}), \bibinfo{title}{{Search for CP Violation in the Decay $\tau^- -> \pi^- K^0_S (>= 0 \pi^0) \nu_tau$}}, \bibinfo{journal}{Phys. Rev. D} \bibinfo{volume}{85} (\bibinfo{year}{2012}) \bibinfo{pages}{031102}, \bibinfo{doi}{\doi{10.1103/PhysRevD.85.031102}}, \bibinfo{note}{[Erratum: Phys.Rev.D 85, 099904 (2012)]}, \eprint{1109.1527}.

\bibtype{Article}%
\bibitem{Grossman:2011zk}
\bibinfo{author}{Yuval Grossman}, \bibinfo{author}{Yosef Nir}, \bibinfo{title}{{CP Violation in $\tau^\pm \to \pi^\pm K_S\nu$ and $D^\pm \to \pi^\pm K_S$: The Importance of $K_S - K_L$ Interference}}, \bibinfo{journal}{JHEP} \bibinfo{volume}{04} (\bibinfo{year}{2012}) \bibinfo{pages}{002}, \bibinfo{doi}{\doi{10.1007/JHEP04(2012)002}}, \eprint{1110.3790}.

\bibtype{Article}%
\bibitem{Descotes-Genon:2014tla}
\bibinfo{author}{S. Descotes-Genon}, \bibinfo{author}{B. Moussallam}, \bibinfo{title}{{Analyticity of $\eta \pi $ isospin-violating form factors and the $\tau \rightarrow \eta \pi \nu $ second-class decay}}, \bibinfo{journal}{Eur. Phys. J. C} \bibinfo{volume}{74} (\bibinfo{year}{2014}) \bibinfo{pages}{2946}, \bibinfo{doi}{\doi{10.1140/epjc/s10052-014-2946-8}}, \eprint{1404.0251}.

\bibtype{Article}%
\bibitem{Escribano:2016ntp}
\bibinfo{author}{Rafel Escribano}, \bibinfo{author}{Sergi Gonzalez-Solis}, \bibinfo{author}{Pablo Roig}, \bibinfo{title}{{Predictions on the second-class current decays $\tau^{-}\to\pi^{-}\eta^{(\prime)}\nu_{\tau}$}}, \bibinfo{journal}{Phys. Rev. D} \bibinfo{volume}{94} (\bibinfo{number}{3}) (\bibinfo{year}{2016}) \bibinfo{pages}{034008}, \bibinfo{doi}{\doi{10.1103/PhysRevD.94.034008}}, \eprint{1601.03989}.

\bibtype{Inproceedings}%
\bibitem{Moussallam:2021flg}
\bibinfo{author}{B. Moussallam}, \bibinfo{title}{{Deriving experimental constraints on the scalar form factor in the second-class $\tau \to\eta \pi \nu$ mode}}, in: \bibinfo{booktitle}{{16th International Workshop on Tau Lepton Physics~}} \bibinfo{year}{2021}, \eprint{2112.04429}.

\bibtype{Article}%
\bibitem{Hayasaka:2009zz}
\bibinfo{author}{K. Hayasaka} (\bibinfo{collaboration}{Belle}), \bibinfo{title}{{Second class current in tau ---\ensuremath{>} pi eta nu analysis and measurement of tau ---\ensuremath{>} h h-prime h'' nu from Belle: electroweak physics from Belle}}, \bibinfo{journal}{PoS} \bibinfo{volume}{EPS-HEP2009} (\bibinfo{year}{2009}) \bibinfo{pages}{374}, \bibinfo{doi}{\doi{10.22323/1.084.0374}}.

\bibtype{Article}%
\bibitem{Escribano:2013bca}
\bibinfo{author}{R. Escribano}, \bibinfo{author}{S. Gonzalez-Solis}, \bibinfo{author}{P. Roig}, \bibinfo{title}{{$\tau^-\to K^-\eta^{(\prime)}\nu_\tau$ decays in Chiral Perturbation Theory with Resonances}}, \bibinfo{journal}{JHEP} \bibinfo{volume}{10} (\bibinfo{year}{2013}) \bibinfo{pages}{039}, \bibinfo{doi}{\doi{10.1007/JHEP10(2013)039}}, \eprint{1307.7908}.

\bibtype{Article}%
\bibitem{Gonzalez-Solis:2019iod}
\bibinfo{author}{Sergi Gonz\`alez-Sol\'\i{}s}, \bibinfo{author}{Pablo Roig}, \bibinfo{title}{{A dispersive analysis of the pion vector form factor and $\tau ^{-}\rightarrow K^{-}K_{S}\nu _{\tau }$ decay}}, \bibinfo{journal}{Eur. Phys. J. C} \bibinfo{volume}{79} (\bibinfo{number}{5}) (\bibinfo{year}{2019}) \bibinfo{pages}{436}, \bibinfo{doi}{\doi{10.1140/epjc/s10052-019-6943-9}}, \eprint{1902.02273}.

\bibtype{Article}%
\bibitem{Belle:2008jjb}
\bibinfo{author}{K. Inami}, et al. (\bibinfo{collaboration}{Belle}), \bibinfo{title}{{Precise measurement of hadronic tau-decays with an eta meson}}, \bibinfo{journal}{Phys. Lett. B} \bibinfo{volume}{672} (\bibinfo{year}{2009}) \bibinfo{pages}{209--218}, \bibinfo{doi}{\doi{10.1016/j.physletb.2009.01.047}}, \eprint{0811.0088}.

\bibtype{Article}%
\bibitem{BaBar:2010bul}
\bibinfo{author}{P. del Amo~Sanchez}, et al. (\bibinfo{collaboration}{BaBar}), \bibinfo{title}{{Studies of tau- ---\ensuremath{>} eta K-nu and tau- ---\ensuremath{>} eta pi- nu(tau) at BaBar and a search for a second-class current}}, \bibinfo{journal}{Phys. Rev. D} \bibinfo{volume}{83} (\bibinfo{year}{2011}) \bibinfo{pages}{032002}, \bibinfo{doi}{\doi{10.1103/PhysRevD.83.032002}}, \eprint{1011.3917}.

\bibtype{Article}%
\bibitem{BaBar:2018qry}
\bibinfo{author}{J.~P. Lees}, et al. (\bibinfo{collaboration}{BaBar}), \bibinfo{title}{{Measurement of the spectral function for the $\tau^-\to K^-K_S\nu_{\tau}$ decay}}, \bibinfo{journal}{Phys. Rev. D} \bibinfo{volume}{98} (\bibinfo{number}{3}) (\bibinfo{year}{2018}) \bibinfo{pages}{032010}, \bibinfo{doi}{\doi{10.1103/PhysRevD.98.032010}}, \eprint{1806.10280}.

\bibtype{Article}%
\bibitem{Jadach:1993hs}
\bibinfo{author}{S. Jadach}, \bibinfo{author}{Z. Was}, \bibinfo{author}{R. Decker}, \bibinfo{author}{Johann~H. Kuhn}, \bibinfo{title}{{The tau decay library TAUOLA: Version 2.4}}, \bibinfo{journal}{Comput. Phys. Commun.} \bibinfo{volume}{76} (\bibinfo{year}{1993}) \bibinfo{pages}{361--380}, \bibinfo{doi}{\doi{10.1016/0010-4655(93)90061-G}}.

\bibtype{Article}%
\bibitem{Kuhn:1990ad}
\bibinfo{author}{Johann~H. Kuhn}, \bibinfo{author}{A. Santamaria}, \bibinfo{title}{{Tau decays to pions}}, \bibinfo{journal}{Z. Phys. C} \bibinfo{volume}{48} (\bibinfo{year}{1990}) \bibinfo{pages}{445--452}, \bibinfo{doi}{\doi{10.1007/BF01572024}}.

\bibtype{Article}%
\bibitem{Colangelo:1996hs}
\bibinfo{author}{Gilberto Colangelo}, \bibinfo{author}{Markus Finkemeier}, \bibinfo{author}{Res Urech}, \bibinfo{title}{{Tau decays and chiral perturbation theory}}, \bibinfo{journal}{Phys. Rev. D} \bibinfo{volume}{54} (\bibinfo{year}{1996}) \bibinfo{pages}{4403--4418}, \bibinfo{doi}{\doi{10.1103/PhysRevD.54.4403}}, \eprint{hep-ph/9604279}.

\bibtype{Article}%
\bibitem{GomezDumm:2003ku}
\bibinfo{author}{D. Gomez~Dumm}, \bibinfo{author}{A. Pich}, \bibinfo{author}{J. Portoles}, \bibinfo{title}{{tau ---\ensuremath{>} pi pi pi nu(tau) decays in the resonance effective theory}}, \bibinfo{journal}{Phys. Rev. D} \bibinfo{volume}{69} (\bibinfo{year}{2004}) \bibinfo{pages}{073002}, \bibinfo{doi}{\doi{10.1103/PhysRevD.69.073002}}, \eprint{hep-ph/0312183}.

\bibtype{Article}%
\bibitem{Dumm:2009va}
\bibinfo{author}{D.~Gomez Dumm}, \bibinfo{author}{P. Roig}, \bibinfo{author}{A. Pich}, \bibinfo{author}{J. Portoles}, \bibinfo{title}{{tau ---\ensuremath{>} pi pi pi nu(tau) decays and the a(1)(1260) off-shell width revisited}}, \bibinfo{journal}{Phys. Lett. B} \bibinfo{volume}{685} (\bibinfo{year}{2010}) \bibinfo{pages}{158--164}, \bibinfo{doi}{\doi{10.1016/j.physletb.2010.01.059}}, \eprint{0911.4436}.

\bibtype{Article}%
\bibitem{Girlanda:1999fu}
\bibinfo{author}{Luca Girlanda}, \bibinfo{author}{Jan Stern}, \bibinfo{title}{{The Decay tau ---\ensuremath{>} 3 pi + nu(tau) as a probe of the mechanism of dynamical chiral symmetry breaking}}, \bibinfo{journal}{Nucl. Phys. B} \bibinfo{volume}{575} (\bibinfo{year}{2000}) \bibinfo{pages}{285--312}, \bibinfo{doi}{\doi{10.1016/S0550-3213(00)00068-7}}, \eprint{hep-ph/9906489}.

\bibtype{Article}%
\bibitem{Sanz-Cillero:2017fvr}
\bibinfo{author}{Juan~Jose Sanz-Cillero}, \bibinfo{author}{Olga Shekhovtsova}, \bibinfo{title}{{Refining the scalar and tensor contributions in $\tau\to \pi\pi\pi\nu_\tau$ decays}}, \bibinfo{journal}{JHEP} \bibinfo{volume}{12} (\bibinfo{year}{2017}) \bibinfo{pages}{080}, \bibinfo{doi}{\doi{10.1007/JHEP12(2017)080}}, \eprint{1707.01137}.

\bibtype{Article}%
\bibitem{ALEPH:2005qgp}
\bibinfo{author}{S. Schael}, et al. (\bibinfo{collaboration}{ALEPH}), \bibinfo{title}{{Branching ratios and spectral functions of tau decays: Final ALEPH measurements and physics implications}}, \bibinfo{journal}{Phys. Rept.} \bibinfo{volume}{421} (\bibinfo{year}{2005}) \bibinfo{pages}{191--284}, \bibinfo{doi}{\doi{10.1016/j.physrep.2005.06.007}}, \eprint{hep-ex/0506072}.

\bibtype{Article}%
\bibitem{GomezDumm:2012dpx}
\bibinfo{author}{Daniel Gomez~Dumm}, \bibinfo{author}{Pablo Roig}, \bibinfo{title}{{Resonance Chiral Lagrangian analysis of $\tau^- \to \eta^{(\prime)} \pi^- \pi^0 \nu_\tau$ decays}}, \bibinfo{journal}{Phys. Rev. D} \bibinfo{volume}{86} (\bibinfo{year}{2012}) \bibinfo{pages}{076009}, \bibinfo{doi}{\doi{10.1103/PhysRevD.86.076009}}, \eprint{1208.1212}.

\bibtype{Article}%
\bibitem{Dumm:2009kj}
\bibinfo{author}{D.~Gomez Dumm}, \bibinfo{author}{P. Roig}, \bibinfo{author}{A. Pich}, \bibinfo{author}{J. Portoles}, \bibinfo{title}{{Hadron structure in tau ---\ensuremath{>} KK pi nu (tau) decays}}, \bibinfo{journal}{Phys. Rev. D} \bibinfo{volume}{81} (\bibinfo{year}{2010}) \bibinfo{pages}{034031}, \bibinfo{doi}{\doi{10.1103/PhysRevD.81.034031}}, \eprint{0911.2640}.

\bibtype{Article}%
\bibitem{Arteaga:2022xxy}
\bibinfo{author}{Saray Arteaga}, \bibinfo{author}{Ling-Yun Dai}, \bibinfo{author}{Adolfo Guevara}, \bibinfo{author}{Pablo Roig}, \bibinfo{title}{{Tension between e+e-\textrightarrow{}\ensuremath{\eta}\ensuremath{\pi}-\ensuremath{\pi}+ and \ensuremath{\tau}-\textrightarrow{}\ensuremath{\eta}\ensuremath{\pi}-\ensuremath{\pi}0\ensuremath{\nu}\ensuremath{\tau} data and nonstandard interactions}}, \bibinfo{journal}{Phys. Rev. D} \bibinfo{volume}{106} (\bibinfo{number}{9}) (\bibinfo{year}{2022}) \bibinfo{pages}{096016}, \bibinfo{doi}{\doi{10.1103/PhysRevD.106.096016}}, \eprint{2209.15537}.

\bibtype{Article}%
\bibitem{Ecker:2002cw}
\bibinfo{author}{G. Ecker}, \bibinfo{author}{R. Unterdorfer}, \bibinfo{title}{{Four pion production in e+ e- annihilation}}, \bibinfo{journal}{Eur. Phys. J. C} \bibinfo{volume}{24} (\bibinfo{year}{2002}) \bibinfo{pages}{535--545}, \bibinfo{doi}{\doi{10.1007/s10052-002-0960-8}}, \eprint{hep-ph/0203075}.

\bibtype{Article}%
\bibitem{Kallen:1952zz}
\bibinfo{author}{Gunnar Källén}, \bibinfo{title}{{On the definition of the Renormalization Constants in Quantum Electrodynamics}}, \bibinfo{journal}{Helv. Phys. Acta} \bibinfo{volume}{25} (\bibinfo{number}{4}) (\bibinfo{year}{1952}) \bibinfo{pages}{417}, \bibinfo{doi}{\doi{10.1007/978-3-319-00627-7_90}}.

\bibtype{Article}%
\bibitem{Lehmann:1954xi}
\bibinfo{author}{H. Lehmann}, \bibinfo{title}{{On the Properties of propagation functions and renormalization constants of quantized fields}}, \bibinfo{journal}{Nuovo Cim.} \bibinfo{volume}{11} (\bibinfo{year}{1954}) \bibinfo{pages}{342--357}, \bibinfo{doi}{\doi{10.1007/BF02783624}}.

\bibtype{Book}%
\bibitem{Weinberg:1995mt}
\bibinfo{author}{Steven Weinberg}, \bibinfo{title}{{The Quantum theory of fields. Vol. 1: Foundations}}, \bibinfo{publisher}{Cambridge University Press} \bibinfo{year}{2005}, ISBN \bibinfo{isbn}{978-0-521-67053-1, 978-0-511-25204-4}, \bibinfo{doi}{\doi{10.1017/CBO9781139644167}}.

\bibtype{Inproceedings}%
\bibitem{deRafael:1997ea}
\bibinfo{author}{Eduardo de Rafael}, \bibinfo{title}{{An Introduction to sum rules in QCD: Course}}, in: \bibinfo{booktitle}{{Les Houches Summer School in Theoretical Physics, Session 68: Probing the Standard Model of Particle Interactions}} \bibinfo{year}{1997}, pp. \bibinfo{pages}{1171--1218}, \eprint{hep-ph/9802448}.

\bibtype{Inproceedings}%
\bibitem{Zwicky:2016lka}
\bibinfo{author}{Roman Zwicky}, \bibinfo{title}{{A brief Introduction to Dispersion Relations and Analyticity}}, in: \bibinfo{booktitle}{{Quantum Field Theory at the Limits}: {from Strong Fields to Heavy Quarks}} \bibinfo{year}{2017}, pp. \bibinfo{pages}{93--120}, \bibinfo{doi}{\doi{10.3204/DESY-PROC-2016-04/Zwicky}}, \eprint{1610.06090}.

\bibtype{Phdthesis}%
\bibitem{Gonzalez-Alonso:2010vnm}
\bibinfo{author}{Martin Gonzalez-Alonso}, \bibinfo{title}{{Low-energy tests of the Standard Model}}, \bibinfo{comment}{Ph.D. thesis}, \bibinfo{school}{U. Valencia (main)} \bibinfo{year}{2010}.

\bibtype{Article}%
\bibitem{Davier:2013sfa}
\bibinfo{author}{Michel Davier}, \bibinfo{author}{Andreas H\"ocker}, \bibinfo{author}{Bogdan Malaescu}, \bibinfo{author}{Chang-Zheng Yuan}, \bibinfo{author}{Zhiqing Zhang}, \bibinfo{title}{{Update of the ALEPH non-strange spectral functions from hadronic $\tau$ decays}}, \bibinfo{journal}{Eur. Phys. J. C} \bibinfo{volume}{74} (\bibinfo{number}{3}) (\bibinfo{year}{2014}) \bibinfo{pages}{2803}, \bibinfo{doi}{\doi{10.1140/epjc/s10052-014-2803-9}}, \eprint{1312.1501}.

\bibtype{Article}%
\bibitem{Braaten:1991qm}
\bibinfo{author}{E. Braaten}, \bibinfo{author}{Stephan Narison}, \bibinfo{author}{A. Pich}, \bibinfo{title}{{QCD analysis of the tau hadronic width}}, \bibinfo{journal}{Nucl. Phys. B} \bibinfo{volume}{373} (\bibinfo{year}{1992}) \bibinfo{pages}{581--612}, \bibinfo{doi}{\doi{10.1016/0550-3213(92)90267-F}}.

\bibtype{Article}%
\bibitem{Shifman:1978bx}
\bibinfo{author}{Mikhail~A. Shifman}, \bibinfo{author}{A.~I. Vainshtein}, \bibinfo{author}{Valentin~I. Zakharov}, \bibinfo{title}{{QCD and Resonance Physics. Theoretical Foundations}}, \bibinfo{journal}{Nucl. Phys. B} \bibinfo{volume}{147} (\bibinfo{year}{1979}) \bibinfo{pages}{385--447}, \bibinfo{doi}{\doi{10.1016/0550-3213(79)90022-1}}.

\bibtype{Article}%
\bibitem{Baikov:2008jh}
\bibinfo{author}{P.~A. Baikov}, \bibinfo{author}{K.~G. Chetyrkin}, \bibinfo{author}{Johann~H. Kuhn}, \bibinfo{title}{{Order alpha**4(s) QCD Corrections to Z and tau Decays}}, \bibinfo{journal}{Phys. Rev. Lett.} \bibinfo{volume}{101} (\bibinfo{year}{2008}) \bibinfo{pages}{012002}, \bibinfo{doi}{\doi{10.1103/PhysRevLett.101.012002}}, \eprint{0801.1821}.

\bibtype{Article}%
\bibitem{Trueman:1979en}
\bibinfo{author}{T.~L. Trueman}, \bibinfo{title}{{Chiral Symmetry in Perturbative {QCD}}}, \bibinfo{journal}{Phys. Lett. B} \bibinfo{volume}{88} (\bibinfo{year}{1979}) \bibinfo{pages}{331--334}, \bibinfo{doi}{\doi{10.1016/0370-2693(79)90480-5}}.

\bibtype{Article}%
\bibitem{Pich:2022tca}
\bibinfo{author}{Antonio Pich}, \bibinfo{author}{Antonio Rodr\'\i{}guez-S\'anchez}, \bibinfo{title}{{Violations of quark-hadron duality in low-energy determinations of \ensuremath{\alpha}$_{s}$}}, \bibinfo{journal}{JHEP} \bibinfo{volume}{07} (\bibinfo{year}{2022}) \bibinfo{pages}{145}, \bibinfo{doi}{\doi{10.1007/JHEP07(2022)145}}, \eprint{2205.07587}.

\bibtype{Article}%
\bibitem{DallaBrida:2022eua}
\bibinfo{author}{Mattia Dalla~Brida}, \bibinfo{author}{Roman H\"ollwieser}, \bibinfo{author}{Francesco Knechtli}, \bibinfo{author}{Tomasz Korzec}, \bibinfo{author}{Alessandro Nada}, \bibinfo{author}{Alberto Ramos}, \bibinfo{author}{Stefan Sint}, \bibinfo{author}{Rainer Sommer} (\bibinfo{collaboration}{ALPHA}), \bibinfo{title}{{Determination of $\alpha _s(m_Z)$ by the non-perturbative decoupling method}}, \bibinfo{journal}{Eur. Phys. J. C} \bibinfo{volume}{82} (\bibinfo{number}{12}) (\bibinfo{year}{2022}) \bibinfo{pages}{1092}, \bibinfo{doi}{\doi{10.1140/epjc/s10052-022-10998-3}}, \eprint{2209.14204}.

\bibtype{Article}%
\bibitem{Brida:2025gii}
\bibinfo{author}{Mattia~Dalla Brida}, \bibinfo{author}{Roman H\"ollwieser}, \bibinfo{author}{Francesco Knechtli}, \bibinfo{author}{Tomasz Korzec}, \bibinfo{author}{Alberto Ramos}, \bibinfo{author}{Stefan Sint}, \bibinfo{author}{Rainer Sommer}, \bibinfo{title}{{The strength of the interaction between quarks and gluons}}  (\bibinfo{year}{2025}), \eprint{2501.06633}.

\bibtype{Article}%
\bibitem{LeDiberder:1992jjr}
\bibinfo{author}{F. Le~Diberder}, \bibinfo{author}{A. Pich}, \bibinfo{title}{{The perturbative QCD prediction to R(tau) revisited}}, \bibinfo{journal}{Phys. Lett. B} \bibinfo{volume}{286} (\bibinfo{year}{1992}) \bibinfo{pages}{147--152}, \bibinfo{doi}{\doi{10.1016/0370-2693(92)90172-Z}}.

\bibtype{Book}%
\bibitem{Pascual:1984zb}
\bibinfo{author}{P. Pascual}, \bibinfo{author}{R. Tarrach}, \bibinfo{title}{{QCD: RENORMALIZATION FOR THE PRACTITIONER}}, \bibinfo{comment}{vol.} \bibinfo{volume}{194} \bibinfo{year}{1984}.

\bibtype{Article}%
\bibitem{Chibisov:1996wf}
\bibinfo{author}{Boris Chibisov}, \bibinfo{author}{R.~David Dikeman}, \bibinfo{author}{Mikhail~A. Shifman}, \bibinfo{author}{N. Uraltsev}, \bibinfo{title}{{Operator product expansion, heavy quarks, QCD duality and its violations}}, \bibinfo{journal}{Int. J. Mod. Phys. A} \bibinfo{volume}{12} (\bibinfo{year}{1997}) \bibinfo{pages}{2075--2133}, \bibinfo{doi}{\doi{10.1142/S0217751X97001316}}, \eprint{hep-ph/9605465}.

\bibtype{Article}%
\bibitem{Cata:2008ye}
\bibinfo{author}{Oscar Cata}, \bibinfo{author}{Maarten Golterman}, \bibinfo{author}{Santi Peris}, \bibinfo{title}{{Unraveling duality violations in hadronic tau decays}}, \bibinfo{journal}{Phys. Rev. D} \bibinfo{volume}{77} (\bibinfo{year}{2008}) \bibinfo{pages}{093006}, \bibinfo{doi}{\doi{10.1103/PhysRevD.77.093006}}, \eprint{0803.0246}.

\bibtype{Article}%
\bibitem{Gonzalez-Alonso:2010kpl}
\bibinfo{author}{Martin Gonzalez-Alonso}, \bibinfo{author}{Antonio Pich}, \bibinfo{author}{Joaquim Prades}, \bibinfo{title}{{Violation of Quark-Hadron Duality and Spectral Chiral Moments in QCD}}, \bibinfo{journal}{Phys. Rev. D} \bibinfo{volume}{81} (\bibinfo{year}{2010}) \bibinfo{pages}{074007}, \bibinfo{doi}{\doi{10.1103/PhysRevD.81.074007}}, \eprint{1001.2269}.

\bibtype{Article}%
\bibitem{Boito:2017cnp}
\bibinfo{author}{Diogo Boito}, \bibinfo{author}{Irinel Caprini}, \bibinfo{author}{Maarten Golterman}, \bibinfo{author}{Kim Maltman}, \bibinfo{author}{Santiago Peris}, \bibinfo{title}{{Hyperasymptotics and quark-hadron duality violations in QCD}}, \bibinfo{journal}{Phys. Rev. D} \bibinfo{volume}{97} (\bibinfo{number}{5}) (\bibinfo{year}{2018}) \bibinfo{pages}{054007}, \bibinfo{doi}{\doi{10.1103/PhysRevD.97.054007}}, \eprint{1711.10316}.

\bibtype{Article}%
\bibitem{Pich:2016bdg}
\bibinfo{author}{Antonio Pich}, \bibinfo{author}{Antonio Rodr\'\i{}guez-S\'anchez}, \bibinfo{title}{{Determination of the QCD coupling from ALEPH $\tau$ decay data}}, \bibinfo{journal}{Phys. Rev. D} \bibinfo{volume}{94} (\bibinfo{number}{3}) (\bibinfo{year}{2016}) \bibinfo{pages}{034027}, \bibinfo{doi}{\doi{10.1103/PhysRevD.94.034027}}, \eprint{1605.06830}.

\bibtype{Article}%
\bibitem{Boito:2020xli}
\bibinfo{author}{Diogo Boito}, \bibinfo{author}{Maarten Golterman}, \bibinfo{author}{Kim Maltman}, \bibinfo{author}{Santiago Peris}, \bibinfo{author}{Marcus~V. Rodrigues}, \bibinfo{author}{Wilder Schaaf}, \bibinfo{title}{{Strong coupling from an improved $\tau$ vector isovector spectral function}}, \bibinfo{journal}{Phys. Rev. D} \bibinfo{volume}{103} (\bibinfo{number}{3}) (\bibinfo{year}{2021}) \bibinfo{pages}{034028}, \bibinfo{doi}{\doi{10.1103/PhysRevD.103.034028}}, \eprint{2012.10440}.

\bibtype{Article}%
\bibitem{Ayala:2022cxo}
\bibinfo{author}{Cesar Ayala}, \bibinfo{author}{Gorazd Cvetic}, \bibinfo{author}{Diego Teca}, \bibinfo{title}{{Borel\textendash{}Laplace sum rules with \ensuremath{\tau} decay data, using OPE with improved anomalous dimensions}}, \bibinfo{journal}{J. Phys. G} \bibinfo{volume}{50} (\bibinfo{number}{4}) (\bibinfo{year}{2023}) \bibinfo{pages}{045004}, \bibinfo{doi}{\doi{10.1088/1361-6471/acbd65}}, \eprint{2206.05631}.

\bibtype{Article}%
\bibitem{Shifman:1995qj}
\bibinfo{author}{Mikhail~A. Shifman}, \bibinfo{title}{{The Case of alpha-s: Z versus low-energies, or how nature prompts us of new physics}}, \bibinfo{journal}{Int. J. Mod. Phys. A} \bibinfo{volume}{11} (\bibinfo{year}{1996}) \bibinfo{pages}{3195--3226}, \bibinfo{doi}{\doi{10.1142/S0217751X9600153X}}, \eprint{hep-ph/9511469}.

\bibtype{Article}%
\bibitem{Gamiz:2006xx}
\bibinfo{author}{Elvira Gamiz}, \bibinfo{author}{Matthias Jamin}, \bibinfo{author}{Antonio Pich}, \bibinfo{author}{Joaquim Prades}, \bibinfo{author}{Felix Schwab}, \bibinfo{title}{{|V(us)| and m(s) from hadronic tau decays}}, \bibinfo{journal}{Nucl. Phys. B Proc. Suppl.} \bibinfo{volume}{169} (\bibinfo{year}{2007}) \bibinfo{pages}{85--89}, \bibinfo{doi}{\doi{10.1016/j.nuclphysbps.2007.02.053}}, \eprint{hep-ph/0612154}.

\bibtype{Article}%
\bibitem{Evangelista:2023fmt}
\bibinfo{author}{Antonio Evangelista}, \bibinfo{author}{Roberto Frezzotti}, \bibinfo{author}{Nazario Tantalo}, \bibinfo{author}{Giuseppe Gagliardi}, \bibinfo{author}{Francesco Sanfilippo}, \bibinfo{author}{Silvano Simula}, \bibinfo{author}{Vittorio Lubicz} (\bibinfo{collaboration}{Extended Twisted Mass}), \bibinfo{title}{{Inclusive hadronic decay rate of the \ensuremath{\tau} lepton from lattice QCD}}, \bibinfo{journal}{Phys. Rev. D} \bibinfo{volume}{108} (\bibinfo{number}{7}) (\bibinfo{year}{2023}) \bibinfo{pages}{074513}, \bibinfo{doi}{\doi{10.1103/PhysRevD.108.074513}}, \eprint{2308.03125}.

\bibtype{Article}%
\bibitem{ExtendedTwistedMass:2024myu}
\bibinfo{author}{Constantia Alexandrou}, et al. (\bibinfo{collaboration}{Extended Twisted Mass}), \bibinfo{title}{{Inclusive Hadronic Decay Rate of the \ensuremath{\tau} Lepton from Lattice QCD: The u\textasciimacron{}s Flavor Channel and the Cabibbo Angle}}, \bibinfo{journal}{Phys. Rev. Lett.} \bibinfo{volume}{132} (\bibinfo{number}{26}) (\bibinfo{year}{2024}) \bibinfo{pages}{261901}, \bibinfo{doi}{\doi{10.1103/PhysRevLett.132.261901}}, \eprint{2403.05404}.

\bibtype{Article}%
\bibitem{Weinberg:1967kj}
\bibinfo{author}{Steven Weinberg}, \bibinfo{title}{{Precise relations between the spectra of vector and axial vector mesons}}, \bibinfo{journal}{Phys. Rev. Lett.} \bibinfo{volume}{18} (\bibinfo{year}{1967}) \bibinfo{pages}{507--509}, \bibinfo{doi}{\doi{10.1103/PhysRevLett.18.507}}.

\bibtype{Article}%
\bibitem{tHooft:1973alw}
\bibinfo{author}{Gerard 't Hooft}, \bibinfo{title}{{A Planar Diagram Theory for Strong Interactions}}, \bibinfo{journal}{Nucl. Phys. B} \bibinfo{volume}{72} (\bibinfo{year}{1974}) \bibinfo{pages}{461}, \bibinfo{doi}{\doi{10.1016/0550-3213(74)90154-0}}.

\bibtype{Article}%
\bibitem{Witten:1979kh}
\bibinfo{author}{Edward Witten}, \bibinfo{title}{{Baryons in the 1/n Expansion}}, \bibinfo{journal}{Nucl. Phys. B} \bibinfo{volume}{160} (\bibinfo{year}{1979}) \bibinfo{pages}{57--115}, \bibinfo{doi}{\doi{10.1016/0550-3213(79)90232-3}}.

\bibtype{Article}%
\bibitem{Pich:2021yll}
\bibinfo{author}{Antonio Pich}, \bibinfo{author}{Antonio Rodr\'\i{}guez-S\'anchez}, \bibinfo{title}{{SU(3) analysis of four-quark operators: $K\to\pi\pi$ and vacuum matrix elements}}, \bibinfo{journal}{JHEP} \bibinfo{volume}{06} (\bibinfo{year}{2021}) \bibinfo{pages}{005}, \bibinfo{doi}{\doi{10.1007/JHEP06(2021)005}}, \eprint{2102.09308}.

\bibtype{Article}%
\bibitem{Donoghue:1999ku}
\bibinfo{author}{John~F. Donoghue}, \bibinfo{author}{Eugene Golowich}, \bibinfo{title}{{Dispersive calculation of B(3/2)(7) and B(3/2)(8) in the chiral limit}}, \bibinfo{journal}{Phys. Lett. B} \bibinfo{volume}{478} (\bibinfo{year}{2000}) \bibinfo{pages}{172--184}, \bibinfo{doi}{\doi{10.1016/S0370-2693(00)00239-2}}, \eprint{hep-ph/9911309}.

\bibtype{Article}%
\bibitem{Boito:2015fra}
\bibinfo{author}{Diogo Boito}, \bibinfo{author}{Anthony Francis}, \bibinfo{author}{Maarten Golterman}, \bibinfo{author}{Renwick Hudspith}, \bibinfo{author}{Randy Lewis}, \bibinfo{author}{Kim Maltman}, \bibinfo{author}{Santiago Peris}, \bibinfo{title}{{Low-energy constants and condensates from ALEPH hadronic \ensuremath{\tau} decay data}}, \bibinfo{journal}{Phys. Rev. D} \bibinfo{volume}{92} (\bibinfo{number}{11}) (\bibinfo{year}{2015}) \bibinfo{pages}{114501}, \bibinfo{doi}{\doi{10.1103/PhysRevD.92.114501}}, \eprint{1503.03450}.

\bibtype{Article}%
\bibitem{Gonzalez-Alonso:2016ndl}
\bibinfo{author}{Martin Gonz\'alez-Alonso}, \bibinfo{author}{Antonio Pich}, \bibinfo{author}{Antonio Rodr\'\i{}guez-S\'anchez}, \bibinfo{title}{{Updated determination of chiral couplings and vacuum condensates from hadronic $\tau$ decay data}}, \bibinfo{journal}{Phys. Rev. D} \bibinfo{volume}{94} (\bibinfo{number}{1}) (\bibinfo{year}{2016}) \bibinfo{pages}{014017}, \bibinfo{doi}{\doi{10.1103/PhysRevD.94.014017}}, \eprint{1602.06112}.

\bibtype{Article}%
\bibitem{Cirigliano:2009wk}
\bibinfo{author}{Vincenzo Cirigliano}, \bibinfo{author}{James Jenkins}, \bibinfo{author}{Martin Gonzalez-Alonso}, \bibinfo{title}{{Semileptonic decays of light quarks beyond the Standard Model}}, \bibinfo{journal}{Nucl. Phys. B} \bibinfo{volume}{830} (\bibinfo{year}{2010}) \bibinfo{pages}{95--115}, \bibinfo{doi}{\doi{10.1016/j.nuclphysb.2009.12.020}}, \eprint{0908.1754}.

\bibtype{Article}%
\bibitem{Grzadkowski:2010es}
\bibinfo{author}{B. Grzadkowski}, \bibinfo{author}{M. Iskrzynski}, \bibinfo{author}{M. Misiak}, \bibinfo{author}{J. Rosiek}, \bibinfo{title}{{Dimension-Six Terms in the Standard Model Lagrangian}}, \bibinfo{journal}{JHEP} \bibinfo{volume}{10} (\bibinfo{year}{2010}) \bibinfo{pages}{085}, \bibinfo{doi}{\doi{10.1007/JHEP10(2010)085}}, \eprint{1008.4884}.

\bibtype{Article}%
\bibitem{Garces:2017jpz}
\bibinfo{author}{E.~A. Garc\'es}, \bibinfo{author}{M. Hern\'andez~Villanueva}, \bibinfo{author}{G. L\'opez~Castro}, \bibinfo{author}{P. Roig}, \bibinfo{title}{{Effective-field theory analysis of the $\tau^- \to \eta^{(\prime)} \pi^- \nu_\tau$ decays}}, \bibinfo{journal}{JHEP} \bibinfo{volume}{12} (\bibinfo{year}{2017}) \bibinfo{pages}{027}, \bibinfo{doi}{\doi{10.1007/JHEP12(2017)027}}, \eprint{1708.07802}.

\bibtype{Article}%
\bibitem{Miranda:2018cpf}
\bibinfo{author}{J.~A. Miranda}, \bibinfo{author}{P. Roig}, \bibinfo{title}{{Effective-field theory analysis of the $\tau^-\to \pi^-\pi^0\nu_\tau$ decays}}, \bibinfo{journal}{JHEP} \bibinfo{volume}{11} (\bibinfo{year}{2018}) \bibinfo{pages}{038}, \bibinfo{doi}{\doi{10.1007/JHEP11(2018)038}}, \eprint{1806.09547}.

\bibtype{Article}%
\bibitem{Cirigliano:2018dyk}
\bibinfo{author}{Vincenzo Cirigliano}, \bibinfo{author}{Adam Falkowski}, \bibinfo{author}{Mart\'\i{}n Gonz\'alez-Alonso}, \bibinfo{author}{Antonio Rodr\'\i{}guez-S\'anchez}, \bibinfo{title}{{Hadronic \ensuremath{\tau} Decays as New Physics Probes in the LHC Era}}, \bibinfo{journal}{Phys. Rev. Lett.} \bibinfo{volume}{122} (\bibinfo{number}{22}) (\bibinfo{year}{2019}) \bibinfo{pages}{221801}, \bibinfo{doi}{\doi{10.1103/PhysRevLett.122.221801}}, \eprint{1809.01161}.

\bibtype{Article}%
\bibitem{Davidson:2022jai}
\bibinfo{author}{Sacha Davidson}, \bibinfo{author}{Bertrand Echenard}, \bibinfo{author}{Robert~H. Bernstein}, \bibinfo{author}{Julian Heeck}, \bibinfo{author}{David~G. Hitlin}, \bibinfo{title}{{Charged Lepton Flavor Violation}}  (\bibinfo{year}{2022}), \eprint{2209.00142}.

\bibtype{Article}%
\bibitem{Ardu:2022pzk}
\bibinfo{author}{Marco Ardu}, \bibinfo{author}{Sacha Davidson}, \bibinfo{author}{Martin Gorbahn}, \bibinfo{title}{{Sensitivity of \ensuremath{\mu}\textrightarrow{}e processes to \ensuremath{\tau} flavor change}}, \bibinfo{journal}{Phys. Rev. D} \bibinfo{volume}{105} (\bibinfo{number}{9}) (\bibinfo{year}{2022}) \bibinfo{pages}{096040}, \bibinfo{doi}{\doi{10.1103/PhysRevD.105.096040}}, \eprint{2202.09246}.

\bibtype{Article}%
\bibitem{Jenkins:2017jig}
\bibinfo{author}{Elizabeth~E. Jenkins}, \bibinfo{author}{Aneesh~V. Manohar}, \bibinfo{author}{Peter Stoffer}, \bibinfo{title}{{Low-Energy Effective Field Theory below the Electroweak Scale: Operators and Matching}}, \bibinfo{journal}{JHEP} \bibinfo{volume}{03} (\bibinfo{year}{2018}) \bibinfo{pages}{016}, \bibinfo{doi}{\doi{10.1007/JHEP03(2018)016}}, \bibinfo{note}{[Erratum: JHEP 12, 043 (2023)]}, \eprint{1709.04486}.

\bibtype{Article}%
\bibitem{Celis:2014asa}
\bibinfo{author}{Alejandro Celis}, \bibinfo{author}{Vincenzo Cirigliano}, \bibinfo{author}{Emilie Passemar}, \bibinfo{title}{{Model-discriminating power of lepton flavor violating $\tau$ decays}}, \bibinfo{journal}{Phys. Rev. D} \bibinfo{volume}{89} (\bibinfo{number}{9}) (\bibinfo{year}{2014}) \bibinfo{pages}{095014}, \bibinfo{doi}{\doi{10.1103/PhysRevD.89.095014}}, \eprint{1403.5781}.

\bibtype{Article}%
\bibitem{Jenkins:2013wua}
\bibinfo{author}{Elizabeth~E. Jenkins}, \bibinfo{author}{Aneesh~V. Manohar}, \bibinfo{author}{Michael Trott}, \bibinfo{title}{{Renormalization Group Evolution of the Standard Model Dimension Six Operators II: Yukawa Dependence}}, \bibinfo{journal}{JHEP} \bibinfo{volume}{01} (\bibinfo{year}{2014}) \bibinfo{pages}{035}, \bibinfo{doi}{\doi{10.1007/JHEP01(2014)035}}, \eprint{1310.4838}.

\bibtype{Article}%
\bibitem{Alonso:2013hga}
\bibinfo{author}{Rodrigo Alonso}, \bibinfo{author}{Elizabeth~E. Jenkins}, \bibinfo{author}{Aneesh~V. Manohar}, \bibinfo{author}{Michael Trott}, \bibinfo{title}{{Renormalization Group Evolution of the Standard Model Dimension Six Operators III: Gauge Coupling Dependence and Phenomenology}}, \bibinfo{journal}{JHEP} \bibinfo{volume}{04} (\bibinfo{year}{2014}) \bibinfo{pages}{159}, \bibinfo{doi}{\doi{10.1007/JHEP04(2014)159}}, \eprint{1312.2014}.

\bibtype{Article}%
\bibitem{Celis:2013xja}
\bibinfo{author}{Alejandro Celis}, \bibinfo{author}{Vincenzo Cirigliano}, \bibinfo{author}{Emilie Passemar}, \bibinfo{title}{{Lepton flavor violation in the Higgs sector and the role of hadronic $\tau$-lepton decays}}, \bibinfo{journal}{Phys. Rev. D} \bibinfo{volume}{89} (\bibinfo{year}{2014}) \bibinfo{pages}{013008}, \bibinfo{doi}{\doi{10.1103/PhysRevD.89.013008}}, \eprint{1309.3564}.

\bibtype{Article}%
\bibitem{Husek:2020fru}
\bibinfo{author}{Tomas Husek}, \bibinfo{author}{Kevin Monsalvez-Pozo}, \bibinfo{author}{Jorge Portoles}, \bibinfo{title}{{Lepton-flavour violation in hadronic tau decays and $\mu-\tau$ conversion in nuclei}}, \bibinfo{journal}{JHEP} \bibinfo{volume}{01} (\bibinfo{year}{2021}) \bibinfo{pages}{059}, \bibinfo{doi}{\doi{10.1007/JHEP01(2021)059}}, \eprint{2009.10428}.

\bibtype{Article}%
\bibitem{Cirigliano:2021img}
\bibinfo{author}{Vincenzo Cirigliano}, \bibinfo{author}{Kaori Fuyuto}, \bibinfo{author}{Christopher Lee}, \bibinfo{author}{Emanuele Mereghetti}, \bibinfo{author}{Bin Yan}, \bibinfo{title}{{Charged Lepton Flavor Violation at the EIC}}, \bibinfo{journal}{JHEP} \bibinfo{volume}{03} (\bibinfo{year}{2021}) \bibinfo{pages}{256}, \bibinfo{doi}{\doi{10.1007/JHEP03(2021)256}}, \eprint{2102.06176}.

\bibtype{Article}%
\bibitem{Plakias:2023esq}
\bibinfo{author}{I. Plakias}, \bibinfo{author}{O. Sumensari}, \bibinfo{title}{{Lepton flavor violation in semileptonic observables}}, \bibinfo{journal}{Phys. Rev. D} \bibinfo{volume}{110} (\bibinfo{number}{3}) (\bibinfo{year}{2024}) \bibinfo{pages}{035016}, \bibinfo{doi}{\doi{10.1103/PhysRevD.110.035016}}, \eprint{2312.14070}.

\end{thebibliography*}

\end{document}